\pdfoutput=1
\documentclass[preprint,amsmath,amssymb,superscriptaddress]{revtex4-1}
\usepackage{xcolor}
\usepackage{slashed}
\usepackage{tabu}
\usepackage{titlesec}
\usepackage[utf8]{inputenc}
\usepackage{graphicx,epsfig,epstopdf}
\usepackage{bm}
\usepackage{epstopdf}
\usepackage{amssymb,amsmath,amsfonts,latexsym}
\usepackage{dcolumn}
\usepackage{bm}
\usepackage{hyperref}
\usepackage{rotating}
\usepackage[utf8]{inputenc}
\newcommand{\ord}{\mathcal{O}}
\newcommand{\bin}{{\rm bin}}
\newcommand{\av}[1]{\langle #1 \rangle}
\newcommand{\nn}{\nonumber\\}

\newcommand{\bea}{\begin{eqnarray}}
\newcommand{\ena}{\end{eqnarray}}

\usepackage{breqn}             
\makeatletter                  
\let\cat@comma@active\@empty   
\makeatother                   
\begin{document}
\title{Decay $B_c^+ \to D_{(s)}^{(*)+} \ell^+\ell^-$ within covariant confined quark model}%

\author{M. A. Ivanov}
\email{ivanovm@theor.jinr.ru}
\affiliation{Bogoliubov Laboratory of Theoretical Physics, Joint Institute for Nuclear Research, 141980 Dubna, Russia}

\author{J. N. Pandya}
\email{jnpandya-phy@spuvvn.edu}
\affiliation{Department of Physics, Sardar Patel University,  Vallabh Vidyanagar 388120, Gujarat,  India.}

\author{P. Santorelli}
\email{pietro.santorelli@na.infn.it}
\affiliation{Dipartimento di Fisica ``E. Pancini”, Universit\`a di Napoli Federico II - Complesso Universitario di Monte S. Angelo Edificio 6, via Cintia, 80126 Napoli, Italy}
\affiliation{INFN sezione di Napoli - Complesso Universitario di Monte S. Angelo Edificio 6, via Cintia,
80126 Napoli, Italy}

\author{N. R. Soni}
\email{nakulphy@gmail.com}
\thanks{corresponding author}
\affiliation{INFN sezione di Napoli - Complesso Universitario di Monte S. Angelo Edificio 6, via Cintia,
80126 Napoli, Italy}
\affiliation{Department of Physics, Faculty of Science,  The Maharaja Sayajirao University of Baroda, Vadodara 390002, Gujarat,  India}

\date{\today}

\begin{abstract}
We study the rare semileptonic decays of $B_c$ mesons within the effective field theoretical framework of covariant confined quark model. The transition form factors corresponding to $B_c^+ \to D^{(*)+}$ and $B_c^+ \to D_s^{(*)+}$ are computed in the entire $q^2$ range. Using form factors, we compute the branching fractions and compare them with the available theoretical results. We also compute various physical observables such as forward-backward asymmetry, longitudinal and transverse polarizations as well as clean angular observables.
\end{abstract}

\maketitle
\section{Introduction}
\label{sec:introduction}
$B_c$ meson is an interesting meson having both heavy quarks with different flavor and the mass below $B\bar{D}$ threshold.  Sometimes it is also considered to be in the heavy quarkonia sector, however unlike charmonia and bottomonia $B_c$ meson can decay through the weak interactions only. This can also be justified by the lifetime of the $B_c$ meson and consequently,  it serves as one of the best candidate for the hunt of new physics beyond standard model.
Semileptonic decay of $B$ meson corresponding to the transition $b \to c\ell\nu_\ell$ is explored in great depth by experimental facilities worldwide and it is observed that their results are deviating from the standard model predictions  \cite{BaBar:2013mob,Belle:2015qfa,Belle:2016ure,LHCb:2015gmp,LHCb:2017smo,LHCb:2023zxo,Mathad:2023zxi}.
LHCb has also reported semileptonic decay of $B_c$ meson and ratio of branching fractions $R(J/\psi) = \mathcal{B}(B_c \to J/\psi \tau^+\nu_\tau)/\mathcal{B}(B_c \to J/\psi \mu^+\nu_\mu) = 0.71 \pm 0.17 (\mathrm{stat}) \pm 0.18 (\mathrm{syst}))$ \cite{LHCb:2017vlu}. Following this observation, HPQCD collaboration performed the computation of observables corresponding to the lepton flavor universality violation from lattice QCD and determined the ratio $R(J/\psi) = 0.2582(38)$ which is found to be smaller than LHCb measurement at $1.8\sigma$ \cite{Harrison:2020nrv}.

However, rare semileptonic decays are yet to be fully explored by the experimental side.
In the past, several anomalies have been reported in $B \to K^{(*)} \ell \ell$ corresponding to the $b \to s \ell \ell$ channel.
Several experimental facilities worldwide including LHCb, Belle,  BaBar have provided information regarding this channels in great detail \cite{DiCanto:2022icc,Crivellin:2022qcj,BaBar:2014omp}.
Further, $B_s \to \phi \ell\ell$ has also been observed by some of these collaborations.
The key observations include the ratio $R_{K^{(*)}} = \mathcal{B}(B \to K^{(*)}\mu^+\mu^-)/ \mathcal{B}(B \to K^{(*)}e^+e^-)$ and some other observables viz. forward-backward asymmetry, polarization and different angular observables that are deviating by $2 - 3~\sigma$ from the standard model predictions \cite{CMS:2024syx,LHCb:2022qnv,LHCb:2021trn,LHCb:2021lvy,Belle:2019oag,BELLE:2019xld,LHCb:2020lmf,LHCb:2019hip,LHCb:2017avl,LHCb:2015svh,LHCb:2014vgu,LHCb:2013ghj,BaBar:2016wgb,BaBar:2012mrf}.
These observations essentially lead to the search for the new physics beyond standard model which are discussed in the literature using the framework of light cone sum rules \cite{Gubernari:2020eft,Khodjamirian:2010vf}, effective theories \cite{Bordone:2024hui,Alguero:2023jeh,Gubernari:2022hxn,Hurth:2020ehu,Ciuchini:2019usw,Alguero:2018nvb,Descotes-Genon:2014uoa} and several other theoretical approaches.
In recent experimental developments,  simultaneous measurements of $R_K$ and $R_{K^*}$ in low and central $q^2$ range corresponding to $q^2 \in [0.1, 1.1]$ GeV$^2$/c$^4$ and $q^2 \in [1.1, 6]$ GeV$^2$/c$^4$ show very good agreement with standard model predictions at $0.2\sigma$ by LHCb collaboration \cite{Seuthe:2023xek,LHCb:2022vje}.

On the same lines, $b \to d\ell\ell$ can also be a potential mode for the search of new physics.
For the transition corresponding to the $b \to d$,  LHCb and Belle collaborations have provided some important data for the channels $B \to (\rho,\omega, \pi, \eta) \ell^+\ell^-$ and $B_s^0 \to \bar{K}^{*0} \mu^+ \mu^- $ \cite{Belle-II:2024pgs,LHCb:2017lpt,LHCb:2015hsa}.
Further, LHCb collaboration has also provided the relative ratios for $b \to d/b \to s$ transition in the channels  $\mathcal{B}(B^+ \to \pi^+ \mu^+\mu^-)/\mathcal{B}(B^+ \to K^+ \mu^+\mu^-)$ and $\mathcal{B}(B_s^0 \to \bar{K}^{*0} \mu^+\mu^-)/\mathcal{B}(\bar{B}^0 \to \bar{K}^{*0} \mu^+ \mu^-)$ \cite{LHCb:2012de,Aaij:2018jhg}.
Several anomalies regarding these studies are reported in the book \cite{Artuso:2022ijh} and in the review article \cite{London:2021lfn} including references therein.
	 	
All these anomalies can also be tested in the rare semileptonic decay of $B_c$ meson where the decay channels $B_c^+ \to D^{(*)+} \ell^+ \ell^-$ and $B_c^+ \to D_s^{(*)+} \ell^+ \ell^-$ can also prove to be promising candidates for the search towards any new physics beyond the standard model.  These modes are yet to be identified precisely by the experimental facilities as well as by the lattice simulations.
Less availability of data on rare semileptonic branching fractions might be due to the difficulty in probing them in presence of background data of other simultaneous prominent decays.
Also, $B_c$ mesons are produced less frequently in comparison to other bottom mesons and so the higher excited states are not very well observed yet.
Very recently, LHCb collaboration has also set an upper limit for $f_c/f_u \times \mathcal{B}(B_c^+ \to D_s^+ \mu^+\mu^-)$  at confidence level of 95\%  where $f_c$ and $f_u$ are fragmentation fractions of a $B$ meson with a $c$ and $u$ quark respectively in proton-proton collisions \cite{LHCb:2023lyb}.
Whereas on the theoretical front,  semileptonic as well as rare semileptonic decays are explored using various approaches as follows.
Geng \textit{et al}., studied rare semileptonic decays using light front quark model and constituent quark model \cite{Geng:2001vy}.
Azizi \textit{et al}., also studied these channels using the three point QCD sum rules \cite{Azizi:2008vv,Azizi:2008vy}.
Faessler \textit{et al}., also studied the exclusive rare decays $B_c \to D^{(*)}\ell\ell$ using the relativistic quark model \cite{Faessler:2002ut}.
Choi H. M. studied the transition form factors and different physical observables using the light front quark model \cite{Choi:2010ha}.
These rare decays are studied using the perturbative QCD approach \cite{Wang:2014yia} as well as using the framework of single universal extra dimension \cite{Yilmaz:2012ah}.
Rare semileptonic decays of $B$ and $B_c$ mesons are also studied using the relativistic quark model RQM \cite{Ebert:2010dv}.
The transition form factors computed in the relativistic quark model \cite{Ebert:2010dv} are also employed for study of rare semileptonic decays with non universal $Z^\prime$ effect \cite{Maji:2020zlq} as well as using two Higgs doublet model \cite{Maji:2020wer}.
Further, these form factors are also employed for the search of new physics in terms of different observables \cite{Dutta:2019wxo,Mohapatra:2021ynn,Zaki:2023mcw}.

In the present work, we study the rare semileptonic decays for the channels $B_c^+ \to D_{(s)}^{(*)+}\ell^+\ell^-$ for $\ell = e, \mu, \tau$ and $\nu$ which are essentially the transitions corresponding to $b \to d(s) \ell^+\ell^-$. The necessary transition form factors are computed in the entire range of momentum transfer squared by employing the covariant confined quark model (CCQM) with built-in infrared confinement leading to computation of branching fractions.
We also compute some more physical observables such as forward-backward asymmetry, longitudinal and transverse polarizations as well as different other angular observables.
We also compare our findings with the available experimental data and  theoretical predictions.
In the past, we have successfully employed CCQM for predicting various transitions for charm as well as bottom hadrons which shows the ingenuity and reliability of the model \cite{Pandya:2023ldv,Soni:2021fky,Soni:2020bvu,Soni:2020sgn,Ivanov:2019nqd,Soni:2018adu,Soni:2017eug}.

This paper is organised in the following way: After the brief introduction to the subject with the recent literature reports in sec. \ref{sec:introduction}, we introduce theoretical model i.e.  covariant confined quark model (CCQM) in sec. \ref{sec:framework}. In this section, we provide the Wilson coefficients and transition form factors and the relations of various observables such as branching fractions, forward-backward asymmetry, longitudinal and transverse polarizations, and angular observables.  Then in sec.  \ref{sec:result}, we list all the numerical results along with the comparison with other theoretical approaches. Finally, we summarise the present work in sec. \ref{sec:summary}.

\section{Theoretical Framework}
\label{sec:framework}
\begin{table*}[!b]
\caption{Values of the input parameters \cite{PhysRevD.110.030001} and Wilson coefficients \cite{Descotes-Genon:2013vna}.}
\begin{tabular*}{\textwidth}{@{\extracolsep{\fill}}ccccccccc@{}}
\hline\hline
  $m_W$ &  $\sin^2\theta_W $ &  $\alpha(M_Z)$ &
$\bar m_c$ &  $\bar m_b$  &  $\bar m_t$ &   &  &\\
\hline
 $80.41$~GeV & $0.2313$ & $1/128.94$ & $1.27$~GeV & $4.68$~GeV & $173.3$~GeV &
  &  & \\
\hline\hline
 $C_1$ &  $C_2$ &  $C_3$ &  $C_4$ & $C_5$ &  $C_6$ &  $C^{\rm eff}_7$ &
$C_9$ &  $C_{10}$ \\
\hline
 $-0.2632$ & $1.0111$ &  $-0.0055$ & $-0.0806$ & 0.0004 & 0.0009 & $-0.2923$ &
  4.0749 & $-4.3085$ \\
\hline\hline
\label{tab:input}
\end{tabular*}
\end{table*}
Within the standard model (SM), the effective Hamiltonian for the $b\to q\ell^+\ell^-$ decay can be written in terms of the following operators \cite{Buras:1994dj,Kruger:1996dt,Buchalla:1995vs}
\begin{eqnarray}
\mathcal{H}^{SM}_{eff}=-\frac{4G_F}{\sqrt{2}}V^*_{tq}V_{tb} \left\{\sum^{10}_{i=1}C_{i}(\mu)\mathcal{O}_i(\mu)
+\frac{V_{ub}^{*}V_{uq}}{V^{*}_{tb}V_{tq}} \sum^{2}_{i=1}C_{i}(\mu)[\mathcal{O}_{i}(\mu)-\mathcal{O}^{u}_i(\mu)]\right\},
\label{eq:hamiltonian}
\end{eqnarray}
where $q = d$ for $b \to d\ell^+\ell^-$ and $q = s$ for $b \to s\ell^+\ell^-$.

In the above equation, $C_i$ are the Wilson coefficients and the set of local operators $\mathcal{O}_i$  obtained within the SM for~$b \to s \ell^+ \ell^-$ as well for $b \to d \ell^+ \ell^-$ transition  using a standard procedure \cite{Kruger:1996dt,Buchalla:1995vs}.  These operators include current-current operators ($\ord_{1,2}$), QCD penguin operators ($\ord_{3-6}$), dipole operators ($\ord_{7,8}$) and electroweak semileptonic penguin operators ($\ord_{9,10}$).
Within the SM,  these operators $\mathcal{O}_i$  and $\mathcal{O}_i^{u}$ defined as
\begin{eqnarray}
\begin{array}{ll}
\ord_{1}^{u}     =  (\bar{q}_{a_1}\gamma^\mu P_L u_{a_2})
              (\bar{u}_{a_2}\gamma_\mu P_L b_{a_1}),                   &
\ord_{2}^{u}     =  (\bar{q}\gamma^\mu P_L u)  (\bar{u}\gamma_\mu P_L b),
\\[2ex]
\ord_{1}     =  (\bar{q}_{a_1}\gamma^\mu P_L c_{a_2})
              (\bar{c}_{a_2}\gamma_\mu P_L b_{a_1}),                   &
\ord_{2}     =  (\bar{q}\gamma^\mu P_L c)  (\bar{c}\gamma_\mu P_L b),
\\[2ex]
\ord_3     =  (\bar{q}\gamma^\mu P_L  b) \sum_{q^\prime}(\bar{q}^\prime\gamma_\mu P_L q^\prime),  &
\ord_4     =  (\bar{q}_{a_1}\gamma^\mu P_L  b_{a_2})
              \sum_{q^\prime} (\bar{q}^\prime_{a_2}\gamma_\mu P_L q_{a_1}^\prime),
\\[2ex]
\ord_5     =  (\bar{q}\gamma^\mu P_L b)
              \sum_{q^\prime}(\bar{q}^\prime\gamma_\mu P_R q^\prime),            &
\ord_6     =  (\bar{q}_{a_1}\gamma^\mu P_L b_{a_2 })
              \sum_{q^\prime}  (\bar{q}^\prime_{a_2} \gamma_\mu P_R q_{a_1}^\prime),
\\[2ex]
\ord_7     =  \frac{e}{16\pi^2} \bar m_b\,
              (\bar{q} \sigma^{\mu\nu} P_R b) F_{\mu\nu},       &
\ord_8    =  \frac{g}{16\pi^2} \bar m_b\,
              (\bar{q}_{a_1} \sigma^{\mu\nu} P_R {\bf T}_{a_1a_2} b_{a_2})
              {\bf G}_{\mu\nu},
\\[2ex]
\ord_9     = \frac{e^2}{16\pi^2}
             (\bar{q} \gamma^\mu P_L b) (\bar\ell\gamma_\mu \ell),     &
\ord_{10}  = \frac{e^2}{16\pi^2}
             (\bar{q} \gamma^\mu P_L b)  (\bar\ell\gamma_\mu\gamma_5 \ell),
\end{array}
\label{eq:operators}
\end{eqnarray}
Here, $G_{\mu\nu}$ is the gluon field strength,  $F_{\mu\nu}$ is the photon field strength,  $a_{1,2}$ denote the color indices,  $T_{a_1,a_2}$ are the SU(3) color generators,  $P_{L,R}$ are the chirality projection operator and $\mu$ is the renormalization scale.
Matrix element for the channels $B_c \to D_{(s)}^{(*)}\ell^+\ell^-$ can be written as \cite{Buras:1994dj,Kruger:1996dt}
\begin{eqnarray}
\mathcal{M}(B_c \to D_{(s)}^{(*)} \ell^+\ell^-) &=& \frac{G_F\alpha}{\sqrt{2}\pi}V^{*}_{tq}V_{tb}
\Biggl\{C^{\mathrm{eff}}_9 \langle D_{(s)}^{(*)} | \bar{q} \gamma_\mu P_L b | B_c \rangle (\bar{\ell}\gamma^\mu
	\ell) +
		C_{10} \langle D_{(s)}^{(*)} | \bar{q} \gamma_\mu P_L b | B_c \rangle (\bar{\ell}\gamma^\mu\gamma_5\ell)\nonumber\\
			&&- \frac{2\bar m_b}{q^2} C^{\mathrm{eff}}_7
			\langle D_{(s)}^{(*)} | \bar{q} i\sigma^{\mu\nu} q_\nu P_R b | B_c \rangle (\bar{\ell}\gamma^\mu \ell)\Biggl\},\label{quarkM}
\end{eqnarray}

Here $C^{\mathrm{eff}}_{9}(\mu)$ contains the corrections to the four-quark operators $\mathcal{O}_{1-6}$ and
$\mathcal{O}^{u}_{1,2}$ in Eq. (\ref{eq:hamiltonian}) in the form of \cite{Deshpande:1988bd,Jezabek:1988ja,Lim:1988yu,Misiak:1992bc,ODonnell:1991cdx,Ali:1991is,Bobeth:1999mk,Chen:2001zc,Wang:2012ab}
\begin{eqnarray}
C^{\mathrm{eff}}_9(\mu)=\xi_1+\lambda_q^* \xi_2,
\end{eqnarray}
with
\begin{eqnarray}
\xi_1 & = & C_9 + C_0 h^{\mathrm{eff}} (\hat{m}_c, \hat{s}) - \frac12 h(1,\hat{s}) (4 C_3 + 4 C_4 +3 C_5 + C_6) \nn
&&- \frac12 h(0, \hat{s}) (C_3 + 3 C_4) + \frac29 (3 C_3 + C_4 + 3 C_5 + C_6)\\
\xi_2 & = & \Big[h^{\mathrm{eff}} (\hat{m}_c, \hat{s}) - h^{\mathrm{eff}} (\hat{m}_u, \hat{s})\Big] (3C_1+C_2)
\end{eqnarray}
where $C_0\equiv 3 C_1 + C_2 + 3 C_3 + C_4+ 3 C_5 + C_6$ and $\lambda_d = (V_{ub}^*V_{ud})/(V_{tb}^*V_{td})$ for the transition corresponding to the $b \to d \ell^+\ell^-$ and $\lambda_s = (V_{ub}^*V_{us})/(V_{tb}^*V_{ts})$ for the transition corresponding to the $b \to s \ell^+\ell^-$.
Further, the quark-loop contribution is given by
\begin{widetext}
\bea
h(\hat m_q,  \hat{s}) & = & - \frac{8}{9}\ln\hat m_q +
\frac{8}{27} + \frac{4}{9} x
- \frac{2}{9} (2+x) |1-x|^{1/2} \left\{
\begin{array}{ll}
\left( \ln\left| \frac{\sqrt{1-x} + 1}{\sqrt{1-x} - 1}\right| - i\pi
\right), &
\mbox{for } x \equiv \frac{4 \hat m_q^2}{\hat{s}} < 1, \nonumber \\
 & \\
2 \arctan \frac{1}{\sqrt{x-1}}, & \mbox{for } x \equiv \frac
{4 \hat m_q^2}{\hat {s}} > 1,
\end{array}
\right.\nonumber
\ena
\end{widetext}
and
\bea
h(0, \hat{s}) & = & \frac{8}{27} - \frac{4}{9} \ln \hat{s} + \frac{4}{9} i\pi,
\nonumber
\ena

and the functions,
\begin{eqnarray}
h^{\mathrm{eff}} (\hat{m}_c, \hat{s}) & = & h(\hat m_c,  \hat{s})  +  \frac{3 \pi}{\alpha^2 C_0} \sum_{V = J/\psi, \psi(2S), ...} \frac{m_V \mathcal{B}(V \to \ell^+ \ell^-) \Gamma_V}{m_V^2 - q^2 - i m_V \Gamma_V} ,
\label{eq:resonance}
\end{eqnarray}
\begin{eqnarray}
h^{\mathrm{eff}} (\hat{m}_u, \hat{s}) & = & h(\hat m_u,  \hat{s}) + \frac{3 \pi}{\alpha^2 C_0} \sum_{V = \rho^0, \omega, \phi} \frac{m_V \mathcal{B}(V \to \ell^+ \ell^-) \Gamma_V}{m_V^2 - q^2 - i m_V \Gamma_V}
\label{eq:resonance_light}
\end{eqnarray}
where $\hat m_q=\bar m_q/m_1$, $\hat s=q^2/m_1^2$ and $\alpha$ is the coupling constant considered  at $Z$-boson mass.
The nonresonant contribution is computed by ignoring the terms containing the vector resonances in Eq. (\ref{eq:resonance}) and (\ref{eq:resonance_light}).
The masses,  total decay widths and dilepton branching fractions of vector mesons are taken from the PDG book \cite{PhysRevD.110.030001}.
For the present computations, we consider the next-to-next leading order Wilson coefficients from the Ref. \cite{Descotes-Genon:2013vna} which are essentially evaluated at renormalization scale $\mu = 2M_W$ and they are computed by renormalization group equation to the hadronic scale $\mu_b = 4.8$ GeV.  The values of input parameters and Wilson coefficients are tabulated in Tab. \ref{tab:input}.

The transition form factors corresponding to the channels $B_c^+ \to D_{(s)}^{(*)+} \ell^+\ell^-$ are computed in the effective field theoretical framework of CCQM \cite{Efimov:1988yd,Efimov:1993,Ivanov:1999ic,Branz:2009cd,Ivanov:2011aa,Gutsche:2012ze,Ivanov:2019nqd}.
The effective Lagrangian for the interaction between meson and constituent quark can be written in the most common form as
\begin{eqnarray}
\mathcal{L}_{\mathrm{int}}  &=& g_M M(x) \int dx_1\int dx_2 F_M(x;x_1,x_2) \bar{q}_2(x_2) \Gamma_M q_1(x_1) + \mathrm{H.c}.
\label{eq:lagrangian}
\end{eqnarray}
Here, $\Gamma_M$ is the Dirac matrix and can take the values according to the type of meson based on the spin.
$F_M$ is the vertex function which describes the effective finite size of the meson given by
\begin{eqnarray}
F_M (x,x_1, x_2) = \delta (x - w_1 x_1 - w_2 x_2) \Phi_M((x_1 - x_2)^2).
\label{eq:vertex}
\end{eqnarray}
Here, $\Phi_M$ is the correlation function for the constituents with masses $m_{q_{1,2}}$ and mass ratios $w_i = m_{q_i}/(m_{q_1} + m_{q_2})$.
Next, we consider the vertex function to be of the form of simple Gaussian function with the effective finite size of the meson ($\Lambda_M$). The vertex function takes the form
\begin{eqnarray}
\tilde{\Phi}_M(-k^2) = e^{k^2 / \Lambda_M^2}
\label{eq:gaussion}
\end{eqnarray}
Here, quark masses ($m_{1,2}$) and size parameters ($\Lambda_M$) are the model parameters and are listed in Tab.  \ref{tab:parameters}.  A key feature of choosing the vertex function to have the Gaussian form is that it will make the analytical computation easier.
$g_M$ in Eq.  (\ref{eq:lagrangian}) is the coupling constant which characterizes the interaction strength between quarks and the meson. In order to justify the fact that quarks are confined within the hadrons, we use compositeness condition  \cite{Salam:1962,Weinberg:1962} for determination of coupling constants. It is given by setting the renormalization constant ($Z_M$) to be equal to zero as,
\begin{eqnarray}
Z_M = 1 - \frac{3g_M^2}{4 \pi^2} \tilde{\Pi}^\prime_M(m_M^2) = 0.
\label{eq:compositeness}
\end{eqnarray}
\begin{figure}
\centering
\includegraphics{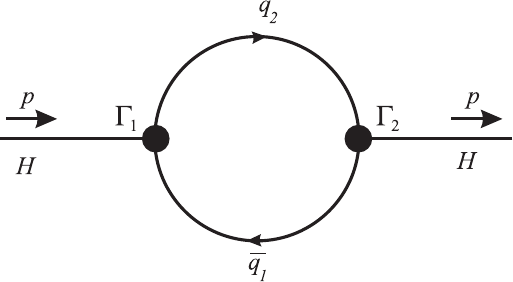}
\caption{Feynman diagram for meson mass operator\label{fig:mass_operator}}
\end{figure}
Here, $\tilde{\Pi}_M$ is the meson mass operator given in Fig.  \ref{fig:mass_operator} defined as
\begin{eqnarray}
\tilde{\Pi}_M(p^2) = N_c g_M^2 \int \frac{d^4k}{(2\pi)^4 i} \tilde{\Phi}_M^2 (-k^2) \mathrm{tr} [\Gamma_M S_1 (k + w_1 p) \Gamma_M S_2(k - w_2p)].
\end{eqnarray}
Here, $N_c = 3$ is the number of colours and $S_{1,2}$ are the free quark propagators that can be written in the form of Fock-Schwinger representation as it provides the additional advantages for the computation of loop integration.
All the necessary analytical computation including trace evaluation, loop integration were performed using the FORM language. At the end, a universal infrared confinement parameter $\lambda = 0.181$ GeV is introduced in order to avoid inclusion of divergences in the quark loop diagrams.
Next, we define the CCQM model parameters such as quark masses ($m_q$) and size parameters $\Lambda_M$.
In present work,  we have utilised the updated least square fit method reported in the past CCQM Ref.  \cite{Ivanov:2015tru,Ganbold:2014pua,Dubnicka:2016nyy,Ivanov:2019nqd}.  In these references, the model parameters are determined by fitting the computed observables such as leptonic decay widths, electromagnetic transition widths and meson masses with the experimental data or lattice simulations and the differences are considered as the absolute uncertainty in the respective size parameters.
It is important to note here that the maximum uncertainty is found to be less than 10\% for the computed form factors at the maximum $q^2$ range.
Further, these uncertainties in the form factors are then transported for the branching fraction computations and other physical observables.

\begin{figure*}
\includegraphics[width=0.45\textwidth]{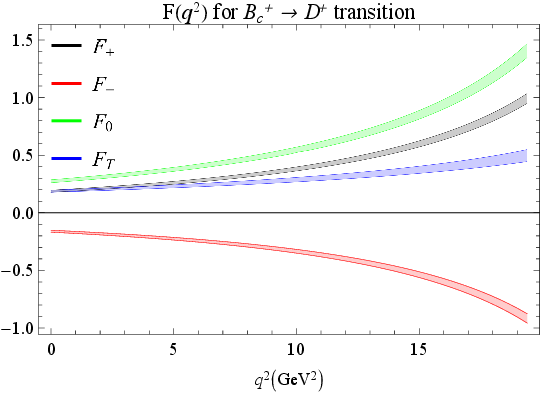}
\hfill\includegraphics[width=0.45\textwidth]{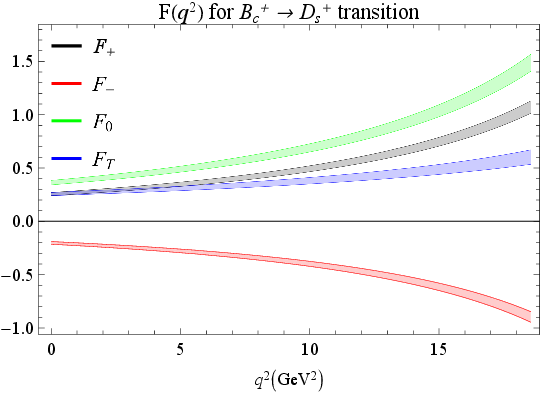}\\
\includegraphics[width=0.45\textwidth]{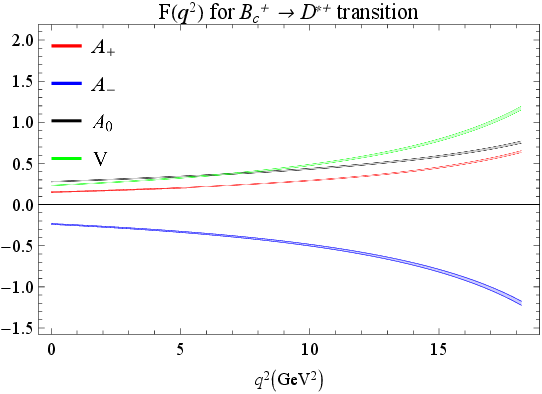}
\hfill\includegraphics[width=0.45\textwidth]{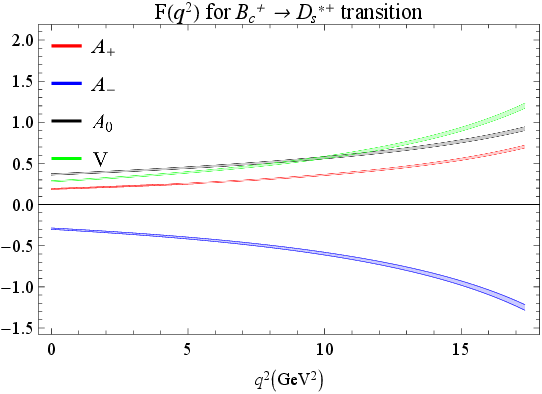}\\
\includegraphics[width=0.45\textwidth]{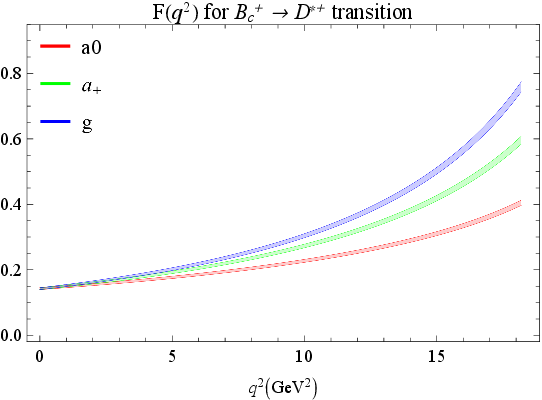}
\hfill\includegraphics[width=0.45\textwidth]{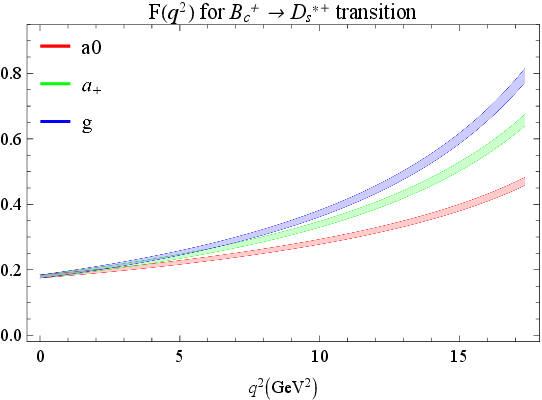}\\
\caption{Form factors\label{fig:form_factor}}
\end{figure*}

With the optimized model parameters and coupling constants, we compute the transition form factors in the whole $q^2$ range.  The transition form factors for the channels $B_c^+ \to D_{(s)}^{(*)+}$ can be written as
\begin{eqnarray}\
\langle D_{(s)} (p_2) | \bar{q} O^\mu b ~|~ B_c (p_1) \rangle
&=& N_c g_{B_c} g_{D_{(s)}} \int \frac{d^4 k}{(2\pi)^4 i} \tilde{\phi}_{B_c} (-(k + w_{13} p_1)^2) \tilde{\phi}_{D_{(s)}}(-(k + w_{23} p_2)^2) \nn & \times & \mathrm{tr}[O^\mu S_1(k + p_1) \gamma^5 S_3(k) \gamma^5 S_2(k + p_2)] \nn
&=& F_+(q^2) P^{\mu} +  F_-(q^2) q^{\mu} ~,\nn
\langle D_{(s)}  (p_2) | \bar{q} \sigma^{\mu\nu} (1 - \gamma^5) b ~|~ B_c (p_1) \rangle  &=&  N_c g_{B_c} g_{D_{(s)}} \int \frac{d^4 k}{(2\pi)^4 i} \tilde{\phi}_{B_c} (-(k + w_{13} p_1)^2) \tilde{\phi}_{D_{(s)}}(-(k + w_{23} p_2)^2) \cr && \times \mathrm{tr}[\sigma^{\mu\nu} (1 - \gamma^5) S_1(k + p_1) \gamma^5 S_3(k) \gamma^5 S_2(k + p_2)] \nn
& = & \frac{i F_T(q^2)}{m_1 + m_2} (P^\mu q^\nu - P^\nu q^\mu + i \varepsilon^{\mu\nu Pq}).
\label{eq:ff_PP}
\end{eqnarray}
\begin{eqnarray}
\langle D_{(s)}^* (p_2, \epsilon)| \bar{q} O^\mu b ~|~ B_c (p_1) \rangle  & = & N_c g_{B_c} g_{D_{(s)}^*} \int \frac{d^4 k}{(2\pi)^4 i} \tilde{\phi}_{B_{c}} (-(k + w_{13} p_1)^2) \tilde{\phi}_{D_{(s)}^*}(-(k + w_{23} p_2)^2) \nn & \times & \mathrm{tr}[O^\mu S_1(k + p_1) \gamma^5 S_3(k) \not\!{\epsilon}_{\nu}^\dag S_2(k + p_2)] \nn & = & \frac{\epsilon_{\nu}^{\dag}}{m_1 + m_2} [ -g^{\mu\nu} P\cdot q A_0(q^2) + P^{\mu} P^{\nu}  A_+(q^2) + q^{\mu} P^{\nu}  A_-(q^2) \nn & + &  i\varepsilon^{\mu\nu\alpha\beta} P_{\alpha} q_{\beta}V(q^2)  ] ~,\nn
\langle D_{(s)}^* (p_2, \epsilon) | \bar{q} \sigma^{\mu\nu} q_\nu (1 + \gamma^5) b ~|~ B_c (p_1) \rangle & = & N_c g_{B_c} g_{D_{(s)}^*} \int \frac{d^4 k}{(2\pi)^4 i} \tilde{\phi}_{B_c} (-(k + w_{13} p_1)^2) \tilde{\phi}_{D_{(s)}^*}(-(k + w_{23} p_2)^2) \nn &\times & \mathrm{tr}[\sigma^{\mu\nu} q_\nu (1 + \gamma^5) S_1(k + p_1) \gamma^5 S_3(k) \not\!{\epsilon}_{\nu}^\dag S_2(k + p_2)] \nn
& = & \epsilon_\nu^\dagger [- (g^{\mu\nu} - q^\mu q^\nu/q^2) P\cdot q a_0(q^2) + i \varepsilon^{\mu\nu\alpha\beta} P_\alpha q_\beta g(q^2)  \nn & + & (P^\mu P^\nu - q^\mu P^\nu P \cdot q/q^2) a_+(q^2)\big].
\label{eq:ff_PV}
\end{eqnarray}
Here, $P$ is the total momentum of the parent and daughter mesons and $q$ is the momentum transfer between them. Polarization vector of the daughter meson is defined in such a way that $\epsilon_\nu^\dag \cdot p_2 = 0$. $p_1^2  = m_{B_c}^2$, $p_2^2  = m_{D_{(s)}^{(*)}}^2$. With all the necessary inputs, the form factors are computed by solving the multidimensional integral using Mathematica in the entire range of momentum transfer squared.
We plot the form factors in Eq. (\ref{eq:ff_PP}) and (\ref{eq:ff_PV}) in Fig. \ref{fig:form_factor} and also represent them in the double pole approximation as
\begin{eqnarray}
F(q^2) = \frac{F(0)}{1 - a \left(\frac{q^2}{m_{1}^2}\right) + b \left(\frac{q^2}{m_{1}^2}\right)^2}
\label{eq:double_pole}
\end{eqnarray}
The form factors at the maximum recoil $F(0)$ and double pole parameters $a$ and $b$ are listed in Tab.  \ref{tab:form_factor}.
It is important to note here that the uncertainty in the form factors for $q^2 = 0$ to $q^2 = q^2_{\mathrm{max}}$ are computed employing the model parameters. Further, uncertainty in the double pole parameters are extracted by performing least square fit with $\chi^2$ to be minimum by employing Minuit algorithm \cite{James:1975dr}.
We also provide the covariance matrix for the coefficients $F(0)$, $a$ and $b$ in Tab.  \ref{tab:covariance_BcD}-\ref{tab:covariance_BcDsv} in Appendix \ref{Appen:covariance} for all the transitions considered here.
\begin{table*}[!htbp]
\caption{Quark masses, meson size parameters and infra-red cut-off parameter (all in GeV)}
\begin{tabular*}{\textwidth}{@{\extracolsep{\fill}}ccccc@{}}
\hline\hline
$\Lambda_{B_c}$ &$\Lambda_{D}$ & $\Lambda_{D^*}$ & $\Lambda_{D_s}$ &$\Lambda_{D_s^*}$ \\
\hline
$2.728 \pm 0.001$ & $1.600 \pm 0.027$ & $1.529 \pm 0.009$ & $1.748 \pm 0.035$ & $1.556 \pm 0.014$  \\
\hline\hline
$m_{u/d}$        &      $m_s$        &      $m_c $      &     $m_b$  & $\lambda$\\
\hline
0.241 & 0.428 & 1.67 & 5.05  &0.181 \\
\hline\hline
\end{tabular*}
\label{tab:parameters}
\end{table*}
\begin{table*}
\caption{Form factors and double pole parameters appeared in Eq. (\ref{eq:double_pole})}
\begin{tabular*}{\textwidth}{@{\extracolsep{\fill}}cccccccc@{}}
\hline\hline
$F$ & $F(0)$ & $a$ & $b$ & $F$ & $F(0)$ & $a$ & $b$\\
\hline
$F_+^{B_c \to D}$ 		& $0.188 \pm 0.003$ & $2.344 \pm 0.012$ & $1.422 \pm 0.027$ & $F_-^{B_c \to D}$ & $-(0.161 \pm 0.002)$  & $2.417 \pm 0.011$ & $1.512 \pm 0.026$ \\
$F_T^{B_c \to D}$ 		& $0.274 \pm 0.004$ & $2.282 \pm 0.012$ & $1.319 \pm 0.028$ &  $F_0^{B_c \to D}$ & $0.187 \pm 0.004$ & $1.487 \pm 0.045$ & $0.458 \pm 0.112$ \\
$A_0^{B_c \to D^*}$ 	& $0.278 \pm 0.002$ & $1.447 \pm 0.010$ & $0.171 \pm 0.026$& $A_+^{B_c \to D^*}$ & $0.152 \pm 0.001$ & $2.155 \pm 0.007$ & $1.089 \pm 0.017$\\
$A_-^{B_c \to D^*}$ 	& $-(0.237 \pm 0.002)$ & $2.409 \pm 0.006$ & $1.459 \pm 0.015$ & $V^{B_c \to D^*}$ & $0.232 \pm 0.002$ & $2.392 \pm 0.006$ & $1.421 \pm 0.015$\\
$a_0^{B_c \to D^*}$ 	& $0.143 \pm 0.001$ & $1.532 \pm 0.010$ & $0.294 \pm 0.025$ & $a_+^{B_c \to D^*}$ & $0.143 \pm 0.009$ & $2.147 \pm 0.007$ & $1.090 \pm 0.018$\\
$g^{B_c \to D^*}$ 		& $0.144 \pm 0.001$ & $2.472 \pm 0.006$ & $1.554 \pm 0.014$\\
\hline
$F_+^{B_c \to D_s}$ 		& $0.256 \pm 0.004$  & $2.204 \pm 0.019$ & $1.253 \pm 0.046$ & $F_-^{B_c \to D_s}$ & $-(0.203 \pm 0.004)$  & $2.265 \pm 0.019$& $1.325 \pm 0.045$ \\
$F_T^{B_c \to D_s}$ 		& $0.363 \pm 0.006$ & $2.152 \pm 0.019$ & $1.167 \pm 0.047$ &  $F_0^{B_c \to D_s}$ & $0.255 \pm 0.006$ & $1.376 \pm 0.057$ & $0.333 \pm 0.149$ \\
$A_0^{B_c \to D_s^*}$ 	& $0.367 \pm 0.003$ & $1.444 \pm 0.016$ & $0.179 \pm 0.041$ & $A_+^{B_c \to D_s^*}$ & $0.191 \pm 0.002$ & $2.116 \pm 0.011$ & $1.049 \pm 0.029$\\
$A_-^{B_c \to D_s^*}$ 	& $-(0.294 \pm 0.003)$ & $2.347 \pm 0.010$ & $1.388 \pm 0.026$ & $V^{B_c \to D_s^*}$ & $0.284 \pm 0.002$ & $2.327 \pm 0.010$ & $1.347 \pm 0.026$\\
$a_0^{B_c \to D_s^*}$ 	& $0.181 \pm 0.002$ & $1.527 \pm 0.015$& $0.295 \pm 0.039$& $a_+^{B_c \to D_s^*}$ & $0.181 \pm 0.002$ & $2.111 \pm 0.011$ & $1.057 \pm 0.029$\\
$g^{B_c \to D_s^*}$ 		& $0.181 \pm 0.002$ & $2.397 \pm 0.010$ & $1.464 \pm 0.025$\\
\hline\hline
\label{tab:form_factor}
\end{tabular*}
\end{table*}

\section{Branching fractions and other physical observables}
Having determined form factors and Wilson coefficients, we compute the differential decay rates for rare semileptonic decays using the relation \cite{Faessler:2002ut}
\begin{eqnarray}
\frac{d\Gamma(B_c \to  D_{(s)}^{(*)} \ell^+ \ell^-)}{dq^2} = \frac{G^2_F}{(2\pi)^3} \left(\frac{\alpha\, V_{tb}^*V_{tq}}{2 \pi}\right)^2  \frac{|{\bf p_2}| q^2 \beta_\ell}{12 m_{1}^2} \mathcal{H}_{\mathrm{tot}}.
\label{eq:branching}
\end{eqnarray}
Here, $\beta_\ell = \sqrt{1 - 4m_\ell^2/q^2}$
and $|{\bf p_2}|=\lambda^{1/2}(m_{1}^2,m_{2}^2,q^2)/(2\,m_{1})$ is the momentum of
the  daughter meson in the rest frame of $B_c$ meson with $\lambda (x,y,z)$ to be the K\"allen function.
Further $m_1 = m_{B_c}$ and $m_2 = m_{D_{(s)}^*}$.
$\mathcal{H}_{\mathrm{tot}}$ is the amplitude given here in terms of the helicity amplitudes:
\begin{eqnarray}
\mathcal{H}_{\mathrm{tot}}  &=&  \frac12 (\mathcal{H}_U^{11} + \mathcal{H}_U^{22} + \mathcal{H}_L^{11} + \mathcal{H}_L^{22}) \nn & + & \left(\frac{2m_\ell^2}{q^2}\right)\left(\frac12 \mathcal{H}_U^{11} - \mathcal{H}_U^{22} + \frac12 \mathcal{H}_L^{11} - \mathcal{H}_L^{22} + \frac32 \mathcal{H}_S^{22} \right).
\label{eq:branching_bilinear}
\end{eqnarray}
The helicity amplitudes here are presented in terms of helicity form factors via the relations for the channels $B_c \to D_{(s)}$ as \cite{Faessler:2002ut,Soni:2020bvu}
\begin{eqnarray}
\mathcal{H}^{ii}_U   &=&  0, \qquad
\mathcal{H}^{ii}_L   = |H^i_{0}|^2, \qquad
\mathcal{H}^{ii}_S   = |H^i_{t0}|^2 .
\end{eqnarray}
with $i = 1,  2$ and these helicity form factors are related to the invariant form factors via
\begin{eqnarray}
H^i_0 & = & \frac{2 m_1 |{\bf p_2}|}{\sqrt{q^2}} \mathcal{F}_+^i, \nn
H^i_{t0} & = & \frac{1}{\sqrt{q^2}} ((m_1^2 - m_2^2) \mathcal{F}_+^i + q^2 \mathcal{F}_-^i).
\end{eqnarray}
The invariant form factors $\mathcal{F}_{+-}^i$ for $i = 1, 2$ are related to the form factors in Eq. (\ref{eq:ff_PP}) as
\begin{eqnarray}
\mathcal{F}_+^1 &=& C_9^{\mathrm{eff}} F_+ + C_7^{\mathrm{eff}} F_T \frac{2 \bar{m_b}}{m_1  +m_2}, \nn
\mathcal{F}_-^1 &=& C_9^{\mathrm{eff}} F_- - C_7^{\mathrm{eff}} F_T \frac{2 \bar{m_b}}{m_1  +m_2} \frac{m_1^2 - m_2^2}{q^2}, \nn
\mathcal{F}_+^2 &=& C_{10} F_+\ \ , \qquad \mathcal{F}_-^2 = C_{10} F_-.
\end{eqnarray}
Similarly,  helicity amplitudes are presented in terms of helicity form factors via relations for the channels $B_c \to D_{(s)}^*$ as \cite{Faessler:2002ut,Soni:2020bvu}
\begin{eqnarray}
\mathcal{H}^{ii}_U   & = &  |H^i_{+1 +1}|^2 +  |H^i_{-1 -1}|^2, \nn
\mathcal{H}^{ii}_L   & = & |H^i_{00}|^2, \qquad
\mathcal{H}^{ii}_S   = |H^i_{t0}|^2 ,
\end{eqnarray}
and these helicity form factors are related to the invariant form factors via
\begin{eqnarray}
H^i_{t0} &=&
\frac{1}{m_1+m_2}\frac{m_1\,|{\bf p_2}|}{m_2\sqrt{q^2}}
         \left(Pq\,(-A^i_0+A^i_+)+q^2 A^i_-\right), \nn
H^i_{\pm1\pm1} &=&
\frac{1}{m_1+m_2}\left(-Pq\, A^i_0\pm 2\,m_1\,|{\bf p_2}|\, V^i \right), \nn
H^i_{00} &=&
\frac{1}{m_1+m_2}\frac{1}{2\,m_2\sqrt{q^2}} \nn & \times &
\left(-Pq\,(m_1^2 - m_2^2 - q^2)\, A^i_0 + 4\,m_1^2\,|{\bf p_2}|^2\, A^i_+\right).
\label{eq:hel_V}
\end{eqnarray}
The invariant form factors $A^i$ and $V^i$ $(i=1,2)$ are related to the form factors in Eq. (\ref{eq:ff_PV}) as
\begin{eqnarray}
V^{(1)} &=&   C_9^{\rm eff}\,V  + C_7^{\rm eff}\,g \,\frac{2\bar m_b(m_1+m_2)}{q^2}\,,
\nn
A_0^{(1)} &=& C_9^{\rm eff}\,A_0
+ C_7^{\rm eff}\,a_0\,\frac{2\bar m_b(m_1+m_2)}{q^2}\,,
\nn
A_+^{(1)} &=& C_9^{\rm eff}\,A_+
+ C_7^{\rm eff}\,a_+\,\frac{2\bar m_b(m_1+m_2)}{q^2}\,,
\nn
A_-^{(1)} &=& C_9^{\rm eff}\,A_-
+ C_7^{\rm eff}\,(a_0-a_+)\,\frac{2\bar m_b(m_1+m_2)}{q^2}\,\frac{Pq}{q^2}\,,
\nn[1.5ex]
V^{(2)}   &=& C_{10}\,V, \qquad A_0^{(2)} = C_{10}\,A_0,\qquad
A_\pm^{(2)} = C_{10}\,A_\pm.
\label{eq:ff-relations}
\end{eqnarray}
Using above relations, we plot differential branching fractions Eq. (\ref{eq:branching}) in Fig. \ref{fig:branching} while computed branching fractions are listed in Tab. \ref{tab:branching}.
We have computed branching fractions corresponding to the dimuon channels also and compared our findings with the results of other reported theoretical results.

The differential decay width for the transition $B_c \to D_{(s)}^{(*)}\nu\bar{\nu}$ can be written as \cite{Faessler:2002ut,Soni:2020bvu}
\begin{eqnarray}
\frac{d\Gamma(B_c^+ \to D_{(s)}^{(*)+} \nu\bar\nu)}{dq^2}
&=& \frac{G_F^2}{(2\pi)^3} \Big(\frac{\alpha \,V_{tb}^*V_{tq}}{2\pi}\Big)^2
\left[\frac{D_\nu(x_t)}{\sin^2\theta_W}\right]^2
\frac{|{\bf p_2}|\, q^2}{4m_1^2}  (H_U+H_L),
\label{eq:rare_nu}
\end{eqnarray}
where $x_t = m_t^2/m_W^2$ and the function $D_\nu(x_t)$ can be written up to NLO correction as \cite{Buchalla:1998ba,Misiak:1999yg,Brod:2010hi,Issadykov:2022iwp}
\begin{eqnarray}
D_\nu(x) = D_0(x) + \frac{\alpha_s}{4\pi} D_1(x)
\end{eqnarray}
with
\begin{eqnarray}
D_0(x) &=& \frac{x}{8}\left(\frac{2+x}{x-1}+\frac{3x-6}{(x-1)^2}\,\ln x\right)\\
D_1(x) &=& - \frac{29 x - x^2 - 4 x^3}{3 (1-x)^2} - \frac{x + 9 x^2 - x^3 - x^4}{(1-x)^3}~ \ln x \nn & + & \frac{8 x + 4 x^2 + x^3 - x^4}{2 (1-x)^3}~\ln^2 x -  \frac{4 x - x^3}{(1-x)^2} \int_1^x dt \frac{\ln t}{1-t} \nn & + &  8 x \frac{\partial D_0(x)}{\partial x} ~\ln \left(\frac{\mu_b^2}{m_W^2}\right).
\end{eqnarray}
As in the previous case, the helicity amplitudes are related to helicity form factors for the channels $B_c^+ \to D_{(s)}^+ \nu\bar{\nu}$ as \cite{Faessler:2002ut,Soni:2020bvu}
\begin{eqnarray}
\mathcal{H}_L = |H_0|^2,  \qquad \mathcal{H}_U = 0
\end{eqnarray}
with
\begin{eqnarray}
H_0 = \frac{2 m_1 |\textbf{p}_2|}{\sqrt{q^2}} F_+.
\label{eq:bilinear_2_PP}
\end{eqnarray}
instead for the channels $B_c^+ \to D_{(s)}^{*+} \nu\bar{\nu}$ as \cite{Faessler:2002ut,Soni:2020bvu}
\begin{eqnarray}
\mathcal{H}_U   &=&  |H_{+1 +1}|^2 +  |H_{-1 -1}|^2, \qquad
\mathcal{H}_L   = |H_{00}|^2,
\label{eq:bilenear_HuHl}
\end{eqnarray}
with
\begin{eqnarray}
H_{\pm1\pm1} &=&
\frac{1}{m_1+m_2}\left(-Pq\, A_0\pm 2\,m_1\,|{\bf p_2}|\, V \right),
\nn
H_{00} &=&
\frac{1}{m_1+m_2}\frac{1}{2\,m_2\sqrt{q^2}}
\left(-Pq\,(m_1^2 - m_2^2 - q^2)\, A_0 + 4\,m_1^2\,|{\bf p_2}|^2\, A_+\right).
\label{eq:bilinear_2_PV}
\end{eqnarray}
The numerical results on the branching fractions are listed in Tab. \ref{tab:branching} in comparison with the results of other theoretical approaches.
\begin{figure*}[htbp]
\includegraphics[width=0.45\textwidth]{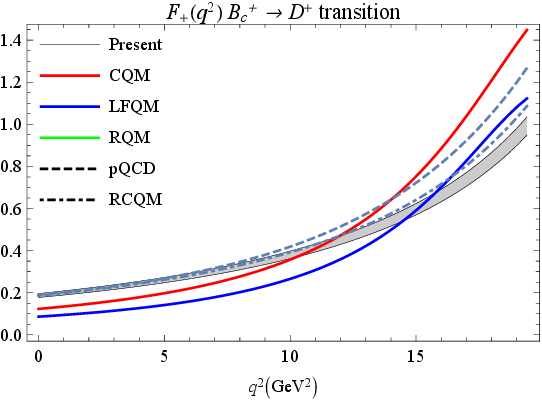}
\hfill\includegraphics[width=0.45\textwidth]{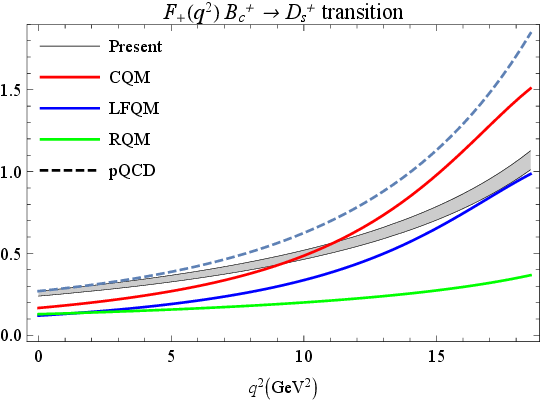}\\
\includegraphics[width=0.45\textwidth]{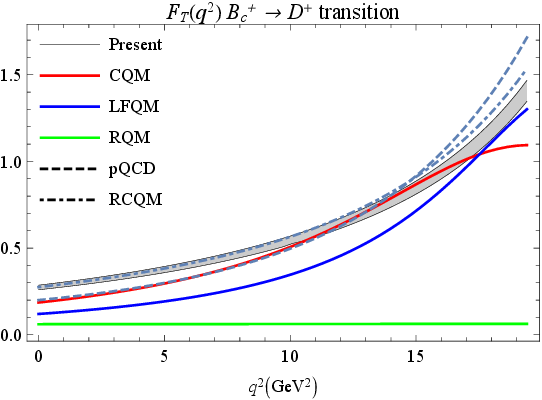}
\hfill\includegraphics[width=0.45\textwidth]{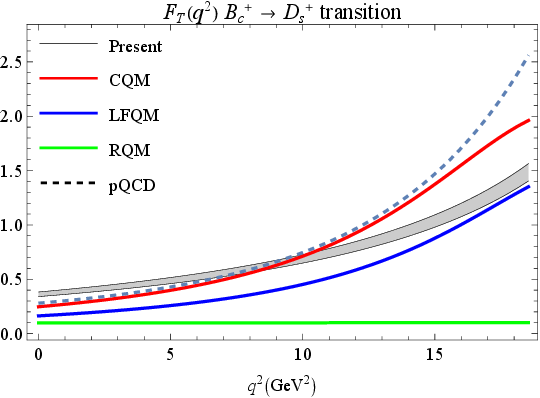}\\
\caption{Form factor comparison for $B_c \to D$ (left) and $B_c \to D_s$ (right) in comparison with relativistic constituent quark model \cite{Faessler:2002ut},  constituent quark model \cite{Geng:2001vy},  light front quark model \cite{Choi:2010ha},  relativistic quark model \cite{Ebert:2010dv} and  perturbative QCD \cite{Wang:2014yia}. \label{fig:fpfT}}
\end{figure*}

\begin{figure*}[htbp]
\includegraphics[width=0.45\textwidth]{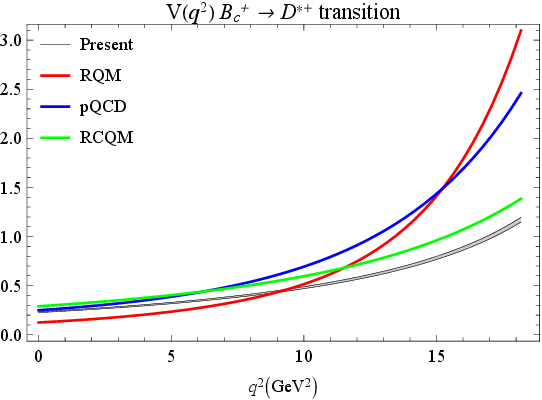}
\hfill\includegraphics[width=0.45\textwidth]{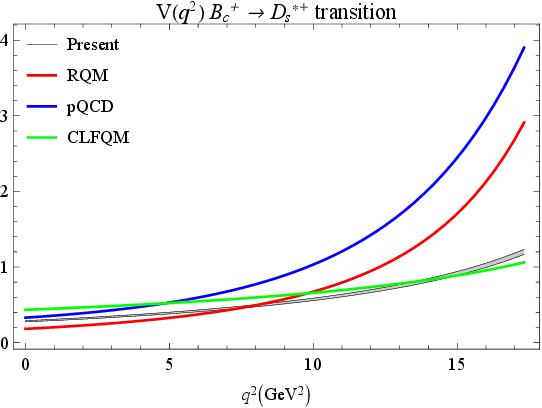}\\
\includegraphics[width=0.45\textwidth]{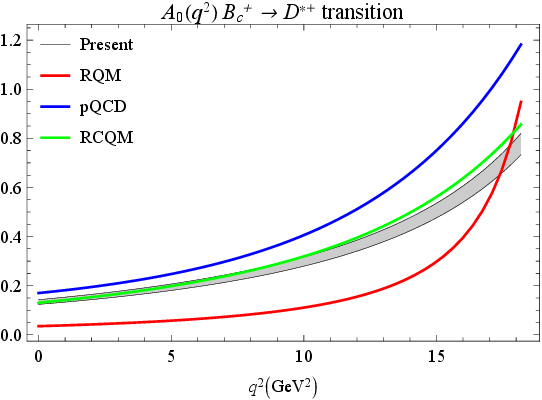}
\hfill\includegraphics[width=0.45\textwidth]{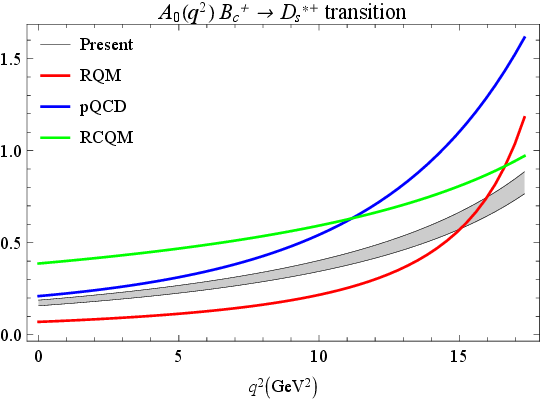}\\
\includegraphics[width=0.45\textwidth]{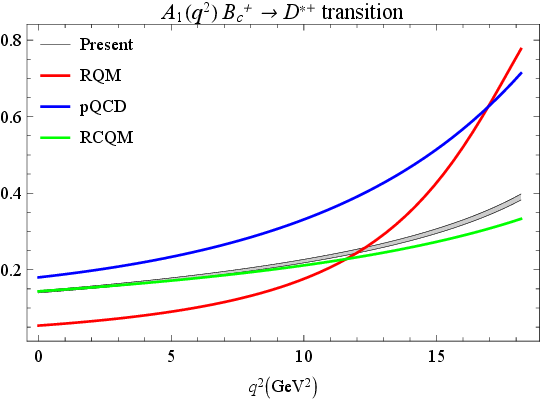}
\hfill\includegraphics[width=0.45\textwidth]{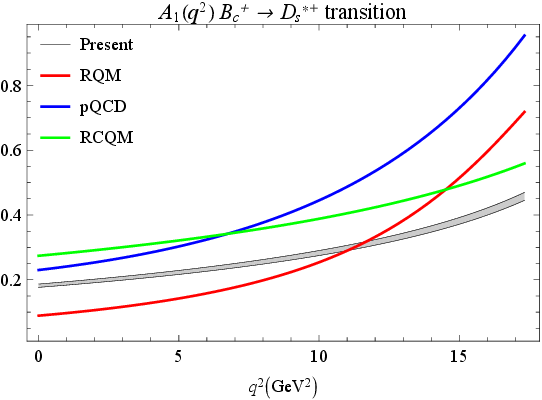}\\
\includegraphics[width=0.45\textwidth]{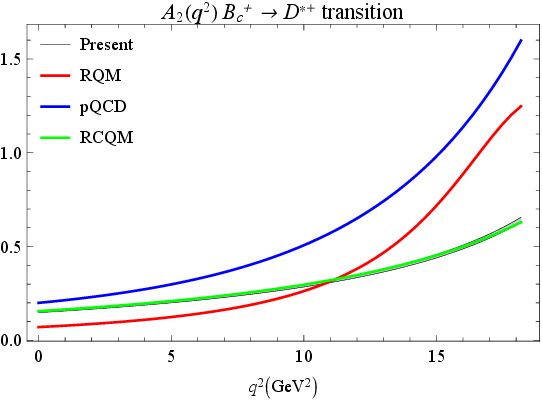}
\hfill\includegraphics[width=0.45\textwidth]{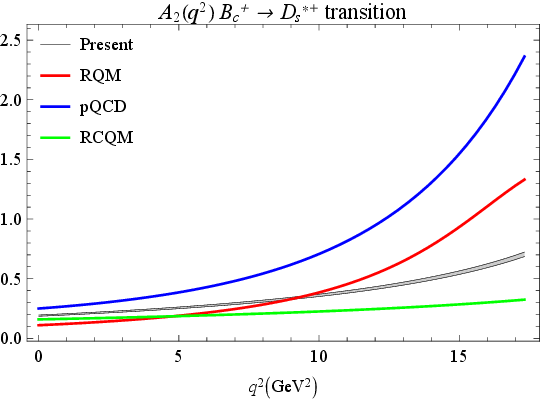}\\
\caption{Form factor comparison for $B_c \to D^*$ (left) and $B_c \to D_s^*$ (right) in comparison with relativistic constituent quark model \cite{Faessler:2002ut},  relativistic quark model \cite{Ebert:2010dv},  perturbative QCD \cite{Wang:2014yia} and covariant light front quark model \cite{Li:2023mrj}.\label{fig:VA1}}
\end{figure*}

\begin{figure*}[htbp]
\includegraphics[width=0.45\textwidth]{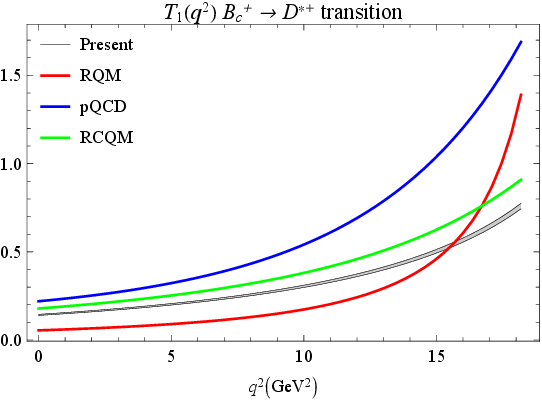}
\hfill\includegraphics[width=0.45\textwidth]{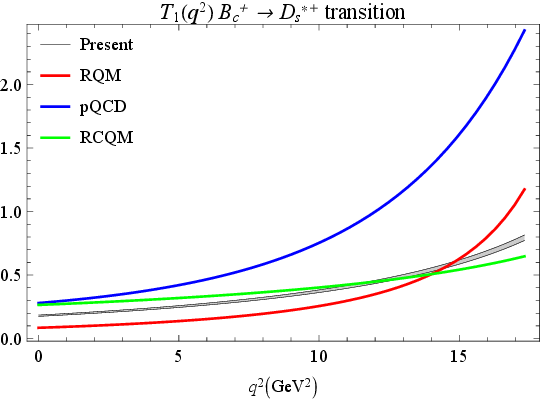}\\
\includegraphics[width=0.45\textwidth]{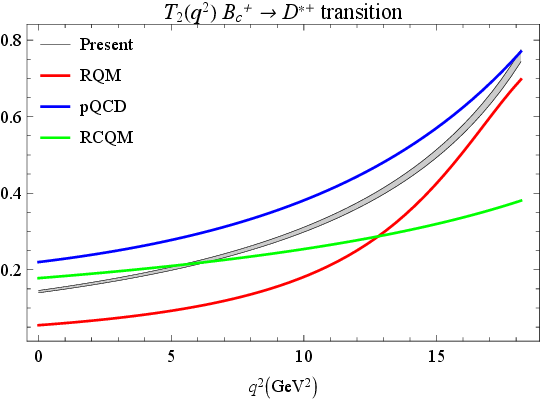}
\hfill\includegraphics[width=0.45\textwidth]{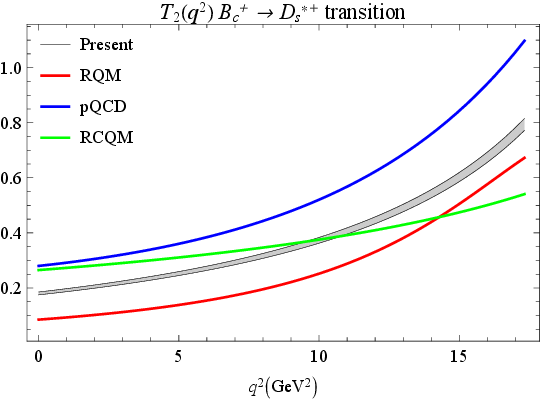}\\
\includegraphics[width=0.45\textwidth]{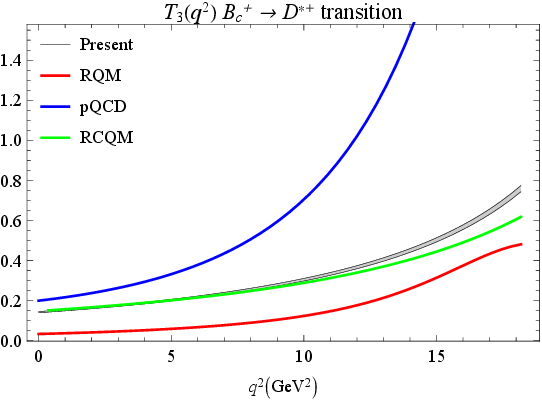}
\hfill\includegraphics[width=0.45\textwidth]{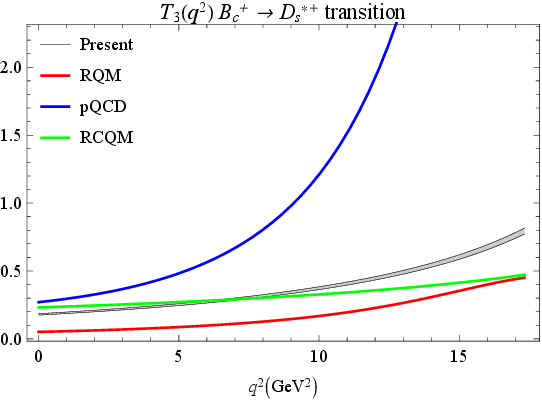}\\
\caption{Form factor comparison for $B_c \to D^*$ (left) and $B_c \to D_s^*$ (right) in comparison with relativistic constituent quark model \cite{Faessler:2002ut},  relativistic quark model \cite{Ebert:2010dv},  perturbative QCD \cite{Wang:2014yia} and covariant light front quark model \cite{Li:2023mrj}.\label{fig:VA2}}
\end{figure*}

We further compute other physical observables such as forward-backward asymmetry, longitudinal and transverse polarizations and other clean observables.
These lepton flavour dependent angular observables are related with the helicity amplitudes as well as corresponding form factors.
Experimental measurements of these observables have played critical role in search for new physics beyond the SM for the transition $b \to s\ell\ell$ corresponding to the channel $B \to K^* \ell\ell$ by LHCb \cite{Aaij:2015oid,LHCb:2020lmf} and Belle collaborations \cite{Wehle:2016yoi}.
Angular analysis are also observed for the channel $B_s \to \phi \mu^+\mu^-$ and $\Lambda_b^0 \to \Lambda \mu^+\mu^-$ by LHCb collaboration \cite{LHCb:2015wdu,LHCb:2018jna}.
However, in other channels, these observables are yet to be identified.
Many of these experimental measurements seem to deviate from the SM predictions and explained using new physics scenario \cite{Arbey:2019duh,Kowalska:2019ley,Alok:2019ufo,Hiller:2014yaa,Hiller:2018wbv,Crivellin:2017zlb,Ko:2017lzd}.
Similar behavior is expected in case of $b \to d \ell\ell$ transitions.
These observables could be helpful in studying the effects of CP violation.
Very recently in Ref. \cite{Li:2023mrj}, these angular observables are computed using the covariant light front quark model where transition form factors are obtained using modified Godfrey-Isgur model for the channel $B_c \to D_s^* \ell\ell$.
The same observables are also studied for the channels $\bar{B}_s \to K^*\ell\ell$ and $\bar{B} \to \rho  \ell\ell$ using light cone sum rule approach \cite{Kindra:2018ayz}.
Recently,  some of us have studied these observables in the $b \to d\ell^+\ell^-$ \cite{Soni:2020bvu} transition.
They are described in terms of four-fold angular distribution and expressed explicitly in terms of helicity form factors.  The detailed description and computation techniques of these observables can be found in the Ref.  \cite{Matias:2012qz,Dubnicka:2015iwg}.
Relation for these observables reads as:
\begin{enumerate}
\item Forward-backward asymmetry
\begin{eqnarray}
A_{\rm FB} &=&
\frac{1}{d\Gamma/dq^2} \left[ \int\limits_0^1 - \int\limits_{-1}^0 \right]
d\!\cos\theta\, \frac{d^2\Gamma}{dq^2 d\!\cos\theta}
= -\frac34\beta_\ell \frac{ {\cal H}_P^{12}} { {\cal H}_{\rm tot} }\,,
\label{eq:AFB}
\end{eqnarray}
\item Longitudinal and Transverse polarization fractions
\begin{eqnarray}
F_L &=&  \frac12 \beta_\ell^2 \frac{ {\cal H}_L^{11} +  {\cal H}_L^{22}}{{\cal H}_{\rm tot} }, \ \ \ \ \ F_T =  \frac12 \beta_\ell^2 \frac{ {\cal H}_U^{11} +  {\cal H}_U^{22}}{{\cal H}_{\rm tot} },.
\end{eqnarray}
\item Clean Observables
\begin{eqnarray}
\av{P_1}_{\rm bin} &=&  -2\, \frac{\int_{{\rm bin}} dq^2\,\beta_\ell^2\,
                                [{\cal H}_T^{11}+ {\cal H}_T^{22}]}
          {\int_{{\rm bin}} dq^2\,\beta_\ell^2\,
                                [{\cal H}_U^{11}+ {\cal H}_U^{22}] }, \nn
\av{P_2}_{\rm bin} &=& -\,\frac{\int_{{\rm bin}} dq^2\,\beta_\ell\, {\cal H}_P^{12}}
          {\int_{{\rm bin}} dq^2\,\beta_\ell^2\,
                                [{\cal H}_U^{11}+ {\cal H}_U^{22}] },
\nn
\av{P_3}_{\rm bin} &=& -\,\frac{\int_{{\rm bin}} dq^2\,\beta_\ell^2\,
                                  [{\cal H}_{IT}^{11}+ {\cal H}_{IT}^{22}]}
          {\int_{{\rm bin}} dq^2\,\beta_\ell^2\,
                                [{\cal H}_U^{11}+ {\cal H}_U^{22}] },\nn
\av{P^\prime_4}_{\rm bin} &=& 2\,\frac{\int_{{\rm bin}} dq^2\,\beta_\ell^2\,
                         [{\cal H}_I^{11}+ {\cal H}_I^{22}]}{N_{\rm bin}},\nn
\av{P^\prime_5}_{\rm bin} &=&  -2\,\frac{\int_{{\rm bin}} dq^2\,\beta_\ell\,
                           [{\cal H}_A^{12}+ {\cal H}_A^{21}]}{N_{\rm bin}},
\nn
\av{P^\prime_6}_{\rm bin} &=&  -2\,\frac{\int_{{\rm bin}} dq^2\,\beta_\ell\,
                         [{\cal H}_{II}^{12}+ {\cal H}_{II}^{21}]}{N_{\rm bin}},
\nn
\av{P^\prime_8}_{\rm bin} &=& 2\,\frac{\int_{{\rm bin}} dq^2\,\beta_\ell^2\,
                         [{\cal H}_{IA}^{11}+ {\cal H}_{IA}^{22}]}{N_{\rm bin}},
\label{eq:binP1-8}
\end{eqnarray}
\end{enumerate}
with the normalization factor $N_{\rm bin}$ defined as
\begin{eqnarray}
{\cal N}_\bin &=& {\textstyle \sqrt{\int_\bin dq^2\,\beta_\ell^2\,
                                 [{\cal H}_U^{11}+{\cal H}_U^{22}]\cdot
                    \int_{{\rm bin}}\,dq^2\,\beta_\ell^2\,
                                 [{\cal H}_L^{11}+{\cal H}_L^{22}]}}.
\end{eqnarray}
For computation of expectation value of these observables over bins, we need to multiply them by the phase space factor $|{\bf p_2}| q^2\beta_\ell$ in corresponding numerator and denominator separately.
The angle appearing in the relation of forward-backward asymmetry is the polar angle between the momentum of parent meson and the momentum transfer.
Further, bins are corresponding to the momentum transfer squared ranges $[1.1, 6.0]$,  $[6.0, 8.0]$, $[11.0, 12.5]$ and $[15.0, 17.0]$ GeV$^2$.

\section{Results and Discussion}
\label{sec:result}
Using covariant confined quark model, we compute the transition form factors using the model parameters in Tab. \ref{tab:parameters} and we plot them in terms of $q^2$ in Fig. \ref{fig:form_factor}.
We also compare our form factors with the results of various other theoretical prediction such as relativistic constituent quark model (RQM) \cite{Faessler:2002ut},  constituent quark model (CQM) \cite{Geng:2001vy}, QCD sum rules (QCDSR) \cite{Azizi:2008vy},  light front quark model (LFQM) \cite{Choi:2010ha},  relativistic quark model (RQM) \cite{Ebert:2010dv},  perturbative QCD (pQCD) \cite{Wang:2014yia}.
However, in order to have the comparison, we need to transform our form factors Eq.  (\ref{eq:ff_PP}) and  (\ref{eq:ff_PV}) with those using BSW form factors \cite{Wirbel:1985ji}. The transformed form factors are denoted using primed symbols.
\begin{eqnarray}
F_0^\prime & = & F_+ + \frac{q^2}{m_1^2 - m_2^2} F_-,\\
A_0 &=& \frac{m_1 + m_2}{m_1 - m_2}\,A_1^\prime\,, \qquad
A_+ = A_2^\prime\,,
\nn
A_- &=&  \frac{2m_2(m_1+m_2)}{q^2}\,(A_3^\prime - A_0^\prime)\,, \qquad
V = V^\prime\,,
\nn[1.2ex]
a_0 &=& T_2^\prime\,, \qquad g = T_1^\prime\,, \qquad
a_+  =  T_2^\prime + \frac{q^2}{m_1^2-m_2^2}\,T_3^\prime\,.
\label{eq:new-ff}
\end{eqnarray}
These form factors also satisfy the constraints
\begin{eqnarray}
 A_0^\prime(0) &=& A_3^\prime(0) \nn
2m_2A_3^\prime(q^2) &=& (m_1+m_2) A_1^\prime(q^2) -(m_1-m_2) A_2^\prime(q^2)\,.
\end{eqnarray}
For the subsequent section, we remove the prime from the form factors in order to avoid confusion.
The graphical comparison between the results can be found in Fig.  \ref{fig:fpfT} - \ref{fig:VA2}.
It is observed that our results of the form factors for the transition $B_c \to D$ are very similar to the pQCD, LFQM, CQM and RCQM predictions.
Whereas, our results are significantly higher with respect to RQM in almost entire $q^2$ range .
For $B_c \to D_s$ transition also, our results are in good agreement with the LFQM and CQM predictions.
Similarly for $B_c \to D_{(s)}^*$ transitions, our form factors are compatible with the RQM and pQCD predictions for the range $q^2 \to q^2_{\mathrm{max}}$, whereas our results are substantially lower for low $q^2$.
The differences in the predictions are mainly attributed to the different methodology employed for the computation of the transition form factors.
\begin{figure*}[htbp]
\includegraphics[width=0.45\textwidth]{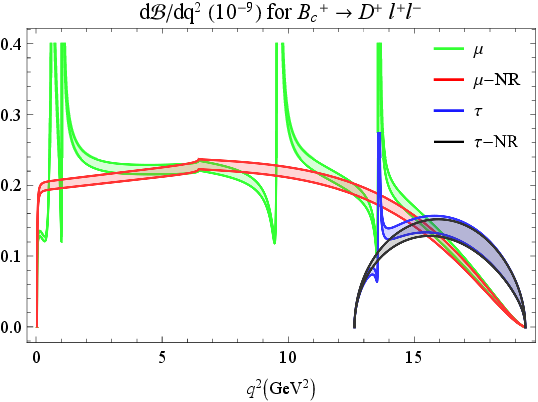}
\hfill\includegraphics[width=0.45\textwidth]{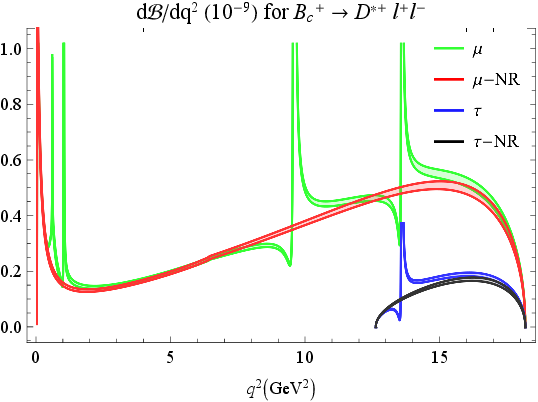}\\
\includegraphics[width=0.45\textwidth]{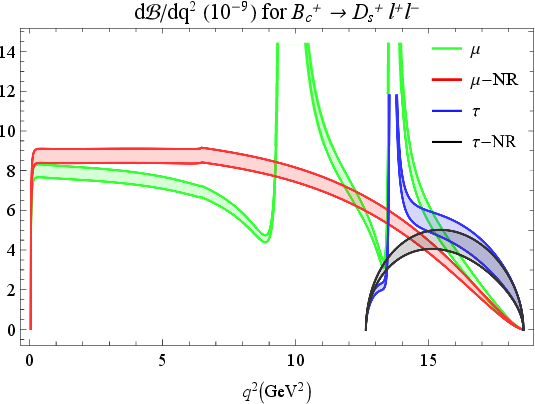}
\hfill\includegraphics[width=0.45\textwidth]{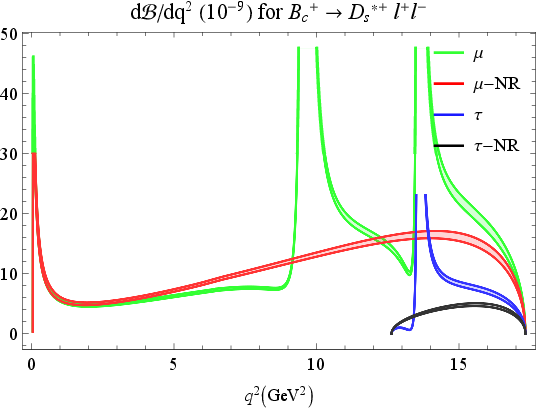}\\
\caption{Differential branching fractions (Green and Blue plots correspond to the inclusion of vector resonance whereas Red and Black plots are without the same (NR).)\label{fig:branching}}
\end{figure*}
\begin{table*}
\caption{Branching fractions of $B_c^+ \to D_{(s)}^{(*)+} \ell^+\ell^-$ for $\ell = e, \mu$ and $\tau$}
\begin{tabular*}{\textwidth}{@{\extracolsep{\fill}}lcccccccccc@{}}
\hline\hline
Channel	&	Without	 & 	With	& 	LFQM \cite{Geng:2001vy}	 & 	CQM \cite{Geng:2001vy}	 & 	pQCD \cite{Wang:2014yia}	 & 	RQM \cite{Ebert:2010dv}				\\
	&	Resonances	&	Resonances	&		&													\\
\hline
$10^9\mathcal{B}(B_c^+ \to D^+ e^+e^-)$	&	$3.457 \pm 0.125$	&	$2.665 \pm 0.065$	&	4.100	&	4.000	&	--	&	--		\\
$10^9\mathcal{B}(B_c^+ \to D^+ \mu^+ \mu^-)$	&	$3.449 \pm 0.125$	&	$2.659 \pm 0.065$	&	4.100	&	4.000	&	3.790	&	3.700		\\
$10^9\mathcal{B}(B_c^+ \to D^+ \tau^+ \tau^-)$	&	$0.730 \pm 0.070$	&	$0.503 \pm 0.056$	&	1.300	&	1.200	&	1.030	&	1.500		\\
$10^8\mathcal{B}(B_c^+ \to D^+ \nu^+ \nu^-)$	&	$1.386 \pm 0.048$	&	--	&	2.770	&	2.740	&	3.130	&	2.160		\\
\hline														
$10^9\mathcal{B}(B_c^+ \to D^{*+} e^+e^-)$	&	$7.031 \pm 0.215$	&	$4.983 \pm 0.095$	&	10.100	&	7.900	&	--	&		--	\\
$10^9\mathcal{B}(B_c^+ \to D^{*+} \mu^+ \mu^-)$	&	$5.934 \pm 0.161$	&	$3.894 \pm 0.070$	&	10.100	&	7.900	&	12.100	&	8.100		\\
$10^9\mathcal{B}(B_c^+ \to D^{*+} \tau^+ \tau^-)$	&	$0.718 \pm 0.025$	&	$0.518 \pm 0.019$	&	1.800	&	1.400	&	1.600	&	1.900		\\
$10^8\mathcal{B}(B_c^+ \to D^{*+} \nu^+ \nu^-)$	&	$2.572 \pm 0.089$	&	--	&	7.640	&	5.990	&	11.000	&	5.120		\\
$10^7\mathcal{B}(B_c^+ \to D^{*+} \gamma)$	& 	$1.235 \pm 0.017$	& 	--	& 	--	& --		& 	--	& 		--	\\
\hline														
$10^7\mathcal{B}(B_c^+ \to D_s^+ e^+e^-)$	&	$1.243 \pm 0.055$			&	$0.797 \pm 0.024$	&	1.360	&	1.330	&	-- 	&	--		\\
$10^7\mathcal{B}(B_c^+ \to D_s^+ \mu^+ \mu^-)$	&	$1.239 \pm 0.055$	&	$0.793 \pm 0.024$	&	1.360	&	1.330	&	1.560	&	1.160		\\
$10^7\mathcal{B}(B_c^+ \to D_s^+ \tau^+ \tau^-)$	&	$0.207 \pm 0.024$	&	$0.136 \pm 0.018$	&	0.340	&	0.370	&	0.380	&	0.330		\\
$10^7\mathcal{B}(B_c^+ \to D_s^+ \nu^+ \nu^-)$	&	$4.979 \pm 0.210$	&	--	&	9.200	&	9.200	&	0.129	&	6.500		\\
\hline														
$10^7\mathcal{B}(B_c^+ \to D_s^{*+} e^+e^-)$	&	$2.311 \pm 0.104$	&	$1.558 \pm 0.055$	&	4.090	&	2.810	&	-- 	&	--		\\
$10^7\mathcal{B}(B_c^+ \to D_s^{*+} \mu^+ \mu^-)$	&	$1.913 \pm 0.070$ 	&	$1.162 \pm 0.028$	&	4.090	&	2.810	&	4.400	&	2.120		\\
$10^7\mathcal{B}(B_c^+ \to D_s^{*+} \tau^+ \tau^-)$	&	$0.173 \pm 0.008$	&	$0.144 \pm 0.007$	&	0.510	&	0.410	&	0.520	&	0.350		\\
$10^7\mathcal{B}(B_c^+ \to D_s^{*+} \nu^+ \nu^-)$	&	$8.340 \pm 0.393$	&	--	&	31.200	&	21.200	&	40.400	&	13.500		\\
$10^6\mathcal{B}(B_c^+ \to D_s^{*+} \gamma)$	&	$4.454 \pm 0.098 $	&	--	&	-- 	&	-- 	&	-- 	&	--		\\\hline\hline																		
\label{tab:branching}
\end{tabular*}
\end{table*}
Using numerical form factors, we have computed branching fractions of rare semileptonic decays using Eq. (\ref{eq:branching}) and in Fig. \ref{fig:branching}, we plot differential branching fractions.
In Tab. \ref{tab:branching}, we provide the branching fractions by numerically integrating area under the differential branching fraction curve in Fig. \ref{fig:branching}.
In the differential branching fraction plots, the peaks near to $q^2 = m_{J/\psi}^2$ and $q^2 = m_{\psi (2S)}^2$ correspond to the charm resonances, whereas in the low $q^2$ range (in the case of $B_c \to D^{(*)}$ transition), they  correspond to the light vector resonances appearing in the effective Wilson coefficients.
In Tab. \ref{tab:branching}, we provide our results considering both with resonance and without resonance contributions.
It is important to note here that in computations of branching fractions, we exclude the experimentally vetoed $q^2$ range corresponding to the charm resonances.
If we include these range, our results are enhanced by an order of magnitude or more.  This is also observed in our recent studies \cite{Soni:2020bvu} as well as in Ref. \cite{Blake:2016olu}.
We also compare our results with theoretical predictions from different quark models, perturbative QCD and QCD sum rule approaches.
Note that for the transitions corresponding to channels $b \to s \ell^+\ell^-$,  we do not include the contribution from the light vector resonances
because of the CKM suppression (of the order of $\lambda^2$ in Wolfenstein representation); thus we neglect the effects of the second term in effective Hamiltonian Eq. (\ref{eq:hamiltonian}). It is observed that our predictions for the non-resonant branching fractions for some channels are in good agreement with the other theoretical predictions. However, if we include the resonant contributions,   our predictions are comparatively lower than the other theoretical approaches.  In all the literature mentioned here, the contributions from the light vector resonances are not included for the transition corresponding to $b \to d \ell^+\ell^-$.
It is to be noted that for the transitions studied here, experimental data as well as lattice simulations results are not available to the best of our knowledge and understanding, and hence it is not logical to comment on the comparison or credibility of the results from the other theoretical approaches including current study.

\begin{figure*}[htbp]
\includegraphics[width=0.45\textwidth]{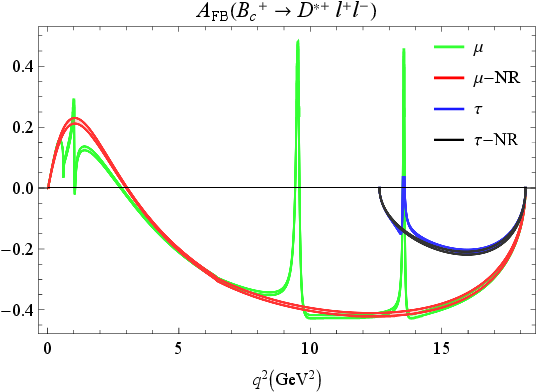}
\hfill\includegraphics[width=0.45\textwidth]{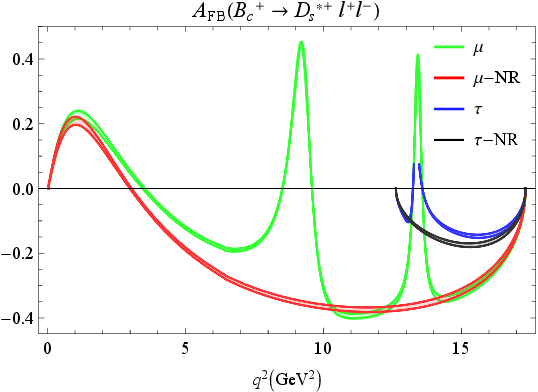}\\
\includegraphics[width=0.45\textwidth]{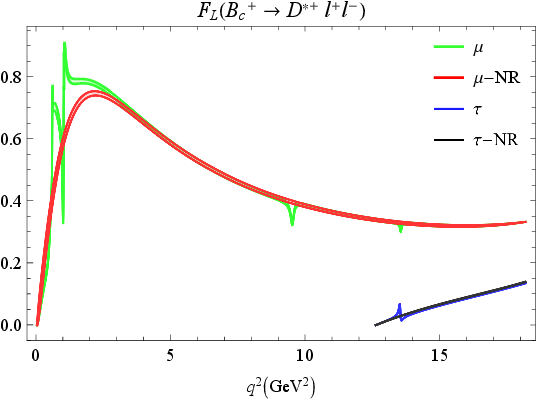}
\hfill\includegraphics[width=0.45\textwidth]{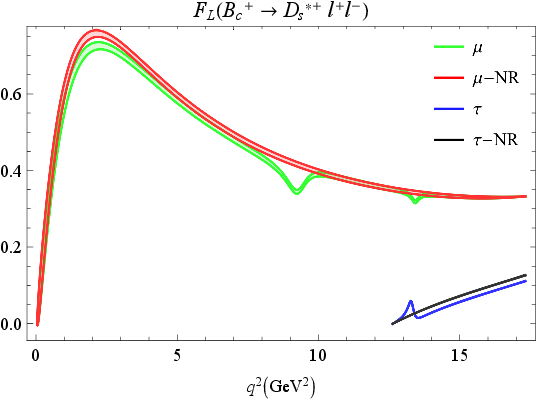}\\
\includegraphics[width=0.45\textwidth]{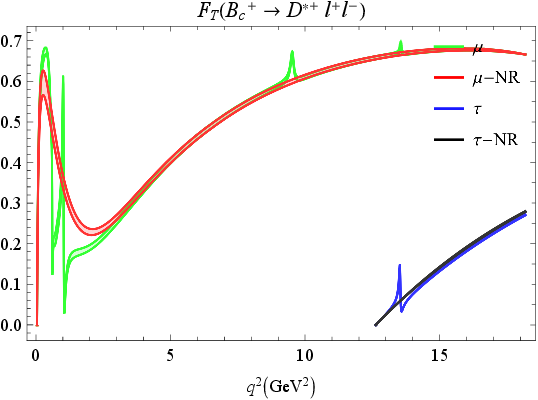}
\hfill\includegraphics[width=0.45\textwidth]{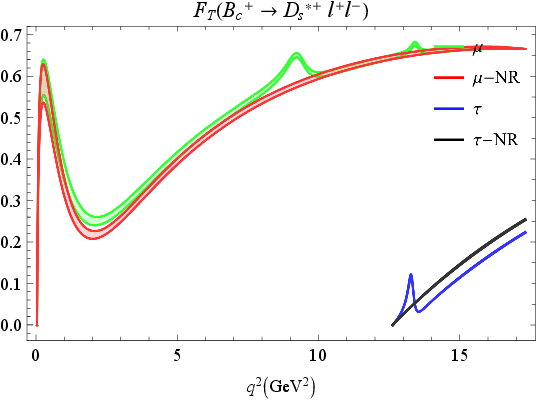}\\
\caption{Forward-backward asymmetry, longitudinal and transverse polarization fractions (Green and Blue plots correspond to the inclusion of vector resonance whereas Red and Black plots are without the same (NR).)\label{fig:asymmetry}}
\end{figure*}

\begin{figure*}[htbp]
\includegraphics[width=0.45\textwidth]{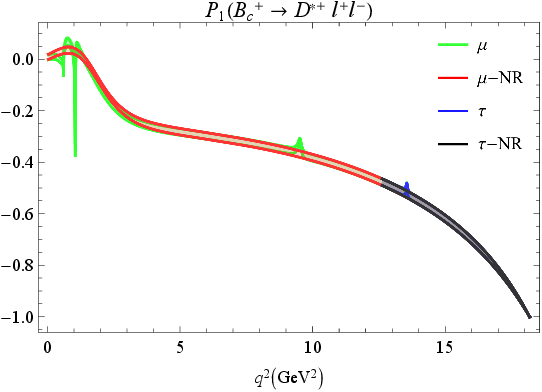}
\hfill\includegraphics[width=0.45\textwidth]{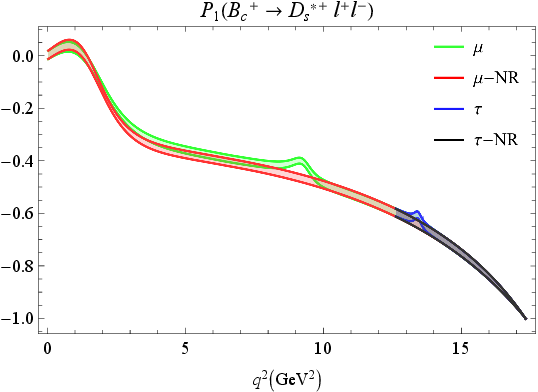}\\
\includegraphics[width=0.45\textwidth]{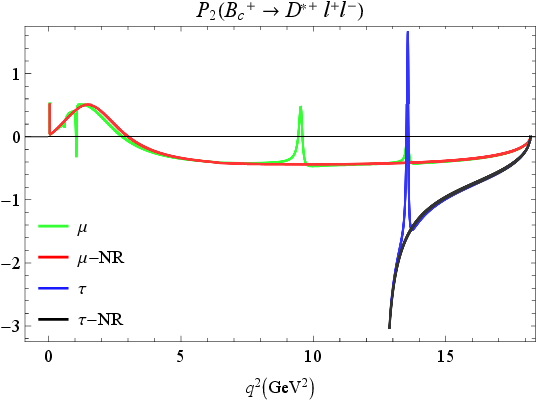}
\hfill\includegraphics[width=0.45\textwidth]{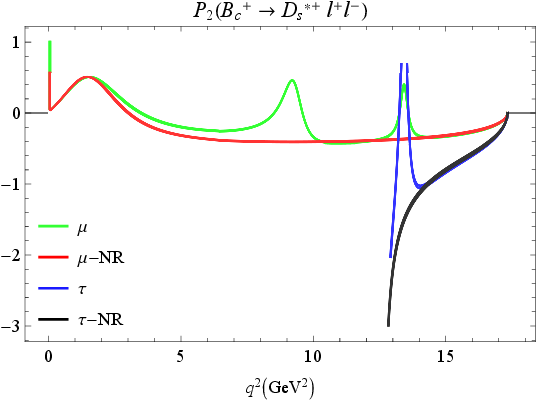}\\
\includegraphics[width=0.45\textwidth]{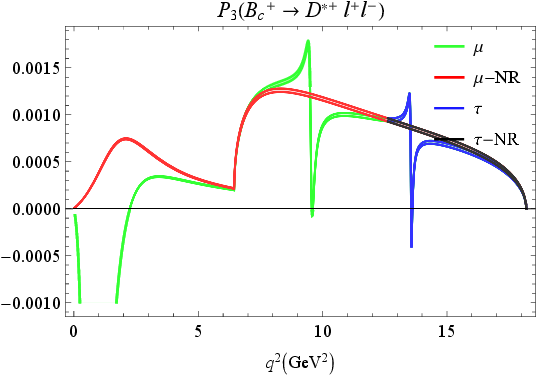}
\hfill\includegraphics[width=0.45\textwidth]{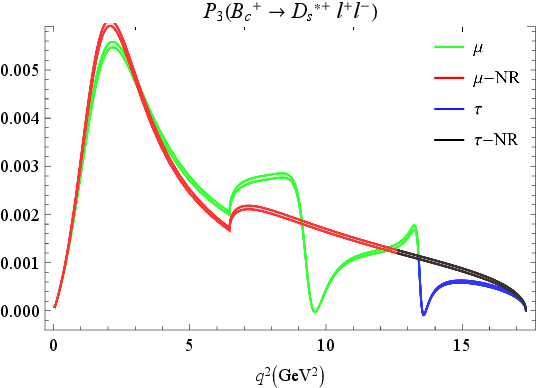}\\
\caption{Clean observables $P_{1,2,3}$ in whole $q^2$ range (Green and Blue plots correspond to the inclusion of vector resonance whereas Red and Black plots are without the same (NR).)\label{fig:p123}}
\end{figure*}

\begin{figure*}[htbp]
\includegraphics[width=0.45\textwidth]{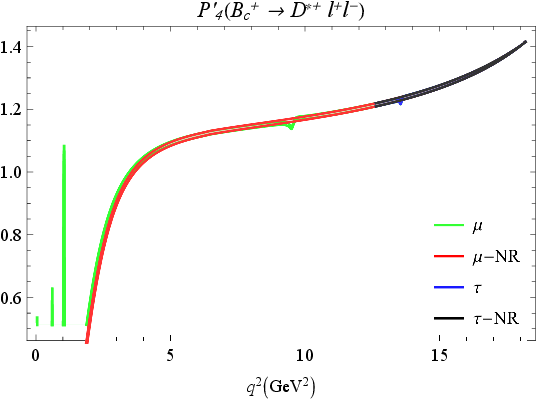}
\hfill\includegraphics[width=0.45\textwidth]{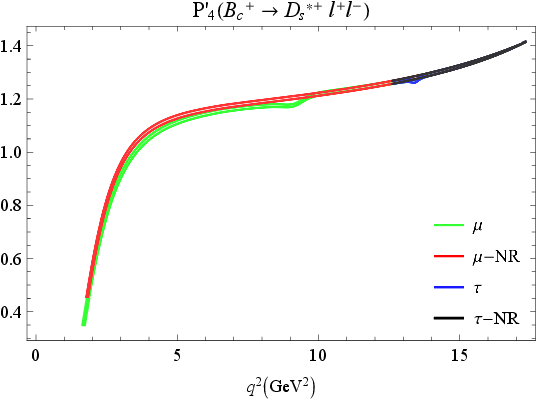}\\
\includegraphics[width=0.45\textwidth]{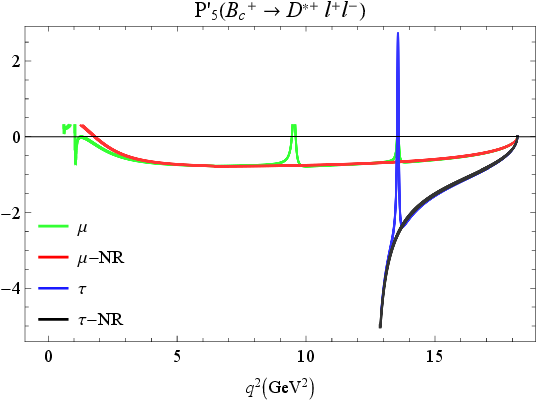}
\hfill\includegraphics[width=0.45\textwidth]{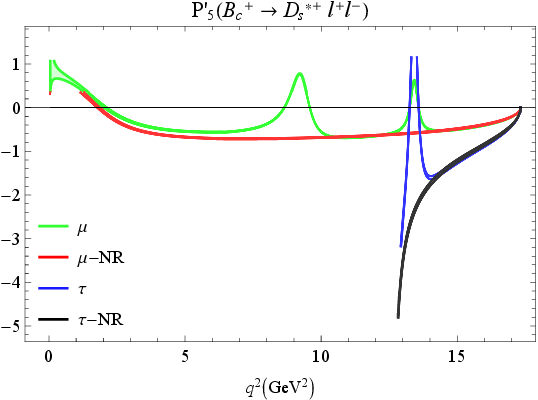}\\
\includegraphics[width=0.45\textwidth]{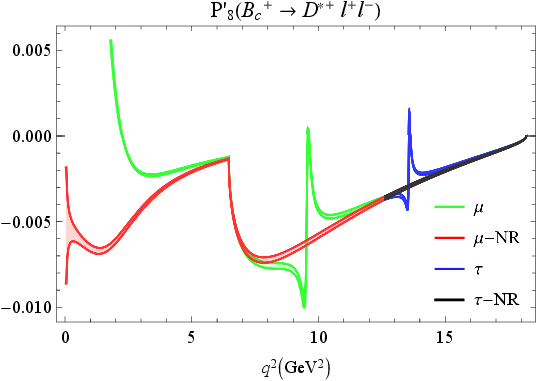}
\hfill\includegraphics[width=0.45\textwidth]{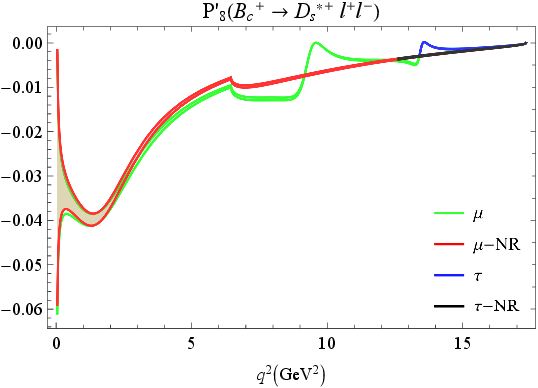}\\
\caption{Clean observables $P_{4,5,8}$ in whole $q^2$ range (Green and Blue plots correspond to the inclusion of vector resonance whereas Red and Black plots are without the same (NR).)\label{fig:p458}}
\end{figure*}

In order to explore further effects of leptons in the final state, we study various observables such as forward-backward asymmetry, longitudinal and transverse polarizations, and angular observables using the relations Eq.  (\ref{eq:AFB}) and (\ref{eq:binP1-8}).
These observables are plotted in the Fig. \ref{fig:asymmetry} - \ref{fig:p458} and their expectation values in the whole $q^2$ range are given in Tab. \ref{tab:obs_DDs}.
In these plots, we include the effects of both nonresonant and resonant contributions, where the peaks correspond to charmonia and light vector resonances.
We also compute the expectation values of these observables in the different $q^2$ bins corresponding to the exclusion of contributions from vector resonances.
Very recently,  Li Y. -L have studied the branching fractions and other observables for the channel $B_c \to D_s^*\ell^+\ell^-$ in the framework of covariant light front quark model (CLFQM) \cite{Li:2023mrj}.
The authors have computed the transition form factors using the modified Isgur-Wise function.  We compare our results of the observables with CLFQM results in Tab. \ref{tab:obsD} - \ref{tab:obsDs2} and it is observed that many of our results are not in agreement with the CLFQM results.
The differences are mainly arising due to their inclusion of the contribution of light vector resonances \cite{Li:2023mrj}.
The light vector resonances have been excluded earlier also for the transition corresponding to the quark channel $b \to s \ell^+\ell^-$ employing CCQM \cite{Dubnicka:2016nyy, Dubnicka:2015iwg}.
To the best of our knowledge and understanding, these observables are yet to be studied using any other theoretical approaches and further these channels are also yet to be explored by the worldwide experimental facilities.
These observables are dependent on the lepton flavours and thus serve as very important probe for the search of new physics beyond the standard model and therefore some insights from the very recent run from the LHCb collaborations as well as from other $B$ factories are expected.

\begin{table*}
\caption{$q^2$- averages of polarization observables over the whole allowed kinematic region for $B_c^+ \to D^{*+} \ell^+ \ell^-$ and $B_c^+ \to D_s^{*+} \ell^+ \ell^-$}
\begin{tabular*}{\textwidth}{@{\extracolsep{\fill}}r|ccc|ccc@{}}
\hline\hline
Obs. & \multicolumn{3}{c|}{$B_c^+ \to D^{*+} \ell^+ \ell^-$} & \multicolumn{3}{c}{$B_c^+ \to D_s^{*+} \ell^+ \ell^-$}  \\
& $e^+ e^-$ & $\mu^+ \mu^-$ & $\tau^+ \tau^-$ & $e^+ e^-$ & $\mu^+ \mu^-$ & $\tau^+ \tau^-$ \\
\hline
$-\langle A_{FB} \rangle$		
& $0.805 \pm 0.018$ & $0.805 \pm 0.018$ & $0.188 \pm 0.008$
& $0.657 \pm 0.021$ & $0.657 \pm 0.021$ & $0.134 \pm 0.008$
\\
$\langle F_L\rangle$				
& $1.028 \pm 0.035$ & $1.215 \pm 0.034$ & $0.095 \pm 0.005$
& $1.003 \pm 0.063$ & $1.199 \pm 0.047$ & $0.080 \pm 0.006$
\\
$\langle F_T\rangle$				
& $1.936 \pm 0.040$ & $1.710 \pm 0.038$ & $0.200 \pm 0.010$
& $1.958 \pm 0.063$ & $1.718 \pm 0.053$ & $0.163 \pm 0.011$
\\
$-\langle P_1 \rangle$	
& $1.274 \pm 0.045$ & $1.403 \pm 0.045$ & $0.778 \pm 0.041$
& $1.490 \pm 0.067$ & $1.644 \pm 0.068$ & $0.856 \pm 0.062$ 		
\\
$-\langle P_2 \rangle$	
& 	$0.819 \pm 0.016$ & $0.937 \pm 0.017$ & $0.627 \pm 0.026$
&	$0.676 \pm 0.020$ & $0.705 \pm 0.021$ & $0.548 \pm 0.032$
\\
$10^{4} \times \langle P_3 \rangle$	
& $5.696 \pm 0.752$ & $5.255 \pm 1.376$ & $4.907 \pm 0.464$
& $26.815 \pm 0.772$ & $42.481 \pm 1.791$ & $4.707 \pm 0.221$
\\
$\langle P_4^\prime \rangle$	
& $2.992 \pm 3.729$ & $3.299 \pm 0.082$ & $1.333 \pm 0.061$
& $3.005 \pm 6.552$ & $3.338 \pm 0.116$ & $1.362 \pm 0.086$
\\
$-\langle P_5^\prime \rangle$
& 	$1.428 \pm 0.030$ & $1.627 \pm 0.032$ & $0.950 \pm 0.036$
& $1.112 \pm 0.034$ & $1.242 \pm 0.037$ & $0.810 \pm 0.042$
\\
$10^{2} \times \langle P_8^\prime \rangle$	
& $3.554 \pm 0.194$ & $5.052 \pm 0.208$ & $-0.106 \pm 0.007$
& $-1.923 \pm 0.137$ & $2.491 \pm 0.098$ & $0.087 \pm 0.008$
\\
\hline\hline
\end{tabular*}
\label{tab:obs_DDs}
\end{table*}

\begin{table*}
\caption{Angular observables in bins for the channel $B_c^+ \to D^{*+}$}
\begin{tabular*}{\textwidth}{@{\extracolsep{\fill}}lcccc@{}}
\hline\hline
Obs.  & bin & $e^+e^-$ & $\mu^+\mu^-$ & $\tau^+\tau^-$ \\
\hline
$ \mathcal{B}$	&	[1.1, 6.0]	&	$0.845 \pm 0.026$	&	$0.840 \pm 0.025$	&		\\
$\times 10^9$	&	[6.0, 8.0]	&	$0.521 \pm 0.011$	&	$0.519 \pm 0.011$	&	\\
	&	[11.0, 12.5]	&	$0.681 \pm 0.015$	&	$0.680 \pm 0.015$	&				\\
	&	[15.0,17.0]	&	$1.025 \pm 0.032$	&	$1.023 \pm 0.032$	&	$0.367 \pm 0.013$			\\
\hline																			
$-\langle A_{FB} \rangle$	&	[1.1, 6.0]	&	$0.087 \pm 0.004$	&	$0.087 \pm 0.004$	&		\\
	&	[6.0, 8.0]	&	$0.317 \pm 0.009$	&	$0.316 \pm 0.009$	&		\\
	&	[11.0, 12.5]	&	$0.421 \pm 0.011$	&	$0.421 \pm 0.011$	&			\\
	&	[15.0,17.0]	&	$0.353 \pm 0.014$	&	$0.353 \pm 0.014$	&	$0.203 \pm 0.009$			\\
\hline																			
$\langle F_L \rangle$	&	[1.1, 6.0]	&	$0.685 \pm 0.034$	&	$0.672 \pm 0.034$	&		\\
	&	[6.0, 8.0]	&	$0.485 \pm 0.019$	&	$0.482 \pm 0.019$	&					\\
	&	[11.0, 12.5]	&	$0.353 \pm 0.013$	&	$0.352 \pm 0.013$	&					\\
	&	[15.0,17.0]	&	$0.321 \pm 0.016$	&	$0.321 \pm 0.016$	&	$0.086 \pm 0.004$				\\
\hline																			
$\langle F_T \rangle$	&	[1.1, 6.0]	&	$0.315 \pm 0.010$	&	$0.311 \pm 0.010$	&		\\
	&	[6.0, 8.0]	&	$0.515 \pm 0.013$	&	$0.513 \pm 0.013$	&				\\
	&	[11.0, 12.5]	&	$0.647 \pm 0.018$	&	$0.645 \pm 0.018$	&				\\
	&	[15.0,17.0]	&	$0.679 \pm 0.029$	&	$0.678 \pm 0.029$	&	$0.181 \pm 0.008$			\\
\hline\hline
\end{tabular*}\label{tab:obsD}
\end{table*}

\begin{table*}
\caption{Same as Tab.  \ref{tab:obsD}}
\begin{tabular*}{\textwidth}{@{\extracolsep{\fill}}lcccc@{}}
\hline\hline
Obs.  & bin & $e^+e^-$ & $\mu^+\mu^-$ & $\tau^+\tau^-$ \\
\hline
$-\langle P_1 \rangle$	&	[1.1, 6.0]	&	$0.241  \pm 0.013$	&	$0.242 \pm 0.013$	&		\\
	&	[6.0, 8.0]	&	$0.315 \pm 0.015$	&	$0.315 \pm 0.015$	&				\\
	&	[11.0, 12.5]	&	$0.441 \pm 0.019$	&	$0.441 \pm 0.019$	&			\\
	&	[15.0,17.0]	&	$0.705 \pm 0.035$	&	$0.705 \pm 0.035$	&	$0.716 \pm 0.036$			\\
\hline
$-\langle P_2 \rangle$	&	[1.1, 6.0]	&	$0.184 \pm 0.008$	&	$0.186 \pm 0.008$	&		\\
	& [6.0, 8.0]	 &	$0.410 \pm 0.009$	&	$0.412 \pm 0.009$	&		\\
	&	[11.0, 12.5]	&	$0.434 \pm 0.010$	&	$0.435 \pm 0.010$	&		\\
	&	[15.0,17.0]	&	$0.347 \pm 0.013$	&	$0.347 \pm 0.013$	&	$0.746 \pm 0.029$	\\
\hline
$\langle P_3 \rangle$	&	[1.1, 6.0]	&	$-(0.058 \pm 1.772)$	&	$0.028 \pm 1.771$	&		\\
	$\times 10^4$	&	[6.0, 8.0]	&	$9.206 \pm 0.141$	&	$9.209 \pm 0.141$	&		\\
	&	[11.0, 12.5]	&	$9.752 \pm 0.165$	&	$9.752 \pm 0.165$	&		\\
	&	[15.0,17.0]	&	$5.854 \pm 0.169$	&	$5.854 \pm 0.168$	&	$5.730 \pm 0.168$	\\
\hline
$\langle P_4^\prime \rangle$	&	[1.1, 6.0]	&	$0.878 \pm 0.031$	&	$0.881 \pm 0.031$	&		\\
	& [6.0, 8.0]	&	$1.131 \pm 0.034$	&	$1.131 \pm 0.034$	&		\\
	& [11.0, 12.5]	&	$1.199 \pm 0.038$	&	$1.199 \pm 0.038$	&		\\
	& [15.0,17.0]	&	$1.306 \pm 0.056$	&	$1.306 \pm 0.056$	&	$1.310 \pm 0.057$	\\
\hline
$-\langle P_5^\prime \rangle$	&	[1.1, 6.0]	&	$0.570 \pm 0.022$	&	$0.576 \pm 0.022$	&		\\
	&	[6.0, 8.0]	&	$0.770 \pm 0.021$	&	$0.772 \pm 0.021$	&		\\
	&	[11.0, 12.5]	&	$0.731 \pm 0.020$	&	$0.732 \pm 0.020$	&		\\
	&	[15.0,17.0]	&	$0.533 \pm 0.019$	&	$0.533 \pm 0.019$	&	 $1.144 \pm 0.041$	\\
\hline
$\langle P_8^\prime \rangle$	&	[1.1, 6.0]	& $1.695 \pm 0.455$	&	$1.592 \pm 0.453$	&		\\
$\times 10^{-3}$ &	[6.0, 8.0]	&	$-(5.412 \pm 0.242)$	&	$-(5.414 \pm 0.242)$	&		\\
&	[11.0, 12.5]	&	$-(4.106 \pm 0.134)$	&	$-(4.106 \pm 0.134)$	&		\\
&	[15.0,17.0]	&	$-(1.385 \pm 0.085)$	&	$-(1.385 \pm 0.085)$ & $-(1.327 \pm 0.084)$\\
\hline\hline
\end{tabular*}\label{tab:obsD1}
\end{table*}

\begin{table*}
\caption{Angular observables in bins for the channel $B_c^+ \to D_{s}^{*+}$}
\begin{tabular*}{\textwidth}{@{\extracolsep{\fill}}lcccccccc@{}}
\hline\hline
Obs.  & bin & $e^+e^-$ & \cite{Li:2023mrj}& $\mu^+\mu^-$ & \cite{Li:2023mrj}& $\tau^+\tau^-$ & \cite{Li:2023mrj}\\
\hline
$10^9 \times \mathcal{B}$	&	[1.1, 6.0]	&	$26.998 \pm 0.917$	&	6.240	&	$26.827 \pm 0.915$	&	6.220	&		&				\\
	&	[6.0, 8.0]	&	$14.930 \pm 0.376$	&	3.560	&	$14.854 \pm 0.375$	&	3.550	&		&				\\
	&	[11.0, 12.5]	&	$25.582 \pm 0.754$	&	2.830	&	$25.518 \pm 0.752$	&	2.830	&		&				\\
	&	[15.0,17.0]	&	$34.543 \pm 1.647$	&	2.560	&	$34.478 \pm 1.644$	&	2.560	&	$13.483 \pm 0.685$	&	0.980			\\
\hline																	
$-\langle A_{FB} \rangle$	&	[1.1, 6.0]	&	$0.002 \pm 0.004$	&	0.061	&	$0.003 \pm 0.004$	&	0.061	&		&				\\
	&	[6.0, 8.0]	&	$0.178 \pm 0.007$	&	0.243	&	$0.178 \pm 0.007$	&	0.242	&		&				\\
	&	[11.0, 12.5]	&	$0.386 \pm 0.014$	&	0.340	&	$0.385 \pm 0.014$	&	0.339	&		&				\\
	&	[15.0,17.0]	&	$0.270 \pm 0.016$	&	0.254	&	$0.270 \pm 0.016$	&	0.254	&	$0.138 \pm 0.009$	&	0.143			\\
\hline																	
$\langle F_L \rangle$	&	[1.1, 6.0]	&	$0.651 \pm 0.038$	&	0.815	&	$0.640 \pm 0.037$	&	0.817	&		&				\\
	&	[6.0, 8.0]	&	$0.483 \pm 0.021$	&	0.637	&	$0.481 \pm 0.022$	&	0.638	&		&				\\
	&	[11.0, 12.5]	&	$0.366 \pm 0.019$	&	0.446	&	$0.365 \pm 0.018$	&	0.446	&		&				\\
	&	[15.0,17.0]	&	$0.331 \pm 0.024$	&	0.352	&	$0.330 \pm 0.024$	&	0.352	&	$0.079 \pm 0.006$	&	0.410			\\
\hline																	
$\langle F_T \rangle$	&	[1.1, 6.0]	&	$0.349 \pm 0.013$	&	0.185	&	$0.344 \pm 0.013$	&	0.183	&		&				\\
	&	[6.0, 8.0]	&	$0.517 \pm 0.015$	&	0.363	&	$0.515 \pm 0.015$	&	0.362	&		&				\\
	&	[11.0, 12.5]	&	$0.634 \pm 0.024$	&	0.554	&	$0.632 \pm 0.024$	&	0.554	&		&				\\
	&	[15.0,17.0]	&	$0.669 \pm 0.044$	&	0.648	&	$0.667 \pm 0.044$	&	0.648	&	$0.159 \pm 0.011$	&	0.590			\\
\hline\hline
\end{tabular*}\label{tab:obsDs1}
\end{table*}
\begin{table*}
\caption{Same as Tab.  \ref{tab:obsDs1}}
\begin{tabular*}{\textwidth}{@{\extracolsep{\fill}}lccccccc@{}}
\hline\hline
Obs.  & bin & $e^+e^-$ & \cite{Li:2023mrj}& $\mu^+\mu^-$ & \cite{Li:2023mrj}& $\tau^+\tau^-$ & \cite{Li:2023mrj}\\
\hline
$-\langle P_1 \rangle$	&	[1.1, 6.0]	&	$0.282 \pm 0.015$	&	0.281	&	$0.284 \pm 0.015$	&	0.281	&		&				\\
	&	[6.0, 8.0]	&	$0.399 \pm 0.017$	&	0.408	&	$0.399 \pm 0.017$	&	0.408	&		&				\\
	&	[11.0, 12.5]	&	$0.557 \pm 0.027$	&	0.543	&	$0.557 \pm 0.027$	&	0.543	&		&				\\
	&	[15.0,17.0]	&	$0.834 \pm 0.059$	&	0.822	&	$0.834 \pm 0.059$	&	0.822	&	$0.845 \pm 0.060$	&	0.826			\\
\hline																	
$-\langle P_2 \rangle$	&	[1.1, 6.0]	&	$0.003 \pm 0.008$	&	0.125	&	$0.005 \pm 0.008$	&	0.125	&		&				\\
	&	[6.0, 8.0]	&	$0.229 \pm 0.008$	&	0.446	&	$0.230 \pm 0.008$	&	0.446	&		&				\\
	&	[11.0, 12.5]	&	$0.406 \pm 0.013$	&	0.409	&	$0.406 \pm 0.013$	&	0.409	&		&				\\
	&	[15.0,17.0]	&	$0.269 \pm 0.059$	&	0.262	&	$0.270 \pm 0.015$	&	0.262	&	$0.579 \pm 0.034$	&	0.269			\\
\hline																	
$10^4 \times \langle P_3 \rangle$	&	[1.1, 6.0]	&	$35.666 \pm 2.953$	&	1.840	&	$35.587 \pm 2.947$	&	1.840	&		&				\\
	&	[6.0, 8.0]	&	$25.421 \pm 1.062$	&	0.070	&	$25.423 \pm 1.061$	&	0.070	&		&				\\
	&	[11.0, 12.5]	&	$12.046 \pm 0.288$	&	24.210	&	$12.047 \pm 0.288$	&	24.210	&		&				\\
	&	[15.0,17.0]	&	$5.114 \pm 0.230$	&	34.420	&	$5.113	 \pm 0.230$ &	34.420	&	$4.944 \pm 0.228$	&	20.080			\\
\hline																	
$\langle P_4^\prime \rangle$	&	[1.1, 6.0]	&	$0.882 \pm 0.037$	&	0.908	&	$0.885 \pm 0.037$	&	0.898	&		&				\\
	&	[6.0, 8.0]	&	$1.160 \pm 0.040$	&	1.177	&	$1.160 \pm 0.040$	&	1.169	&		&				\\
	&	[11.0, 12.5]	&	$1.247 \pm 0.052$	&	1.240	&	$1.247 \pm 0.052$	&	1.236	&		&				\\
	&	[15.0,17.0]	&	$1.354 \pm 0.084$	&	1.350	&	$1.354 \pm 0.084$	&	1.347	&	$1.358 \pm 0.085$	&	0.561			\\
\hline																	
$-\langle P_5^\prime \rangle$	&	[1.1, 6.0]	&	$0.315 \pm 0.023$	&	0.540	&	$0.319 \pm 0.023$	&	0.534	&		&				\\
	&	[6.0, 8.0]	&	$0.507 \pm 0.019$	&	0.766	&	$0.508 \pm 0.019$	&	0.761	&		&				\\
	&	[11.0, 12.5]	&	$0.654 \pm 0.023$	&	0.664	&	$0.656 \pm 0.023$	&	0.662	&		&				\\
	&	[15.0,17.0]	&	$0.399 \pm 0.020$	&	0.390	&	$0.399 \pm 0.020$	&	0.390	&	$0.855 \pm 0.044$	&	0.163			\\
\hline																	
$-10^3 \times \langle P_8^\prime \rangle$	&	[1.1, 6.0]	&	$22.160 \pm 0.980$	&	1.196	&	$22.052 \pm 0.975$	&	1.164	&		&				\\
	&	[6.0, 8.0]	&	$11.841 \pm 0.473$ &	0.029	&	$11.842 \pm 0.473$	&	0.029	&		&				\\
	&	[11.0, 12.5]	&	$3.866 \pm 0.206$	 &	7.054	&	$3.866 \pm 0.206$	 	&	7.030	&		&				\\
	&	[15.0,17.0]	&	$0.970 \pm 0.088$	&	6.147	&	$0.970 \pm 0.088$		&	6.137	&	$0.921 \pm 0.085$	&	1.467			\\
\hline\hline
\end{tabular*}\label{tab:obsDs2}
\end{table*}

\section{Summary and Conclusion}
\label{sec:summary}
In this article, we systematically study the rare semileptonic decay of $B_c$ meson within the framework of covariant confined quark model.  We have considered the channels $B_c^+ \to D^{+(*)}\ell^+\ell^-$ and $B_c^+ \to D_s^{+(*)}\ell^+\ell^-$ for all the lepton flavours.
The necessary transition form factors are computed in the whole range of momentum transfer squared and compared with different theoretical approaches.
Our results of the form factors are compatible with the other quark models.
Using these form factors and Wilson coefficients, we have computed the rare semileptonic branching fractions and compared our results with other approaches.
We have also computed various observables such as forward-backward asymmetry, longitudinal and transverse polarizations, and angular observables.
In present work, the computation of effective Wilson coefficients includes the contribution of charm resonances as well as light vector resonances in the case of $B_c^+ \to D^{+(*)}\ell^+\ell^-$. For $B_c^+ \to D_s^{+(*)}\ell^+\ell^-$, we only include the contribution of charm vector resonances.
It is observed that our predictions are systematically lower than those reported in literature and the main reason for these differences are the inclusion of the contribution of light vector resonances in definition of effective Wilson coefficients as well as the choice of the numerical values of the Wilson coefficients.

To conclude, we have provided the complete description of rare semileptonic decays of $B_c$ mesons in the framework of CCQM along with the different physical observables which would play an important role for the identification of these observables for future experimental facilities and also to look for the test of new physics beyond standard model, if any.

\section*{ACKNOWLEDGEMENTS}
NRS would like to thank INFN - Section of Naples for the Post-doctoral research grant (July 2023 to July 2024).
\clearpage

\appendix
\section{Covariance among form factor parameters}
\label{Appen:covariance}

Within the framework of CCQM, the theoretical uncertainties are directly tied to the uncertainties in the free parameters of the model. 
To provide the reader with a practical way to utilise our model, we fit the behaviour of the form factors as a function of $q^2$ using the double-pole parametrization which is characterised by three parameters (namely $F(0)$, $a$ and $b$), has been fitted using a standard chi-squared minimization procedure.
Here, we present the covariance matrices for all the transition form factors for the channels $B_c^+ \to D_{(s)}^+$ as well as for the channels $B_c^+ \to D_{(s)}^{*+}$. They demonstrate the covariance among the parameters corresponding to all form factors entering the same decay channel. All the matrices are listed in Tab. \ref{tab:covariance_BcD} - \ref{tab:covariance_BcDsv}.
\begin{table*}[!h]
\caption{Covariance matrix for $B_c^+ \to D^+$ form factors and associated parameters ($\times 10^{-3}$)}
\begin{tabular*}{\textwidth}{@{\extracolsep{\fill}}cccccccccccccc@{}}
\hline\hline
		&		&		&	$F_+$ 	&		&		&	$F_-$ 	&		&	 	&	$F_T$ 	&		&		&	 $F_0$ 	&		\\
	&		&	$F(0)$ 	&	 $a$ 	&	 $b$ 	&	$F(0)$ 	&	 $a$ 	&	 $b$ 	&	$F(0)$ 	&	 $a$ 	&	 $b$ 	&	$F(0)$ 	&	 $a$ 	&	 $b$ 	\\
\hline
	&	$F(0)$ 	&	0.009	&	0.036	&	0.081	&	-0.006	&	0.033	&	0.078	&	0.012	&	0.036	&	0.084	&	0.012	&	0.135	&	0.336	\\
$F_+$ 	&	 $a$ 	&		&	0.144	&	0.324	&	-0.024	&	0.132	&	0.312	&	0.048	&	0.144	&	0.336	&	0.048	&	0.540	&	1.344	\\
	&	 $b$ 	&		&		&	0.729	&	-0.054	&	0.297	&	0.702	&	0.108	&	0.324	&	0.756	&	0.108	&	1.215	&	3.024	\\
\hline
	&	$F(0)$ 	&		&		&		&	0.004	&	-0.022	&	-0.052	&	-0.008	&	-0.024	&	-0.056	&	-0.008	&	-0.090	&	-0.224	\\
$F_-$ 	&	 $a$ 	&		&		&		&		&	0.121	&	0.286	&	0.044	&	0.132	&	0.308	&	0.044	&	0.495	&	1.232	\\
	&	 $b$ 	&		&		&		&		&		&	0.676	&	0.104	&	0.312	&	0.728	&	0.104	&	1.170	&	2.912	\\
\hline
	&	$F(0)$ 	&		&		&		&		&		&		&	0.016	&	0.048	&	0.112	&	0.016	&	0.180	&	0.448	\\
$F_T$ 	&	 $a$ 	&		&		&		&		&		&		&		&	0.144	&	0.336	&	0.048	&	0.540	&	1.344	\\
	&	 $b$ 	&		&		&		&		&		&		&		&		&	0.784	&	0.112	&	1.260	&	3.136	\\
\hline
	&	$F(0)$ 	&		&		&		&		&		&		&		&		&		&	0.016	&	0.180	&	0.448	\\
 $F_0$ 	&	 $a$ 	&		&		&		&		&		&		&		&		&		&		&	2.025	&	5.040	\\
	&	 $b$ 	&		&		&		&		&		&		&		&		&		&		&		&	12.544	\\
\hline\hline
\label{tab:covariance_BcD}
\end{tabular*}
\end{table*}
\begin{table*}[!h]
\caption{Covariance matrix for $B_c^+ \to D_s^+$ form factors and associated parameters ($\times 10^{-3}$)}
\begin{tabular*}{\textwidth}{@{\extracolsep{\fill}}cccccccccccccc@{}}
\hline\hline
		&		&		&	$F_+$ 	&		&		&	$F_-$ 	&		&	 	&	$F_T$ 	&		&		&	 $F_0$ 	&		\\
	&		&	$F(0)$ 	&	 $a$ 	&	 $b$ 	&	$F(0)$ 	&	 $a$ 	&	 $b$ 	&	$F(0)$ 	&	 $a$ 	&	 $b$ 	&	$F(0)$ 	&	 $a$ 	&	 $b$ 	\\
\hline
	&	$F(0)$ 	&	0.016	&	-0.076	&	-0.184	&	-0.016	&	0.038	&	-0.180	&	-0.024	&	-0.076	&	0.188	&	-0.024	&	-0.228	&	-0.596	\\
$F_+$ 	&	 $a$ 	&		&	0.361	&	0.874	&	0.076	&	-0.180	&	0.855	&	0.114	&	0.361	&	-0.893	&	0.114	&	1.083	&	2.831	\\
	&	 $b$ 	&		&		&	2.116	&	0.184	&	-0.437	&	2.070	&	0.276	&	0.874	&	-2.162	&	0.276	&	2.622	&	6.853	\\
\hline
	&	$F(0)$ 	&		&		&		&	0.016	&	-0.038	&	0.180	&	0.024	&	0.076	&	-0.188	&	0.024	&	0.228	&	0.596	\\
$F_-$ 	&	 $a$ 	&		&		&		&		&	0.361	&	-0.427	&	-0.057	&	-0.180	&	0.446	&	-0.057	&	-0.541	&	-1.415	\\
	&	 $b$ 	&		&		&		&		&		&	2.025	&	0.270	&	0.855	&	-2.115	&	0.270	&	2.565	&	6.705	\\
\hline
	&	$F(0)$ 	&		&		&		&		&		&		&	0.036	&	0.114	&	-0.282	&	0.036	&	0.342	&	0.894	\\
$F_T$ 	&	 $a$ 	&		&		&		&		&		&		&		&	0.361	&	-0.893	&	0.114	&	1.083	&	2.831	\\
	&	 $b$ 	&		&		&		&		&		&		&		&		&	2.209	&	-0.282	&	-2.679	&	-7.003	\\
\hline
	&	$F(0)$ 	&		&		&		&		&		&		&		&		&		&	0.036	&	0.342	&	0.894	\\
 $F_0$ 	&	 $a$ 	&		&		&		&		&		&		&		&		&		&		&	3.249	&	8.493	\\
	&	 $b$ 	&		&		&		&		&		&		&		&		&		&		&		&	22.201	\\
\hline\hline
\label{tab:covariance_BcDs}
\end{tabular*}
\end{table*}

\begin{sidewaystable*}
\caption{Covariance matrix for $B_c^+ \to D^{*+}$ form factors and associated parameters ($\times 10^{-3}$)}
\begin{tabular*}{\textwidth}{@{\extracolsep{\fill}}ccccccccccccccccccccccc@{}}
\hline\hline
	&		&		&	$A_0$ 	&		&		&	$A_+$ 	&		&	 	&	 $A_-$ 	&		&		&	 $V$ 	&		&		&	 $a_0$ 	&		&		&	 $a_+$ 	&		&		&	 $g$	&		\\
	&		&	$F(0)$ 	&	 $a$ 	&	 $b$ 	&	$F(0)$ 	&	 $a$ 	&	 $b$ 	&	$F(0)$ 	&	 $a$ 	&	 $b$ 	&	$F(0)$ 	&	 $a$ 	&	 $b$ 	&	$F(0)$ 	&	 $a$ 	&	 $b$ 	&	$F(0)$ 	&	 $a$ 	&	 $b$ 	&	$F(0)$ 	&	 $a$ 	&	 $b$ 	\\
\hline
	&	$F(0)$ 	&	0.004	&	0.020	&	0.052	&	0.002	&	-0.014	&	-0.034	&	-0.004	&	-0.012	&	-0.030	&	0.004	&	-0.012	&	-0.030	&	-0.002	&	-0.020	&	-0.050	&	-0.018	&	-0.014	&	-0.036	&	-0.002	&	-0.012	&	-0.028	\\
$A_0$ 	&	 $a$ 	&		&	0.100	&	0.260	&	0.010	&	-0.070	&	-0.170	&	-0.020	&	-0.060	&	-0.150	&	0.020	&	-0.060	&	-0.150	&	-0.010	&	-0.100	&	-0.250	&	-0.090	&	-0.070	&	-0.180	&	-0.010	&	-0.060	&	-0.140	\\
	&	 $b$ 	&		&		&	0.676	&	0.026	&	-0.182	&	-0.442	&	-0.052	&	-0.156	&	-0.390	&	0.052	&	-0.156	&	-0.390	&	-0.026	&	-0.260	&	-0.650	&	-0.234	&	-0.182	&	-0.468	&	-0.026	&	-0.156	&	-0.364	\\
\hline
	&	$F(0)$ 	&		&		&		&	0.001	&	-0.007	&	-0.017	&	-0.002	&	-0.006	&	-0.015	&	0.002	&	-0.006	&	-0.015	&	-0.001	&	-0.010	&	-0.025	&	-0.009	&	-0.007	&	-0.018	&	-0.001	&	-0.006	&	-0.014	\\
$A_+$ 	&	 $a$ 	&		&		&		&		&	0.049	&	0.119	&	0.014	&	0.042	&	0.105	&	-0.014	&	0.042	&	0.105	&	0.007	&	0.070	&	0.175	&	0.063	&	0.049	&	0.126	&	0.007	&	0.042	&	0.098	\\
	&	 $b$ 	&		&		&		&		&		&	0.289	&	0.034	&	0.102	&	0.255	&	-0.034	&	0.102	&	0.255	&	0.017	&	0.170	&	0.425	&	0.153	&	0.119	&	0.306	&	17.000	&	0.102	&	0.238	\\
\hline
	&	$F(0)$ 	&		&		&		&		&		&		&	0.004	&	0.012	&	0.030	&	-0.004	&	0.012	&	0.030	&	0.002	&	0.020	&	0.050	&	0.002	&	0.020	&	0.050	&	0.002	&	0.012	&	0.028	\\
 $A_-$ 	&	 $a$ 	&		&		&		&		&		&		&		&	0.036	&	0.090	&	-0.012	&	0.036	&	0.090	&	0.006	&	0.060	&	0.150	&	0.054	&	0.042	&	0.108	&	0.006	&	0.036	&	0.084	\\
	&	 $b$ 	&		&		&		&		&		&		&		&		&	0.225	&	-0.030	&	0.090	&	0.225	&	0.015	&	0.150	&	0.375	&	0.135	&	0.105	&	0.270	&	0.015	&	0.090	&	0.210	\\
\hline
	&	$F(0)$ 	&		&		&		&		&		&		&		&		&		&	0.004	&	-0.012	&	-0.030	&	-0.002	&	-0.020	&	-0.050	&	-0.018	&	-0.014	&	-0.036	&	-0.002	&	-0.012	&	-0.028	\\
$V$ 	&	 $a$ 	&		&		&		&		&		&		&		&		&		&		&	0.036	&	0.090	&	0.006	&	0.060	&	0.140	&	0.054	&	0.042	&	0.108	&	0.006	&	0.036	&	0.084	\\
	&	 $b$ 	&		&		&		&		&		&		&		&		&		&		&		&	0.225	&	0.015	&	0.150	&	0.375	&	0.135	&	0.105	&	0.270	&	0.015	&	0.090	&	0.210	\\
\hline
	&	$F(0)$ 	&		&		&		&		&		&		&		&		&		&		&		&		&	0.001	&	0.010	&	0.025	&	0.009	&	0.007	&	0.018	&	0.001	&	0.006	&	0.014	\\
 $a_0$ 	&	 $a$ 	&		&		&		&		&		&		&		&		&		&		&		&		&		&	0.100	&	0.250	&	0.090	&	0.070	&	0.180	&	0.010	&	0.060	&	0.140	\\
	&	 $b$ 	&		&		&		&		&		&		&		&		&		&		&		&		&		&		&	0.625	&	0.225	&	0.175	&	0.450	&	0.025	&	0.150	&	0.350	\\
\hline
	&	$F(0)$ 	&		&		&		&		&		&		&		&		&		&		&		&		&		&		&		&	0.081	&	0.063	&	0.162	&	0.009	&	0.054	&	0.126	\\
 $a_+$ 	&	 $a$ 	&		&		&		&		&		&		&		&		&		&		&		&		&		&		&		&		&	0.049	&	0.126	&	0.007	&	0.042	&	0.098	\\
	&	 $b$ 	&		&		&		&		&		&		&		&		&		&		&		&		&		&		&		&		&		&	0.324	&	0.018	&	0.108	&	0.252	\\
\hline
	&	$F(0)$ 	&		&		&		&		&		&		&		&		&		&		&		&		&		&		&		&		&		&		&	0.001	&	0.006	&	0.014	\\
$g$ 	&	 $a$ 	&		&		&		&		&		&		&		&		&		&		&		&		&		&		&		&		&		&		&		&	0.036	&	0.084	\\
	&	 $b$ 	&		&		&		&		&		&		&		&		&		&		&		&		&		&		&		&		&		&		&		&		&	0.196	\\
\hline\hline
\label{tab:covariance_BcDv}
\end{tabular*}
\end{sidewaystable*}
\begin{sidewaystable*}
\caption{Covariance matrix for $B_c^+ \to D_s^{*+}$ form factors and associated parameters ($\times 10^{-3}$)}
\begin{tabular*}{\textwidth}{@{\extracolsep{\fill}}ccccccccccccccccccccccc@{}}
\hline\hline
	&		&		&	$A_0$ 	&		&		&	$A_+$ 	&		&	 	&	 $A_-$ 	&		&		&	 $V$ 	&		&		&	 $a_0$ 	&		&		&	 $a_+$ 	&		&		&	 $g$	&		\\
	&		&	$F(0)$ 	&	 $a$ 	&	 $b$ 	&	$F(0)$ 	&	 $a$ 	&	 $b$ 	&	$F(0)$ 	&	 $a$ 	&	 $b$ 	&	$F(0)$ 	&	 $a$ 	&	 $b$ 	&	$F(0)$ 	&	 $a$ 	&	 $b$ 	&	$F(0)$ 	&	 $a$ 	&	 $b$ 	&	$F(0)$ 	&	 $a$ 	&	 $b$ 	\\
\hline
	&	$F(0)$ 	&	0.009	&	0.048	&	0.123	&	0.006	&	-0.033	&	-0.087	&	-0.009	&	-0.030	&	-0.078	&	0.006	&	-0.030	&	-0.078	&	-0.006	&	-0.045	&	-0.117	&	-0.006	&	-0.033	&	-0.087	&	-0.006	&	-0.030	&	-0.075	\\
$A_0$ 	&	 $a$ 	&		&	0.256	&	0.656	&	0.032	&	-0.176	&	-0.464	&	-0.048	&	-0.160	&	-0.416	&	0.032	&	-0.160	&	-0.416	&	-0.032	&	-0.240	&	-0.624	&	-0.032	&	-0.176	&	-0.464	&	-0.032	&	-0.160	&	-0.400	\\
	&	 $b$ 	&		&		&	1.681	&	0.082	&	-0.451	&	-1.189	&	-0.123	&	-0.410	&	-1.066	&	0.082	&	-0.410	&	-1.066	&	-0.082	&	-0.615	&	-1.599	&	-0.082	&	-0.451	&	-1.189	&	-0.082	&	-0.410	&	-1.025	\\
\hline
	&	$F(0)$ 	&		&		&		&	0.004	&	-0.022	&	-0.058	&	-0.006	&	-0.020	&	-0.052	&	0.004	&	-0.020	&	-0.052	&	-0.004	&	-0.030	&	-0.078	&	-0.004	&	-0.022	&	-0.058	&	-0.004	&	-0.020	&	-0.050	\\
$A_+$ 	&	 $a$ 	&		&		&		&		&	0.121	&	0.319	&	0.033	&	0.110	&	0.286	&	-0.022	&	0.110	&	0.286	&	0.022	&	0.165	&	0.429	&	0.022	&	0.121	&	0.319	&	0.022	&	0.110	&	0.275	\\
	&	 $b$ 	&		&		&		&		&		&	0.841	&	0.087	&	0,29	&	0.754	&	-0.058	&	0.290	&	0.754	&	0.058	&	0.435	&	1.131	&	0.058	&	0.319	&	0.841	&	0.058	&	0.290	&	0.725	\\
\hline
	&	$F(0)$ 	&		&		&		&		&		&		&	0.009	&	0.030	&	0.078	&	-0.006	&	0.030	&	0.078	&	0.006	&	0.045	&	0.117	&	0.006	&	0.045	&	0.117	&	0.006	&	0.030	&	0.075	\\
 $A_-$ 	&	 $a$ 	&		&		&		&		&		&		&		&	0.100	&	0.260	&	-0.020	&	0.100	&	0.260	&	0.020	&	0.150	&	0.390	&	0.020	&	0.110	&	0.290	&	0.020	&	0.100	&	0.250	\\
	&	 $b$ 	&		&		&		&		&		&		&		&		&	0.676	&	-0.052	&	0.260	&	0.676	&	0.052	&	0.390	&	1.014	&	0.052	&	0.286	&	0.754	&	0.052	&	0.260	&	0.650	\\
\hline
	&	$F(0)$ 	&		&		&		&		&		&		&		&		&		&	0.004	&	-0.020	&	-0.052	&	-0.004	&	-0.030	&	-0.078	&	-0.004	&	-0.022	&	-0.058	&	-0.004	&	-0.020	&	-0.050	\\
$V$ 	&	 $a$ 	&		&		&		&		&		&		&		&		&		&		&	0.100	&	0.260	&	0.020	&	0.150	&	0.390	&	0.020	&	0.110	&	0.290	&	0.020	&	0.100	&	0.250	\\
	&	 $b$ 	&		&		&		&		&		&		&		&		&		&		&		&	0.676	&	0.052	&	0.390	&	1.014	&	0.052	&	0.286	&	0.754	&	0.052	&	0.260	&	0.650	\\
\hline
	&	$F(0)$ 	&		&		&		&		&		&		&		&		&		&		&		&		&	0.004	&	0.030	&	0.078	&	0.004	&	0.022	&	0.058	&	0.004	&	0.020	&	0.050	\\
 $a_0$ 	&	 $a$ 	&		&		&		&		&		&		&		&		&		&		&		&		&		&	0.225	&	0.585	&	0.030	&	0.165	&	0.435	&	0.030	&	0.150	&	0.375	\\
	&	 $b$ 	&		&		&		&		&		&		&		&		&		&		&		&		&		&		&	1.521	&	0.078	&	0.429	&	1.131	&	0.078	&	0.390	&	0.975	\\
\hline
	&	$F(0)$ 	&		&		&		&		&		&		&		&		&		&		&		&		&		&		&		&	0.004	&	0.022	&	0.058	&	0.004	&	0.020	&	0.050	\\
 $a_+$ 	&	 $a$ 	&		&		&		&		&		&		&		&		&		&		&		&		&		&		&		&		&	0.121	&	0.319	&	0.022	&	0.110	&	0.275	\\
	&	 $b$ 	&		&		&		&		&		&		&		&		&		&		&		&		&		&		&		&		&		&	0.841	&	0.058	&	0.290	&	0.725	\\
\hline
	&	$F(0)$ 	&		&		&		&		&		&		&		&		&		&		&		&		&		&		&		&		&		&		&	0.004	&	0.020	&	0.050	\\
$g$ 	&	 $a$ 	&		&		&		&		&		&		&		&		&		&		&		&		&		&		&		&		&		&		&		&	0.100	&	0.250	\\
	&	 $b$ 	&		&		&		&		&		&		&		&		&		&		&		&		&		&		&		&		&		&		&		&		&	0.625	\\
\hline\hline
\label{tab:covariance_BcDsv}
\end{tabular*}
\end{sidewaystable*}

\clearpage
\bibliography{apssamp}

\begin{thebibliography}{112}%
\makeatletter
\providecommand \@ifxundefined [1]{%
 \@ifx{#1\undefined}
}%
\providecommand \@ifnum [1]{%
 \ifnum #1\expandafter \@firstoftwo
 \else \expandafter \@secondoftwo
 \fi
}%
\providecommand \@ifx [1]{%
 \ifx #1\expandafter \@firstoftwo
 \else \expandafter \@secondoftwo
 \fi
}%
\providecommand \natexlab [1]{#1}%
\providecommand \enquote  [1]{``#1''}%
\providecommand \bibnamefont  [1]{#1}%
\providecommand \bibfnamefont [1]{#1}%
\providecommand \citenamefont [1]{#1}%
\providecommand \href@noop [0]{\@secondoftwo}%
\providecommand \href [0]{\begingroup \@sanitize@url \@href}%
\providecommand \@href[1]{\@@startlink{#1}\@@href}%
\providecommand \@@href[1]{\endgroup#1\@@endlink}%
\providecommand \@sanitize@url [0]{\catcode `\\12\catcode `\$12\catcode
  `\&12\catcode `\#12\catcode `\^12\catcode `\_12\catcode `\%12\relax}%
\providecommand \@@startlink[1]{}%
\providecommand \@@endlink[0]{}%
\providecommand \url  [0]{\begingroup\@sanitize@url \@url }%
\providecommand \@url [1]{\endgroup\@href {#1}{\urlprefix }}%
\providecommand \urlprefix  [0]{URL }%
\providecommand \Eprint [0]{\href }%
\providecommand \doibase [0]{http://dx.doi.org/}%
\providecommand \selectlanguage [0]{\@gobble}%
\providecommand \bibinfo  [0]{\@secondoftwo}%
\providecommand \bibfield  [0]{\@secondoftwo}%
\providecommand \translation [1]{[#1]}%
\providecommand \BibitemOpen [0]{}%
\providecommand \bibitemStop [0]{}%
\providecommand \bibitemNoStop [0]{.\EOS\space}%
\providecommand \EOS [0]{\spacefactor3000\relax}%
\providecommand \BibitemShut  [1]{\csname bibitem#1\endcsname}%
\let\auto@bib@innerbib\@empty
\bibitem [{\citenamefont {Lees}\ \emph {et~al.}(2013)\citenamefont {Lees} \emph
  {et~al.}}]{BaBar:2013mob}%
  \BibitemOpen
  \bibfield  {author} {\bibinfo {author} {\bibfnamefont {J.~P.}\ \bibnamefont
  {Lees}} \emph {et~al.} (\bibinfo {collaboration} {BaBar}),\ }\href {\doibase
  10.1103/PhysRevD.88.072012} {\bibfield  {journal} {\bibinfo  {journal} {Phys.
  Rev. D}\ }\textbf {\bibinfo {volume} {88}},\ \bibinfo {pages} {072012}
  (\bibinfo {year} {2013})},\ \Eprint {http://arxiv.org/abs/1303.0571}
  {arXiv:1303.0571 [hep-ex]} \BibitemShut {NoStop}%
\bibitem [{\citenamefont {Huschle}\ \emph {et~al.}(2015)\citenamefont {Huschle}
  \emph {et~al.}}]{Belle:2015qfa}%
  \BibitemOpen
  \bibfield  {author} {\bibinfo {author} {\bibfnamefont {M.}~\bibnamefont
  {Huschle}} \emph {et~al.} (\bibinfo {collaboration} {Belle}),\ }\href
  {\doibase 10.1103/PhysRevD.92.072014} {\bibfield  {journal} {\bibinfo
  {journal} {Phys. Rev. D}\ }\textbf {\bibinfo {volume} {92}},\ \bibinfo
  {pages} {072014} (\bibinfo {year} {2015})},\ \Eprint
  {http://arxiv.org/abs/1507.03233} {arXiv:1507.03233 [hep-ex]} \BibitemShut
  {NoStop}%
\bibitem [{\citenamefont {Sato}\ \emph {et~al.}(2016)\citenamefont {Sato} \emph
  {et~al.}}]{Belle:2016ure}%
  \BibitemOpen
  \bibfield  {author} {\bibinfo {author} {\bibfnamefont {Y.}~\bibnamefont
  {Sato}} \emph {et~al.} (\bibinfo {collaboration} {Belle}),\ }\href {\doibase
  10.1103/PhysRevD.94.072007} {\bibfield  {journal} {\bibinfo  {journal} {Phys.
  Rev. D}\ }\textbf {\bibinfo {volume} {94}},\ \bibinfo {pages} {072007}
  (\bibinfo {year} {2016})},\ \Eprint {http://arxiv.org/abs/1607.07923}
  {arXiv:1607.07923 [hep-ex]} \BibitemShut {NoStop}%
\bibitem [{\citenamefont {Aaij}\ \emph
  {et~al.}(2015{\natexlab{a}})\citenamefont {Aaij} \emph
  {et~al.}}]{LHCb:2015gmp}%
  \BibitemOpen
  \bibfield  {author} {\bibinfo {author} {\bibfnamefont {R.}~\bibnamefont
  {Aaij}} \emph {et~al.} (\bibinfo {collaboration} {LHCb}),\ }\href {\doibase
  10.1103/PhysRevLett.115.111803} {\bibfield  {journal} {\bibinfo  {journal}
  {Phys. Rev. Lett.}\ }\textbf {\bibinfo {volume} {115}},\ \bibinfo {pages}
  {111803} (\bibinfo {year} {2015}{\natexlab{a}})},\ \bibinfo {note} {[Erratum:
  Phys.Rev.Lett. 115, 159901 (2015)]},\ \Eprint
  {http://arxiv.org/abs/1506.08614} {arXiv:1506.08614 [hep-ex]} \BibitemShut
  {NoStop}%
\bibitem [{\citenamefont {Aaij}\ \emph
  {et~al.}(2018{\natexlab{a}})\citenamefont {Aaij} \emph
  {et~al.}}]{LHCb:2017smo}%
  \BibitemOpen
  \bibfield  {author} {\bibinfo {author} {\bibfnamefont {R.}~\bibnamefont
  {Aaij}} \emph {et~al.} (\bibinfo {collaboration} {LHCb}),\ }\href {\doibase
  10.1103/PhysRevLett.120.171802} {\bibfield  {journal} {\bibinfo  {journal}
  {Phys. Rev. Lett.}\ }\textbf {\bibinfo {volume} {120}},\ \bibinfo {pages}
  {171802} (\bibinfo {year} {2018}{\natexlab{a}})},\ \Eprint
  {http://arxiv.org/abs/1708.08856} {arXiv:1708.08856 [hep-ex]} \BibitemShut
  {NoStop}%
\bibitem [{\citenamefont {Aaij}\ \emph
  {et~al.}(2023{\natexlab{a}})\citenamefont {Aaij} \emph
  {et~al.}}]{LHCb:2023zxo}%
  \BibitemOpen
  \bibfield  {author} {\bibinfo {author} {\bibfnamefont {R.}~\bibnamefont
  {Aaij}} \emph {et~al.} (\bibinfo {collaboration} {LHCb}),\ }\href {\doibase
  10.1103/PhysRevLett.131.111802} {\bibfield  {journal} {\bibinfo  {journal}
  {Phys. Rev. Lett.}\ }\textbf {\bibinfo {volume} {131}},\ \bibinfo {pages}
  {111802} (\bibinfo {year} {2023}{\natexlab{a}})},\ \Eprint
  {http://arxiv.org/abs/2302.02886} {arXiv:2302.02886 [hep-ex]} \BibitemShut
  {NoStop}%
\bibitem [{\citenamefont {Mathad}(2023)}]{Mathad:2023zxi}%
  \BibitemOpen
  \bibfield  {author} {\bibinfo {author} {\bibfnamefont {A.}~\bibnamefont
  {Mathad}} (\bibinfo {collaboration} {LHCb}),\ }in\ \href@noop {} {\emph
  {\bibinfo {booktitle} {{57th Rencontres de Moriond on Electroweak
  Interactions and Unified Theories}}}}\ (\bibinfo {year} {2023})\ \Eprint
  {http://arxiv.org/abs/2305.08133} {arXiv:2305.08133 [hep-ex]} \BibitemShut
  {NoStop}%
\bibitem [{\citenamefont {Aaij}\ \emph
  {et~al.}(2018{\natexlab{b}})\citenamefont {Aaij} \emph
  {et~al.}}]{LHCb:2017vlu}%
  \BibitemOpen
  \bibfield  {author} {\bibinfo {author} {\bibfnamefont {R.}~\bibnamefont
  {Aaij}} \emph {et~al.} (\bibinfo {collaboration} {LHCb}),\ }\href {\doibase
  10.1103/PhysRevLett.120.121801} {\bibfield  {journal} {\bibinfo  {journal}
  {Phys. Rev. Lett.}\ }\textbf {\bibinfo {volume} {120}},\ \bibinfo {pages}
  {121801} (\bibinfo {year} {2018}{\natexlab{b}})},\ \Eprint
  {http://arxiv.org/abs/1711.05623} {arXiv:1711.05623 [hep-ex]} \BibitemShut
  {NoStop}%
\bibitem [{\citenamefont {Harrison}\ \emph {et~al.}(2020)\citenamefont
  {Harrison}, \citenamefont {Davies},\ and\ \citenamefont
  {Lytle}}]{Harrison:2020nrv}%
  \BibitemOpen
  \bibfield  {author} {\bibinfo {author} {\bibfnamefont {J.}~\bibnamefont
  {Harrison}}, \bibinfo {author} {\bibfnamefont {C.~T.~H.}\ \bibnamefont
  {Davies}}, \ and\ \bibinfo {author} {\bibfnamefont {A.}~\bibnamefont {Lytle}}
  (\bibinfo {collaboration} {LATTICE-HPQCD}),\ }\href {\doibase
  10.1103/PhysRevLett.125.222003} {\bibfield  {journal} {\bibinfo  {journal}
  {Phys. Rev. Lett.}\ }\textbf {\bibinfo {volume} {125}},\ \bibinfo {pages}
  {222003} (\bibinfo {year} {2020})},\ \Eprint
  {http://arxiv.org/abs/2007.06956} {arXiv:2007.06956 [hep-lat]} \BibitemShut
  {NoStop}%
\bibitem [{\citenamefont {Di~Canto}\ and\ \citenamefont
  {Meinel}(2022)}]{DiCanto:2022icc}%
  \BibitemOpen
  \bibfield  {author} {\bibinfo {author} {\bibfnamefont {A.}~\bibnamefont
  {Di~Canto}}\ and\ \bibinfo {author} {\bibfnamefont {S.}~\bibnamefont
  {Meinel}},\ }\href@noop {} {\  (\bibinfo {year} {2022})},\ \Eprint
  {http://arxiv.org/abs/2208.05403} {arXiv:2208.05403 [hep-ex]} \BibitemShut
  {NoStop}%
\bibitem [{\citenamefont {Crivellin}\ and\ \citenamefont
  {Matias}(2022)}]{Crivellin:2022qcj}%
  \BibitemOpen
  \bibfield  {author} {\bibinfo {author} {\bibfnamefont {A.}~\bibnamefont
  {Crivellin}}\ and\ \bibinfo {author} {\bibfnamefont {J.}~\bibnamefont
  {Matias}},\ }in\ \href@noop {} {\emph {\bibinfo {booktitle} {{1st Pan-African
  Astro-Particle and Collider Physics Workshop}}}}\ (\bibinfo {year} {2022})\
  \Eprint {http://arxiv.org/abs/2204.12175} {arXiv:2204.12175 [hep-ph]}
  \BibitemShut {NoStop}%
\bibitem [{\citenamefont {Bevan}\ \emph {et~al.}(2014)\citenamefont {Bevan}
  \emph {et~al.}}]{BaBar:2014omp}%
  \BibitemOpen
  \bibfield  {author} {\bibinfo {author} {\bibfnamefont {A.~J.}\ \bibnamefont
  {Bevan}} \emph {et~al.} (\bibinfo {collaboration} {BaBar, Belle}),\ }\href
  {\doibase 10.1140/epjc/s10052-014-3026-9} {\bibfield  {journal} {\bibinfo
  {journal} {Eur. Phys. J. C}\ }\textbf {\bibinfo {volume} {74}},\ \bibinfo
  {pages} {3026} (\bibinfo {year} {2014})},\ \Eprint
  {http://arxiv.org/abs/1406.6311} {arXiv:1406.6311 [hep-ex]} \BibitemShut
  {NoStop}%
\bibitem [{\citenamefont {Hayrapetyan}\ \emph {et~al.}(2024)\citenamefont
  {Hayrapetyan} \emph {et~al.}}]{CMS:2024syx}%
  \BibitemOpen
  \bibfield  {author} {\bibinfo {author} {\bibfnamefont {A.}~\bibnamefont
  {Hayrapetyan}} \emph {et~al.} (\bibinfo {collaboration} {CMS,
  EMAIL:cms-publication-committee-chair@cern.ch}),\ }\href {\doibase
  10.1088/1361-6633/ad4e65} {\bibfield  {journal} {\bibinfo  {journal} {Rept.
  Prog. Phys.}\ }\textbf {\bibinfo {volume} {87}},\ \bibinfo {pages} {077802}
  (\bibinfo {year} {2024})},\ \Eprint {http://arxiv.org/abs/2401.07090}
  {arXiv:2401.07090 [hep-ex]} \BibitemShut {NoStop}%
\bibitem [{\citenamefont {Aaij}\ \emph
  {et~al.}(2023{\natexlab{b}})\citenamefont {Aaij} \emph
  {et~al.}}]{LHCb:2022qnv}%
  \BibitemOpen
  \bibfield  {author} {\bibinfo {author} {\bibfnamefont {R.}~\bibnamefont
  {Aaij}} \emph {et~al.} (\bibinfo {collaboration} {LHCb}),\ }\href {\doibase
  10.1103/PhysRevLett.131.051803} {\bibfield  {journal} {\bibinfo  {journal}
  {Phys. Rev. Lett.}\ }\textbf {\bibinfo {volume} {131}},\ \bibinfo {pages}
  {051803} (\bibinfo {year} {2023}{\natexlab{b}})},\ \Eprint
  {http://arxiv.org/abs/2212.09152} {arXiv:2212.09152 [hep-ex]} \BibitemShut
  {NoStop}%
\bibitem [{\citenamefont {Aaij}\ \emph
  {et~al.}(2022{\natexlab{a}})\citenamefont {Aaij} \emph
  {et~al.}}]{LHCb:2021trn}%
  \BibitemOpen
  \bibfield  {author} {\bibinfo {author} {\bibfnamefont {R.}~\bibnamefont
  {Aaij}} \emph {et~al.} (\bibinfo {collaboration} {LHCb}),\ }\href {\doibase
  10.1038/s41567-023-02095-3} {\bibfield  {journal} {\bibinfo  {journal}
  {Nature Phys.}\ }\textbf {\bibinfo {volume} {18}},\ \bibinfo {pages} {277}
  (\bibinfo {year} {2022}{\natexlab{a}})},\ \bibinfo {note} {[Addendum: Nature
  Phys. 19, (2023)]},\ \Eprint {http://arxiv.org/abs/2103.11769}
  {arXiv:2103.11769 [hep-ex]} \BibitemShut {NoStop}%
\bibitem [{\citenamefont {Aaij}\ \emph
  {et~al.}(2022{\natexlab{b}})\citenamefont {Aaij} \emph
  {et~al.}}]{LHCb:2021lvy}%
  \BibitemOpen
  \bibfield  {author} {\bibinfo {author} {\bibfnamefont {R.}~\bibnamefont
  {Aaij}} \emph {et~al.} (\bibinfo {collaboration} {LHCb}),\ }\href {\doibase
  10.1103/PhysRevLett.128.191802} {\bibfield  {journal} {\bibinfo  {journal}
  {Phys. Rev. Lett.}\ }\textbf {\bibinfo {volume} {128}},\ \bibinfo {pages}
  {191802} (\bibinfo {year} {2022}{\natexlab{b}})},\ \Eprint
  {http://arxiv.org/abs/2110.09501} {arXiv:2110.09501 [hep-ex]} \BibitemShut
  {NoStop}%
\bibitem [{\citenamefont {Abdesselam}\ \emph {et~al.}(2021)\citenamefont
  {Abdesselam} \emph {et~al.}}]{Belle:2019oag}%
  \BibitemOpen
  \bibfield  {author} {\bibinfo {author} {\bibfnamefont {A.}~\bibnamefont
  {Abdesselam}} \emph {et~al.} (\bibinfo {collaboration} {Belle}),\ }\href
  {\doibase 10.1103/PhysRevLett.126.161801} {\bibfield  {journal} {\bibinfo
  {journal} {Phys. Rev. Lett.}\ }\textbf {\bibinfo {volume} {126}},\ \bibinfo
  {pages} {161801} (\bibinfo {year} {2021})},\ \Eprint
  {http://arxiv.org/abs/1904.02440} {arXiv:1904.02440 [hep-ex]} \BibitemShut
  {NoStop}%
\bibitem [{\citenamefont {Choudhury}\ \emph {et~al.}(2021)\citenamefont
  {Choudhury} \emph {et~al.}}]{BELLE:2019xld}%
  \BibitemOpen
  \bibfield  {author} {\bibinfo {author} {\bibfnamefont {S.}~\bibnamefont
  {Choudhury}} \emph {et~al.} (\bibinfo {collaboration} {BELLE}),\ }\href
  {\doibase 10.1007/JHEP03(2021)105} {\bibfield  {journal} {\bibinfo  {journal}
  {JHEP}\ }\textbf {\bibinfo {volume} {03}},\ \bibinfo {pages} {105} (\bibinfo
  {year} {2021})},\ \Eprint {http://arxiv.org/abs/1908.01848} {arXiv:1908.01848
  [hep-ex]} \BibitemShut {NoStop}%
\bibitem [{\citenamefont {Aaij}\ \emph {et~al.}(2020)\citenamefont {Aaij} \emph
  {et~al.}}]{LHCb:2020lmf}%
  \BibitemOpen
  \bibfield  {author} {\bibinfo {author} {\bibfnamefont {R.}~\bibnamefont
  {Aaij}} \emph {et~al.} (\bibinfo {collaboration} {LHCb}),\ }\href {\doibase
  10.1103/PhysRevLett.125.011802} {\bibfield  {journal} {\bibinfo  {journal}
  {Phys. Rev. Lett.}\ }\textbf {\bibinfo {volume} {125}},\ \bibinfo {pages}
  {011802} (\bibinfo {year} {2020})},\ \Eprint
  {http://arxiv.org/abs/2003.04831} {arXiv:2003.04831 [hep-ex]} \BibitemShut
  {NoStop}%
\bibitem [{\citenamefont {Aaij}\ \emph {et~al.}(2019)\citenamefont {Aaij} \emph
  {et~al.}}]{LHCb:2019hip}%
  \BibitemOpen
  \bibfield  {author} {\bibinfo {author} {\bibfnamefont {R.}~\bibnamefont
  {Aaij}} \emph {et~al.} (\bibinfo {collaboration} {LHCb}),\ }\href {\doibase
  10.1103/PhysRevLett.122.191801} {\bibfield  {journal} {\bibinfo  {journal}
  {Phys. Rev. Lett.}\ }\textbf {\bibinfo {volume} {122}},\ \bibinfo {pages}
  {191801} (\bibinfo {year} {2019})},\ \Eprint
  {http://arxiv.org/abs/1903.09252} {arXiv:1903.09252 [hep-ex]} \BibitemShut
  {NoStop}%
\bibitem [{\citenamefont {Aaij}\ \emph
  {et~al.}(2017{\natexlab{a}})\citenamefont {Aaij} \emph
  {et~al.}}]{LHCb:2017avl}%
  \BibitemOpen
  \bibfield  {author} {\bibinfo {author} {\bibfnamefont {R.}~\bibnamefont
  {Aaij}} \emph {et~al.} (\bibinfo {collaboration} {LHCb}),\ }\href {\doibase
  10.1007/JHEP08(2017)055} {\bibfield  {journal} {\bibinfo  {journal} {JHEP}\
  }\textbf {\bibinfo {volume} {08}},\ \bibinfo {pages} {055} (\bibinfo {year}
  {2017}{\natexlab{a}})},\ \Eprint {http://arxiv.org/abs/1705.05802}
  {arXiv:1705.05802 [hep-ex]} \BibitemShut {NoStop}%
\bibitem [{\citenamefont {Aaij}\ \emph
  {et~al.}(2016{\natexlab{a}})\citenamefont {Aaij} \emph
  {et~al.}}]{LHCb:2015svh}%
  \BibitemOpen
  \bibfield  {author} {\bibinfo {author} {\bibfnamefont {R.}~\bibnamefont
  {Aaij}} \emph {et~al.} (\bibinfo {collaboration} {LHCb}),\ }\href {\doibase
  10.1007/JHEP02(2016)104} {\bibfield  {journal} {\bibinfo  {journal} {JHEP}\
  }\textbf {\bibinfo {volume} {02}},\ \bibinfo {pages} {104} (\bibinfo {year}
  {2016}{\natexlab{a}})},\ \Eprint {http://arxiv.org/abs/1512.04442}
  {arXiv:1512.04442 [hep-ex]} \BibitemShut {NoStop}%
\bibitem [{\citenamefont {Aaij}\ \emph {et~al.}(2014)\citenamefont {Aaij} \emph
  {et~al.}}]{LHCb:2014vgu}%
  \BibitemOpen
  \bibfield  {author} {\bibinfo {author} {\bibfnamefont {R.}~\bibnamefont
  {Aaij}} \emph {et~al.} (\bibinfo {collaboration} {LHCb}),\ }\href {\doibase
  10.1103/PhysRevLett.113.151601} {\bibfield  {journal} {\bibinfo  {journal}
  {Phys. Rev. Lett.}\ }\textbf {\bibinfo {volume} {113}},\ \bibinfo {pages}
  {151601} (\bibinfo {year} {2014})},\ \Eprint {http://arxiv.org/abs/1406.6482}
  {arXiv:1406.6482 [hep-ex]} \BibitemShut {NoStop}%
\bibitem [{\citenamefont {Aaij}\ \emph {et~al.}(2013)\citenamefont {Aaij} \emph
  {et~al.}}]{LHCb:2013ghj}%
  \BibitemOpen
  \bibfield  {author} {\bibinfo {author} {\bibfnamefont {R.}~\bibnamefont
  {Aaij}} \emph {et~al.} (\bibinfo {collaboration} {LHCb}),\ }\href {\doibase
  10.1103/PhysRevLett.111.191801} {\bibfield  {journal} {\bibinfo  {journal}
  {Phys. Rev. Lett.}\ }\textbf {\bibinfo {volume} {111}},\ \bibinfo {pages}
  {191801} (\bibinfo {year} {2013})},\ \Eprint {http://arxiv.org/abs/1308.1707}
  {arXiv:1308.1707 [hep-ex]} \BibitemShut {NoStop}%
\bibitem [{\citenamefont {Lees}\ \emph {et~al.}(2017)\citenamefont {Lees} \emph
  {et~al.}}]{BaBar:2016wgb}%
  \BibitemOpen
  \bibfield  {author} {\bibinfo {author} {\bibfnamefont {J.~P.}\ \bibnamefont
  {Lees}} \emph {et~al.} (\bibinfo {collaboration} {BaBar}),\ }\href {\doibase
  10.1103/PhysRevLett.118.031802} {\bibfield  {journal} {\bibinfo  {journal}
  {Phys. Rev. Lett.}\ }\textbf {\bibinfo {volume} {118}},\ \bibinfo {pages}
  {031802} (\bibinfo {year} {2017})},\ \Eprint
  {http://arxiv.org/abs/1605.09637} {arXiv:1605.09637 [hep-ex]} \BibitemShut
  {NoStop}%
\bibitem [{\citenamefont {Lees}\ \emph {et~al.}(2012)\citenamefont {Lees} \emph
  {et~al.}}]{BaBar:2012mrf}%
  \BibitemOpen
  \bibfield  {author} {\bibinfo {author} {\bibfnamefont {J.~P.}\ \bibnamefont
  {Lees}} \emph {et~al.} (\bibinfo {collaboration} {BaBar}),\ }\href {\doibase
  10.1103/PhysRevD.86.032012} {\bibfield  {journal} {\bibinfo  {journal} {Phys.
  Rev. D}\ }\textbf {\bibinfo {volume} {86}},\ \bibinfo {pages} {032012}
  (\bibinfo {year} {2012})},\ \Eprint {http://arxiv.org/abs/1204.3933}
  {arXiv:1204.3933 [hep-ex]} \BibitemShut {NoStop}%
\bibitem [{\citenamefont {Gubernari}\ \emph {et~al.}(2021)\citenamefont
  {Gubernari}, \citenamefont {van Dyk},\ and\ \citenamefont
  {Virto}}]{Gubernari:2020eft}%
  \BibitemOpen
  \bibfield  {author} {\bibinfo {author} {\bibfnamefont {N.}~\bibnamefont
  {Gubernari}}, \bibinfo {author} {\bibfnamefont {D.}~\bibnamefont {van Dyk}},
  \ and\ \bibinfo {author} {\bibfnamefont {J.}~\bibnamefont {Virto}},\ }\href
  {\doibase 10.1007/JHEP02(2021)088} {\bibfield  {journal} {\bibinfo  {journal}
  {JHEP}\ }\textbf {\bibinfo {volume} {02}},\ \bibinfo {pages} {088} (\bibinfo
  {year} {2021})},\ \Eprint {http://arxiv.org/abs/2011.09813} {arXiv:2011.09813
  [hep-ph]} \BibitemShut {NoStop}%
\bibitem [{\citenamefont {Khodjamirian}\ \emph {et~al.}(2010)\citenamefont
  {Khodjamirian}, \citenamefont {Mannel}, \citenamefont {Pivovarov},\ and\
  \citenamefont {Wang}}]{Khodjamirian:2010vf}%
  \BibitemOpen
  \bibfield  {author} {\bibinfo {author} {\bibfnamefont {A.}~\bibnamefont
  {Khodjamirian}}, \bibinfo {author} {\bibfnamefont {T.}~\bibnamefont
  {Mannel}}, \bibinfo {author} {\bibfnamefont {A.~A.}\ \bibnamefont
  {Pivovarov}}, \ and\ \bibinfo {author} {\bibfnamefont {Y.~M.}\ \bibnamefont
  {Wang}},\ }\href {\doibase 10.1007/JHEP09(2010)089} {\bibfield  {journal}
  {\bibinfo  {journal} {JHEP}\ }\textbf {\bibinfo {volume} {09}},\ \bibinfo
  {pages} {089} (\bibinfo {year} {2010})},\ \Eprint
  {http://arxiv.org/abs/1006.4945} {arXiv:1006.4945 [hep-ph]} \BibitemShut
  {NoStop}%
\bibitem [{\citenamefont {Bordone}\ \emph {et~al.}(2024)\citenamefont
  {Bordone}, \citenamefont {isidori}, \citenamefont {M\"achler},\ and\
  \citenamefont {Tinari}}]{Bordone:2024hui}%
  \BibitemOpen
  \bibfield  {author} {\bibinfo {author} {\bibfnamefont {M.}~\bibnamefont
  {Bordone}}, \bibinfo {author} {\bibfnamefont {G.}~\bibnamefont {isidori}},
  \bibinfo {author} {\bibfnamefont {S.}~\bibnamefont {M\"achler}}, \ and\
  \bibinfo {author} {\bibfnamefont {A.}~\bibnamefont {Tinari}},\ }\href@noop {}
  {\  (\bibinfo {year} {2024})},\ \Eprint {http://arxiv.org/abs/2401.18007}
  {arXiv:2401.18007 [hep-ph]} \BibitemShut {NoStop}%
\bibitem [{\citenamefont {Alguer\'o}\ \emph {et~al.}(2023)\citenamefont
  {Alguer\'o}, \citenamefont {Biswas}, \citenamefont {Capdevila}, \citenamefont
  {Descotes-Genon}, \citenamefont {Matias},\ and\ \citenamefont
  {Novoa-Brunet}}]{Alguero:2023jeh}%
  \BibitemOpen
  \bibfield  {author} {\bibinfo {author} {\bibfnamefont {M.}~\bibnamefont
  {Alguer\'o}}, \bibinfo {author} {\bibfnamefont {A.}~\bibnamefont {Biswas}},
  \bibinfo {author} {\bibfnamefont {B.}~\bibnamefont {Capdevila}}, \bibinfo
  {author} {\bibfnamefont {S.}~\bibnamefont {Descotes-Genon}}, \bibinfo
  {author} {\bibfnamefont {J.}~\bibnamefont {Matias}}, \ and\ \bibinfo {author}
  {\bibfnamefont {M.}~\bibnamefont {Novoa-Brunet}},\ }\href {\doibase
  10.1140/epjc/s10052-023-11824-0} {\bibfield  {journal} {\bibinfo  {journal}
  {Eur. Phys. J. C}\ }\textbf {\bibinfo {volume} {83}},\ \bibinfo {pages} {648}
  (\bibinfo {year} {2023})},\ \Eprint {http://arxiv.org/abs/2304.07330}
  {arXiv:2304.07330 [hep-ph]} \BibitemShut {NoStop}%
\bibitem [{\citenamefont {Gubernari}\ \emph {et~al.}(2022)\citenamefont
  {Gubernari}, \citenamefont {Reboud}, \citenamefont {van Dyk},\ and\
  \citenamefont {Virto}}]{Gubernari:2022hxn}%
  \BibitemOpen
  \bibfield  {author} {\bibinfo {author} {\bibfnamefont {N.}~\bibnamefont
  {Gubernari}}, \bibinfo {author} {\bibfnamefont {M.}~\bibnamefont {Reboud}},
  \bibinfo {author} {\bibfnamefont {D.}~\bibnamefont {van Dyk}}, \ and\
  \bibinfo {author} {\bibfnamefont {J.}~\bibnamefont {Virto}},\ }\href
  {\doibase 10.1007/JHEP09(2022)133} {\bibfield  {journal} {\bibinfo  {journal}
  {JHEP}\ }\textbf {\bibinfo {volume} {09}},\ \bibinfo {pages} {133} (\bibinfo
  {year} {2022})},\ \Eprint {http://arxiv.org/abs/2206.03797} {arXiv:2206.03797
  [hep-ph]} \BibitemShut {NoStop}%
\bibitem [{\citenamefont {Hurth}\ \emph {et~al.}(2021)\citenamefont {Hurth},
  \citenamefont {Mahmoudi},\ and\ \citenamefont {Neshatpour}}]{Hurth:2020ehu}%
  \BibitemOpen
  \bibfield  {author} {\bibinfo {author} {\bibfnamefont {T.}~\bibnamefont
  {Hurth}}, \bibinfo {author} {\bibfnamefont {F.}~\bibnamefont {Mahmoudi}}, \
  and\ \bibinfo {author} {\bibfnamefont {S.}~\bibnamefont {Neshatpour}},\
  }\href {\doibase 10.1103/PhysRevD.103.095020} {\bibfield  {journal} {\bibinfo
   {journal} {Phys. Rev. D}\ }\textbf {\bibinfo {volume} {103}},\ \bibinfo
  {pages} {095020} (\bibinfo {year} {2021})},\ \Eprint
  {http://arxiv.org/abs/2012.12207} {arXiv:2012.12207 [hep-ph]} \BibitemShut
  {NoStop}%
\bibitem [{\citenamefont {Ciuchini}\ \emph {et~al.}(2019)\citenamefont
  {Ciuchini}, \citenamefont {Coutinho}, \citenamefont {Fedele}, \citenamefont
  {Franco}, \citenamefont {Paul}, \citenamefont {Silvestrini},\ and\
  \citenamefont {Valli}}]{Ciuchini:2019usw}%
  \BibitemOpen
  \bibfield  {author} {\bibinfo {author} {\bibfnamefont {M.}~\bibnamefont
  {Ciuchini}}, \bibinfo {author} {\bibfnamefont {A.~M.}\ \bibnamefont
  {Coutinho}}, \bibinfo {author} {\bibfnamefont {M.}~\bibnamefont {Fedele}},
  \bibinfo {author} {\bibfnamefont {E.}~\bibnamefont {Franco}}, \bibinfo
  {author} {\bibfnamefont {A.}~\bibnamefont {Paul}}, \bibinfo {author}
  {\bibfnamefont {L.}~\bibnamefont {Silvestrini}}, \ and\ \bibinfo {author}
  {\bibfnamefont {M.}~\bibnamefont {Valli}},\ }\href {\doibase
  10.1140/epjc/s10052-019-7210-9} {\bibfield  {journal} {\bibinfo  {journal}
  {Eur. Phys. J. C}\ }\textbf {\bibinfo {volume} {79}},\ \bibinfo {pages} {719}
  (\bibinfo {year} {2019})},\ \Eprint {http://arxiv.org/abs/1903.09632}
  {arXiv:1903.09632 [hep-ph]} \BibitemShut {NoStop}%
\bibitem [{\citenamefont {Alguer\'o}\ \emph {et~al.}(2019)\citenamefont
  {Alguer\'o}, \citenamefont {Capdevila}, \citenamefont {Descotes-Genon},
  \citenamefont {Masjuan},\ and\ \citenamefont {Matias}}]{Alguero:2018nvb}%
  \BibitemOpen
  \bibfield  {author} {\bibinfo {author} {\bibfnamefont {M.}~\bibnamefont
  {Alguer\'o}}, \bibinfo {author} {\bibfnamefont {B.}~\bibnamefont
  {Capdevila}}, \bibinfo {author} {\bibfnamefont {S.}~\bibnamefont
  {Descotes-Genon}}, \bibinfo {author} {\bibfnamefont {P.}~\bibnamefont
  {Masjuan}}, \ and\ \bibinfo {author} {\bibfnamefont {J.}~\bibnamefont
  {Matias}},\ }\href {\doibase 10.1103/PhysRevD.99.075017} {\bibfield
  {journal} {\bibinfo  {journal} {Phys. Rev. D}\ }\textbf {\bibinfo {volume}
  {99}},\ \bibinfo {pages} {075017} (\bibinfo {year} {2019})},\ \Eprint
  {http://arxiv.org/abs/1809.08447} {arXiv:1809.08447 [hep-ph]} \BibitemShut
  {NoStop}%
\bibitem [{\citenamefont {Descotes-Genon}\ \emph {et~al.}(2014)\citenamefont
  {Descotes-Genon}, \citenamefont {Hofer}, \citenamefont {Matias},\ and\
  \citenamefont {Virto}}]{Descotes-Genon:2014uoa}%
  \BibitemOpen
  \bibfield  {author} {\bibinfo {author} {\bibfnamefont {S.}~\bibnamefont
  {Descotes-Genon}}, \bibinfo {author} {\bibfnamefont {L.}~\bibnamefont
  {Hofer}}, \bibinfo {author} {\bibfnamefont {J.}~\bibnamefont {Matias}}, \
  and\ \bibinfo {author} {\bibfnamefont {J.}~\bibnamefont {Virto}},\ }\href
  {\doibase 10.1007/JHEP12(2014)125} {\bibfield  {journal} {\bibinfo  {journal}
  {JHEP}\ }\textbf {\bibinfo {volume} {12}},\ \bibinfo {pages} {125} (\bibinfo
  {year} {2014})},\ \Eprint {http://arxiv.org/abs/1407.8526} {arXiv:1407.8526
  [hep-ph]} \BibitemShut {NoStop}%
\bibitem [{\citenamefont {Seuthe}(2023)}]{Seuthe:2023xek}%
  \BibitemOpen
  \bibfield  {author} {\bibinfo {author} {\bibfnamefont {A.}~\bibnamefont
  {Seuthe}} (\bibinfo {collaboration} {LHCb}),\ }in\ \href@noop {} {\emph
  {\bibinfo {booktitle} {{57th Rencontres de Moriond on QCD and High Energy
  Interactions}}}}\ (\bibinfo {year} {2023})\ \Eprint
  {http://arxiv.org/abs/2305.08216} {arXiv:2305.08216 [hep-ex]} \BibitemShut
  {NoStop}%
\bibitem [{\citenamefont {Aaij}\ \emph
  {et~al.}(2023{\natexlab{c}})\citenamefont {Aaij} \emph
  {et~al.}}]{LHCb:2022vje}%
  \BibitemOpen
  \bibfield  {author} {\bibinfo {author} {\bibfnamefont {R.}~\bibnamefont
  {Aaij}} \emph {et~al.} (\bibinfo {collaboration} {LHCb}),\ }\href {\doibase
  10.1103/PhysRevD.108.032002} {\bibfield  {journal} {\bibinfo  {journal}
  {Phys. Rev. D}\ }\textbf {\bibinfo {volume} {108}},\ \bibinfo {pages}
  {032002} (\bibinfo {year} {2023}{\natexlab{c}})},\ \Eprint
  {http://arxiv.org/abs/2212.09153} {arXiv:2212.09153 [hep-ex]} \BibitemShut
  {NoStop}%
\bibitem [{\citenamefont {Adachi}\ \emph {et~al.}(2024)\citenamefont {Adachi}
  \emph {et~al.}}]{Belle-II:2024pgs}%
  \BibitemOpen
  \bibfield  {author} {\bibinfo {author} {\bibfnamefont {I.}~\bibnamefont
  {Adachi}} \emph {et~al.} (\bibinfo {collaboration} {Belle-II, Belle}),\
  }\href@noop {} {\  (\bibinfo {year} {2024})},\ \Eprint
  {http://arxiv.org/abs/2404.08133} {arXiv:2404.08133 [hep-ex]} \BibitemShut
  {NoStop}%
\bibitem [{\citenamefont {Aaij}\ \emph
  {et~al.}(2017{\natexlab{b}})\citenamefont {Aaij} \emph
  {et~al.}}]{LHCb:2017lpt}%
  \BibitemOpen
  \bibfield  {author} {\bibinfo {author} {\bibfnamefont {R.}~\bibnamefont
  {Aaij}} \emph {et~al.} (\bibinfo {collaboration} {LHCb}),\ }\href {\doibase
  10.1007/JHEP04(2017)029} {\bibfield  {journal} {\bibinfo  {journal} {JHEP}\
  }\textbf {\bibinfo {volume} {04}},\ \bibinfo {pages} {029} (\bibinfo {year}
  {2017}{\natexlab{b}})},\ \Eprint {http://arxiv.org/abs/1701.08705}
  {arXiv:1701.08705 [hep-ex]} \BibitemShut {NoStop}%
\bibitem [{\citenamefont {Aaij}\ \emph
  {et~al.}(2015{\natexlab{b}})\citenamefont {Aaij} \emph
  {et~al.}}]{LHCb:2015hsa}%
  \BibitemOpen
  \bibfield  {author} {\bibinfo {author} {\bibfnamefont {R.}~\bibnamefont
  {Aaij}} \emph {et~al.} (\bibinfo {collaboration} {LHCb}),\ }\href {\doibase
  10.1007/JHEP10(2015)034} {\bibfield  {journal} {\bibinfo  {journal} {JHEP}\
  }\textbf {\bibinfo {volume} {10}},\ \bibinfo {pages} {034} (\bibinfo {year}
  {2015}{\natexlab{b}})},\ \Eprint {http://arxiv.org/abs/1509.00414}
  {arXiv:1509.00414 [hep-ex]} \BibitemShut {NoStop}%
\bibitem [{\citenamefont {Aaij}\ \emph {et~al.}(2012)\citenamefont {Aaij} \emph
  {et~al.}}]{LHCb:2012de}%
  \BibitemOpen
  \bibfield  {author} {\bibinfo {author} {\bibfnamefont {R.}~\bibnamefont
  {Aaij}} \emph {et~al.} (\bibinfo {collaboration} {LHCb Collaboration}),\
  }\href {\doibase 10.1007/JHEP12(2012)125} {\bibfield  {journal} {\bibinfo
  {journal} {JHEP}\ }\textbf {\bibinfo {volume} {12}},\ \bibinfo {pages} {125}
  (\bibinfo {year} {2012})},\ \Eprint {http://arxiv.org/abs/1210.2645}
  {arXiv:1210.2645 [hep-ex]} \BibitemShut {NoStop}%
\bibitem [{\citenamefont {Aaij}\ \emph
  {et~al.}(2018{\natexlab{c}})\citenamefont {Aaij} \emph
  {et~al.}}]{Aaij:2018jhg}%
  \BibitemOpen
  \bibfield  {author} {\bibinfo {author} {\bibfnamefont {R.}~\bibnamefont
  {Aaij}} \emph {et~al.} (\bibinfo {collaboration} {LHCb Collaboration}),\
  }\href {\doibase 10.1007/JHEP07(2018)020} {\bibfield  {journal} {\bibinfo
  {journal} {JHEP}\ }\textbf {\bibinfo {volume} {07}},\ \bibinfo {pages} {020}
  (\bibinfo {year} {2018}{\natexlab{c}})},\ \Eprint
  {http://arxiv.org/abs/1804.07167} {arXiv:1804.07167 [hep-ex]} \BibitemShut
  {NoStop}%
\bibitem [{\citenamefont {Artuso}\ \emph {et~al.}(2022)\citenamefont {Artuso},
  \citenamefont {Isidori},\ and\ \citenamefont {Stone}}]{Artuso:2022ijh}%
  \BibitemOpen
  \bibfield  {author} {\bibinfo {author} {\bibfnamefont {M.}~\bibnamefont
  {Artuso}}, \bibinfo {author} {\bibfnamefont {G.}~\bibnamefont {Isidori}}, \
  and\ \bibinfo {author} {\bibfnamefont {S.}~\bibnamefont {Stone}},\ }\href
  {\doibase 10.1142/12696} {\emph {\bibinfo {title} {{New Physics in b
  Decays}}}}\ (\bibinfo  {publisher} {World Scientific},\ \bibinfo {year}
  {2022})\BibitemShut {NoStop}%
\bibitem [{\citenamefont {London}\ and\ \citenamefont
  {Matias}(2022)}]{London:2021lfn}%
  \BibitemOpen
  \bibfield  {author} {\bibinfo {author} {\bibfnamefont {D.}~\bibnamefont
  {London}}\ and\ \bibinfo {author} {\bibfnamefont {J.}~\bibnamefont
  {Matias}},\ }\href {\doibase 10.1146/annurev-nucl-102020-090209} {\bibfield
  {journal} {\bibinfo  {journal} {Ann. Rev. Nucl. Part. Sci.}\ }\textbf
  {\bibinfo {volume} {72}},\ \bibinfo {pages} {37} (\bibinfo {year} {2022})},\
  \Eprint {http://arxiv.org/abs/2110.13270} {arXiv:2110.13270 [hep-ph]}
  \BibitemShut {NoStop}%
\bibitem [{\citenamefont {Aaij}\ \emph {et~al.}(2024)\citenamefont {Aaij} \emph
  {et~al.}}]{LHCb:2023lyb}%
  \BibitemOpen
  \bibfield  {author} {\bibinfo {author} {\bibfnamefont {R.}~\bibnamefont
  {Aaij}} \emph {et~al.} (\bibinfo {collaboration} {LHCb}),\ }\href {\doibase
  10.1007/JHEP02(2024)032} {\bibfield  {journal} {\bibinfo  {journal} {JHEP}\
  }\textbf {\bibinfo {volume} {02}},\ \bibinfo {pages} {032} (\bibinfo {year}
  {2024})},\ \Eprint {http://arxiv.org/abs/2308.06162} {arXiv:2308.06162
  [hep-ex]} \BibitemShut {NoStop}%
\bibitem [{\citenamefont {Geng}\ \emph {et~al.}(2002)\citenamefont {Geng},
  \citenamefont {Hwang},\ and\ \citenamefont {Liu}}]{Geng:2001vy}%
  \BibitemOpen
  \bibfield  {author} {\bibinfo {author} {\bibfnamefont {C.~Q.}\ \bibnamefont
  {Geng}}, \bibinfo {author} {\bibfnamefont {C.-W.}\ \bibnamefont {Hwang}}, \
  and\ \bibinfo {author} {\bibfnamefont {C.~C.}\ \bibnamefont {Liu}},\ }\href
  {\doibase 10.1103/PhysRevD.65.094037} {\bibfield  {journal} {\bibinfo
  {journal} {Phys. Rev. D}\ }\textbf {\bibinfo {volume} {65}},\ \bibinfo
  {pages} {094037} (\bibinfo {year} {2002})},\ \Eprint
  {http://arxiv.org/abs/hep-ph/0110376} {arXiv:hep-ph/0110376} \BibitemShut
  {NoStop}%
\bibitem [{\citenamefont {Azizi}\ \emph {et~al.}(2008)\citenamefont {Azizi},
  \citenamefont {Falahati}, \citenamefont {Bashiry},\ and\ \citenamefont
  {Zebarjad}}]{Azizi:2008vv}%
  \BibitemOpen
  \bibfield  {author} {\bibinfo {author} {\bibfnamefont {K.}~\bibnamefont
  {Azizi}}, \bibinfo {author} {\bibfnamefont {F.}~\bibnamefont {Falahati}},
  \bibinfo {author} {\bibfnamefont {V.}~\bibnamefont {Bashiry}}, \ and\
  \bibinfo {author} {\bibfnamefont {S.~M.}\ \bibnamefont {Zebarjad}},\ }\href
  {\doibase 10.1103/PhysRevD.77.114024} {\bibfield  {journal} {\bibinfo
  {journal} {Phys. Rev. D}\ }\textbf {\bibinfo {volume} {77}},\ \bibinfo
  {pages} {114024} (\bibinfo {year} {2008})},\ \Eprint
  {http://arxiv.org/abs/0806.0583} {arXiv:0806.0583 [hep-ph]} \BibitemShut
  {NoStop}%
\bibitem [{\citenamefont {Azizi}\ and\ \citenamefont
  {Khosravi}(2008)}]{Azizi:2008vy}%
  \BibitemOpen
  \bibfield  {author} {\bibinfo {author} {\bibfnamefont {K.}~\bibnamefont
  {Azizi}}\ and\ \bibinfo {author} {\bibfnamefont {R.}~\bibnamefont
  {Khosravi}},\ }\href {\doibase 10.1103/PhysRevD.78.036005} {\bibfield
  {journal} {\bibinfo  {journal} {Phys. Rev. D}\ }\textbf {\bibinfo {volume}
  {78}},\ \bibinfo {pages} {036005} (\bibinfo {year} {2008})},\ \Eprint
  {http://arxiv.org/abs/0806.0590} {arXiv:0806.0590 [hep-ph]} \BibitemShut
  {NoStop}%
\bibitem [{\citenamefont {Faessler}\ \emph {et~al.}(2002)\citenamefont
  {Faessler}, \citenamefont {Gutsche}, \citenamefont {Ivanov}, \citenamefont
  {Korner},\ and\ \citenamefont {Lyubovitskij}}]{Faessler:2002ut}%
  \BibitemOpen
  \bibfield  {author} {\bibinfo {author} {\bibfnamefont {A.}~\bibnamefont
  {Faessler}}, \bibinfo {author} {\bibfnamefont {T.}~\bibnamefont {Gutsche}},
  \bibinfo {author} {\bibfnamefont {M.~A.}\ \bibnamefont {Ivanov}}, \bibinfo
  {author} {\bibfnamefont {J.~G.}\ \bibnamefont {Korner}}, \ and\ \bibinfo
  {author} {\bibfnamefont {V.~E.}\ \bibnamefont {Lyubovitskij}},\ }\href
  {\doibase 10.1007/s1010502c0018} {\bibfield  {journal} {\bibinfo  {journal}
  {Eur. Phys. J. direct}\ }\textbf {\bibinfo {volume} {4}},\ \bibinfo {pages}
  {18} (\bibinfo {year} {2002})},\ \Eprint
  {http://arxiv.org/abs/hep-ph/0205287} {arXiv:hep-ph/0205287} \BibitemShut
  {NoStop}%
\bibitem [{\citenamefont {Choi}(2010)}]{Choi:2010ha}%
  \BibitemOpen
  \bibfield  {author} {\bibinfo {author} {\bibfnamefont {H.-M.}\ \bibnamefont
  {Choi}},\ }\href {\doibase 10.1103/PhysRevD.81.054003} {\bibfield  {journal}
  {\bibinfo  {journal} {Phys. Rev. D}\ }\textbf {\bibinfo {volume} {81}},\
  \bibinfo {pages} {054003} (\bibinfo {year} {2010})},\ \Eprint
  {http://arxiv.org/abs/1001.3432} {arXiv:1001.3432 [hep-ph]} \BibitemShut
  {NoStop}%
\bibitem [{\citenamefont {Wang}\ \emph {et~al.}(2014)\citenamefont {Wang},
  \citenamefont {Yu}, \citenamefont {L\"u},\ and\ \citenamefont
  {Xiao}}]{Wang:2014yia}%
  \BibitemOpen
  \bibfield  {author} {\bibinfo {author} {\bibfnamefont {W.-F.}\ \bibnamefont
  {Wang}}, \bibinfo {author} {\bibfnamefont {X.}~\bibnamefont {Yu}}, \bibinfo
  {author} {\bibfnamefont {C.-D.}\ \bibnamefont {L\"u}}, \ and\ \bibinfo
  {author} {\bibfnamefont {Z.-J.}\ \bibnamefont {Xiao}},\ }\href {\doibase
  10.1103/PhysRevD.90.094018} {\bibfield  {journal} {\bibinfo  {journal} {Phys.
  Rev. D}\ }\textbf {\bibinfo {volume} {90}},\ \bibinfo {pages} {094018}
  (\bibinfo {year} {2014})},\ \Eprint {http://arxiv.org/abs/1401.0391}
  {arXiv:1401.0391 [hep-ph]} \BibitemShut {NoStop}%
\bibitem [{\citenamefont {Yilmaz}(2012)}]{Yilmaz:2012ah}%
  \BibitemOpen
  \bibfield  {author} {\bibinfo {author} {\bibfnamefont {U.~O.}\ \bibnamefont
  {Yilmaz}},\ }\href {\doibase 10.1103/PhysRevD.85.115026} {\bibfield
  {journal} {\bibinfo  {journal} {Phys. Rev. D}\ }\textbf {\bibinfo {volume}
  {85}},\ \bibinfo {pages} {115026} (\bibinfo {year} {2012})},\ \Eprint
  {http://arxiv.org/abs/1204.1261} {arXiv:1204.1261 [hep-ph]} \BibitemShut
  {NoStop}%
\bibitem [{\citenamefont {Ebert}\ \emph {et~al.}(2010)\citenamefont {Ebert},
  \citenamefont {Faustov},\ and\ \citenamefont {Galkin}}]{Ebert:2010dv}%
  \BibitemOpen
  \bibfield  {author} {\bibinfo {author} {\bibfnamefont {D.}~\bibnamefont
  {Ebert}}, \bibinfo {author} {\bibfnamefont {R.~N.}\ \bibnamefont {Faustov}},
  \ and\ \bibinfo {author} {\bibfnamefont {V.~O.}\ \bibnamefont {Galkin}},\
  }\href {\doibase 10.1103/PhysRevD.82.034032} {\bibfield  {journal} {\bibinfo
  {journal} {Phys. Rev. D}\ }\textbf {\bibinfo {volume} {82}},\ \bibinfo
  {pages} {034032} (\bibinfo {year} {2010})},\ \Eprint
  {http://arxiv.org/abs/1006.4231} {arXiv:1006.4231 [hep-ph]} \BibitemShut
  {NoStop}%
\bibitem [{\citenamefont {Maji}\ \emph
  {et~al.}(2020{\natexlab{a}})\citenamefont {Maji}, \citenamefont {Mahata},
  \citenamefont {Nayek}, \citenamefont {Biswas},\ and\ \citenamefont
  {Sahoo}}]{Maji:2020zlq}%
  \BibitemOpen
  \bibfield  {author} {\bibinfo {author} {\bibfnamefont {P.}~\bibnamefont
  {Maji}}, \bibinfo {author} {\bibfnamefont {S.}~\bibnamefont {Mahata}},
  \bibinfo {author} {\bibfnamefont {P.}~\bibnamefont {Nayek}}, \bibinfo
  {author} {\bibfnamefont {S.}~\bibnamefont {Biswas}}, \ and\ \bibinfo {author}
  {\bibfnamefont {S.}~\bibnamefont {Sahoo}},\ }\href {\doibase
  10.1088/1674-1137/44/7/073106} {\bibfield  {journal} {\bibinfo  {journal}
  {Chin. Phys. C}\ }\textbf {\bibinfo {volume} {44}},\ \bibinfo {pages}
  {073106} (\bibinfo {year} {2020}{\natexlab{a}})},\ \Eprint
  {http://arxiv.org/abs/2003.12272} {arXiv:2003.12272 [hep-ph]} \BibitemShut
  {NoStop}%
\bibitem [{\citenamefont {Maji}\ \emph
  {et~al.}(2020{\natexlab{b}})\citenamefont {Maji}, \citenamefont {Biswas},
  \citenamefont {Nayek},\ and\ \citenamefont {Sahoo}}]{Maji:2020wer}%
  \BibitemOpen
  \bibfield  {author} {\bibinfo {author} {\bibfnamefont {P.}~\bibnamefont
  {Maji}}, \bibinfo {author} {\bibfnamefont {S.}~\bibnamefont {Biswas}},
  \bibinfo {author} {\bibfnamefont {P.}~\bibnamefont {Nayek}}, \ and\ \bibinfo
  {author} {\bibfnamefont {S.}~\bibnamefont {Sahoo}},\ }\href {\doibase
  10.1093/ptep/ptaa048} {\bibfield  {journal} {\bibinfo  {journal} {PTEP}\
  }\textbf {\bibinfo {volume} {2020}},\ \bibinfo {pages} {053B07} (\bibinfo
  {year} {2020}{\natexlab{b}})},\ \Eprint {http://arxiv.org/abs/2003.07041}
  {arXiv:2003.07041 [hep-ph]} \BibitemShut {NoStop}%
\bibitem [{\citenamefont {Dutta}(2019)}]{Dutta:2019wxo}%
  \BibitemOpen
  \bibfield  {author} {\bibinfo {author} {\bibfnamefont {R.}~\bibnamefont
  {Dutta}},\ }\href {\doibase 10.1103/PhysRevD.100.075025} {\bibfield
  {journal} {\bibinfo  {journal} {Phys. Rev. D}\ }\textbf {\bibinfo {volume}
  {100}},\ \bibinfo {pages} {075025} (\bibinfo {year} {2019})},\ \Eprint
  {http://arxiv.org/abs/1906.02412} {arXiv:1906.02412 [hep-ph]} \BibitemShut
  {NoStop}%
\bibitem [{\citenamefont {Mohapatra}\ \emph {et~al.}(2022)\citenamefont
  {Mohapatra}, \citenamefont {Rajeev},\ and\ \citenamefont
  {Dutta}}]{Mohapatra:2021ynn}%
  \BibitemOpen
  \bibfield  {author} {\bibinfo {author} {\bibfnamefont {M.~K.}\ \bibnamefont
  {Mohapatra}}, \bibinfo {author} {\bibfnamefont {N.}~\bibnamefont {Rajeev}}, \
  and\ \bibinfo {author} {\bibfnamefont {R.}~\bibnamefont {Dutta}},\ }\href
  {\doibase 10.1103/PhysRevD.105.115022} {\bibfield  {journal} {\bibinfo
  {journal} {Phys. Rev. D}\ }\textbf {\bibinfo {volume} {105}},\ \bibinfo
  {pages} {115022} (\bibinfo {year} {2022})},\ \Eprint
  {http://arxiv.org/abs/2108.10106} {arXiv:2108.10106 [hep-ph]} \BibitemShut
  {NoStop}%
\bibitem [{\citenamefont {Zaki}\ \emph {et~al.}(2023)\citenamefont {Zaki},
  \citenamefont {Paracha},\ and\ \citenamefont {Bhutta}}]{Zaki:2023mcw}%
  \BibitemOpen
  \bibfield  {author} {\bibinfo {author} {\bibfnamefont {M.}~\bibnamefont
  {Zaki}}, \bibinfo {author} {\bibfnamefont {M.~A.}\ \bibnamefont {Paracha}}, \
  and\ \bibinfo {author} {\bibfnamefont {F.~M.}\ \bibnamefont {Bhutta}},\
  }\href {\doibase 10.1016/j.nuclphysb.2023.116236} {\bibfield  {journal}
  {\bibinfo  {journal} {Nucl. Phys. B}\ }\textbf {\bibinfo {volume} {992}},\
  \bibinfo {pages} {116236} (\bibinfo {year} {2023})},\ \Eprint
  {http://arxiv.org/abs/2303.01145} {arXiv:2303.01145 [hep-ph]} \BibitemShut
  {NoStop}%
\bibitem [{\citenamefont {Pandya}\ \emph {et~al.}(2023)\citenamefont {Pandya},
  \citenamefont {Santorelli},\ and\ \citenamefont {Soni}}]{Pandya:2023ldv}%
  \BibitemOpen
  \bibfield  {author} {\bibinfo {author} {\bibfnamefont {J.~N.}\ \bibnamefont
  {Pandya}}, \bibinfo {author} {\bibfnamefont {P.}~\bibnamefont {Santorelli}},
  \ and\ \bibinfo {author} {\bibfnamefont {N.~R.}\ \bibnamefont {Soni}}\
  }(\bibinfo {year} {2023})\ \Eprint {http://arxiv.org/abs/2307.14245}
  {arXiv:2307.14245 [hep-ph]} \BibitemShut {NoStop}%
\bibitem [{\citenamefont {Soni}\ \emph {et~al.}(2023)\citenamefont {Soni},
  \citenamefont {Issadykov}, \citenamefont {Gadaria}, \citenamefont
  {Tyulemissov}, \citenamefont {Patel},\ and\ \citenamefont
  {Pandya}}]{Soni:2021fky}%
  \BibitemOpen
  \bibfield  {author} {\bibinfo {author} {\bibfnamefont {N.~R.}\ \bibnamefont
  {Soni}}, \bibinfo {author} {\bibfnamefont {A.}~\bibnamefont {Issadykov}},
  \bibinfo {author} {\bibfnamefont {A.~N.}\ \bibnamefont {Gadaria}}, \bibinfo
  {author} {\bibfnamefont {Z.}~\bibnamefont {Tyulemissov}}, \bibinfo {author}
  {\bibfnamefont {J.~J.}\ \bibnamefont {Patel}}, \ and\ \bibinfo {author}
  {\bibfnamefont {J.~N.}\ \bibnamefont {Pandya}},\ }\href {\doibase
  10.1140/epjp/s13360-023-03779-8} {\bibfield  {journal} {\bibinfo  {journal}
  {Eur. Phys. J. Plus}\ }\textbf {\bibinfo {volume} {138}},\ \bibinfo {pages}
  {163} (\bibinfo {year} {2023})},\ \Eprint {http://arxiv.org/abs/2110.12740}
  {arXiv:2110.12740 [hep-ph]} \BibitemShut {NoStop}%
\bibitem [{\citenamefont {Soni}\ \emph {et~al.}(2022)\citenamefont {Soni},
  \citenamefont {Issadykov}, \citenamefont {Gadaria}, \citenamefont {Patel},\
  and\ \citenamefont {Pandya}}]{Soni:2020bvu}%
  \BibitemOpen
  \bibfield  {author} {\bibinfo {author} {\bibfnamefont {N.~R.}\ \bibnamefont
  {Soni}}, \bibinfo {author} {\bibfnamefont {A.}~\bibnamefont {Issadykov}},
  \bibinfo {author} {\bibfnamefont {A.~N.}\ \bibnamefont {Gadaria}}, \bibinfo
  {author} {\bibfnamefont {J.~J.}\ \bibnamefont {Patel}}, \ and\ \bibinfo
  {author} {\bibfnamefont {J.~N.}\ \bibnamefont {Pandya}},\ }\href {\doibase
  10.1140/epja/s10050-022-00685-y} {\bibfield  {journal} {\bibinfo  {journal}
  {Eur. Phys. J. A}\ }\textbf {\bibinfo {volume} {58}},\ \bibinfo {pages} {39}
  (\bibinfo {year} {2022})},\ \Eprint {http://arxiv.org/abs/2008.07202}
  {arXiv:2008.07202 [hep-ph]} \BibitemShut {NoStop}%
\bibitem [{\citenamefont {Soni}\ \emph {et~al.}(2020)\citenamefont {Soni},
  \citenamefont {Gadaria}, \citenamefont {Patel},\ and\ \citenamefont
  {Pandya}}]{Soni:2020sgn}%
  \BibitemOpen
  \bibfield  {author} {\bibinfo {author} {\bibfnamefont {N.~R.}\ \bibnamefont
  {Soni}}, \bibinfo {author} {\bibfnamefont {A.~N.}\ \bibnamefont {Gadaria}},
  \bibinfo {author} {\bibfnamefont {J.~J.}\ \bibnamefont {Patel}}, \ and\
  \bibinfo {author} {\bibfnamefont {J.~N.}\ \bibnamefont {Pandya}},\ }\href
  {\doibase 10.1103/PhysRevD.102.016013} {\bibfield  {journal} {\bibinfo
  {journal} {Phys. Rev. D}\ }\textbf {\bibinfo {volume} {102}},\ \bibinfo
  {pages} {016013} (\bibinfo {year} {2020})},\ \Eprint
  {http://arxiv.org/abs/2001.10195} {arXiv:2001.10195 [hep-ph]} \BibitemShut
  {NoStop}%
\bibitem [{\citenamefont {Ivanov}\ \emph {et~al.}(2019)\citenamefont {Ivanov},
  \citenamefont {Körner}, \citenamefont {Pandya}, \citenamefont {Santorelli},
  \citenamefont {Soni},\ and\ \citenamefont {Tran}}]{Ivanov:2019nqd}%
  \BibitemOpen
  \bibfield  {author} {\bibinfo {author} {\bibfnamefont {M.~A.}\ \bibnamefont
  {Ivanov}}, \bibinfo {author} {\bibfnamefont {J.~G.}\ \bibnamefont {Körner}},
  \bibinfo {author} {\bibfnamefont {J.~N.}\ \bibnamefont {Pandya}}, \bibinfo
  {author} {\bibfnamefont {P.}~\bibnamefont {Santorelli}}, \bibinfo {author}
  {\bibfnamefont {N.~R.}\ \bibnamefont {Soni}}, \ and\ \bibinfo {author}
  {\bibfnamefont {C.-T.}\ \bibnamefont {Tran}},\ }\href {\doibase
  10.1007/s11467-019-0908-1} {\bibfield  {journal} {\bibinfo  {journal} {Front.
  Phys. (Beijing)}\ }\textbf {\bibinfo {volume} {14}},\ \bibinfo {pages}
  {64401} (\bibinfo {year} {2019})},\ \Eprint {http://arxiv.org/abs/1904.07740}
  {arXiv:1904.07740 [hep-ph]} \BibitemShut {NoStop}%
\bibitem [{\citenamefont {Soni}\ \emph {et~al.}(2018)\citenamefont {Soni},
  \citenamefont {Ivanov}, \citenamefont {Körner}, \citenamefont {Pandya},
  \citenamefont {Santorelli},\ and\ \citenamefont {Tran}}]{Soni:2018adu}%
  \BibitemOpen
  \bibfield  {author} {\bibinfo {author} {\bibfnamefont {N.~R.}\ \bibnamefont
  {Soni}}, \bibinfo {author} {\bibfnamefont {M.~A.}\ \bibnamefont {Ivanov}},
  \bibinfo {author} {\bibfnamefont {J.~G.}\ \bibnamefont {Körner}}, \bibinfo
  {author} {\bibfnamefont {J.~N.}\ \bibnamefont {Pandya}}, \bibinfo {author}
  {\bibfnamefont {P.}~\bibnamefont {Santorelli}}, \ and\ \bibinfo {author}
  {\bibfnamefont {C.~T.}\ \bibnamefont {Tran}},\ }\href {\doibase
  10.1103/PhysRevD.98.114031} {\bibfield  {journal} {\bibinfo  {journal} {Phys.
  Rev. D}\ }\textbf {\bibinfo {volume} {98}},\ \bibinfo {pages} {114031}
  (\bibinfo {year} {2018})},\ \Eprint {http://arxiv.org/abs/1810.11907}
  {arXiv:1810.11907 [hep-ph]} \BibitemShut {NoStop}%
\bibitem [{\citenamefont {Soni}\ and\ \citenamefont
  {Pandya}(2017)}]{Soni:2017eug}%
  \BibitemOpen
  \bibfield  {author} {\bibinfo {author} {\bibfnamefont {N.~R.}\ \bibnamefont
  {Soni}}\ and\ \bibinfo {author} {\bibfnamefont {J.~N.}\ \bibnamefont
  {Pandya}},\ }\href {\doibase 10.1103/PhysRevD.96.016017} {\bibfield
  {journal} {\bibinfo  {journal} {Phys. Rev. D}\ }\textbf {\bibinfo {volume}
  {96}},\ \bibinfo {pages} {016017} (\bibinfo {year} {2017})},\ \bibinfo {note}
  {[Erratum: Phys.Rev.D 99, 059901 (2019)]},\ \Eprint
  {http://arxiv.org/abs/1706.01190} {arXiv:1706.01190 [hep-ph]} \BibitemShut
  {NoStop}%
\bibitem [{\citenamefont {Navas~\textit{et al}.}(2024)}]{PhysRevD.110.030001}%
  \BibitemOpen
  \bibfield  {author} {\bibinfo {author} {\bibfnamefont {S.}~\bibnamefont
  {Navas~\textit{et al}.}} (\bibinfo {collaboration} {Particle Data Grou}),\
  }\href {\doibase 10.1103/PhysRevD.110.030001} {\bibfield  {journal} {\bibinfo
   {journal} {Phys. Rev. D}\ }\textbf {\bibinfo {volume} {110}},\ \bibinfo
  {pages} {030001} (\bibinfo {year} {2024})}\BibitemShut {NoStop}%
\bibitem [{\citenamefont {Descotes-Genon}\ \emph {et~al.}(2013)\citenamefont
  {Descotes-Genon}, \citenamefont {Hurth}, \citenamefont {Matias},\ and\
  \citenamefont {Virto}}]{Descotes-Genon:2013vna}%
  \BibitemOpen
  \bibfield  {author} {\bibinfo {author} {\bibfnamefont {S.}~\bibnamefont
  {Descotes-Genon}}, \bibinfo {author} {\bibfnamefont {T.}~\bibnamefont
  {Hurth}}, \bibinfo {author} {\bibfnamefont {J.}~\bibnamefont {Matias}}, \
  and\ \bibinfo {author} {\bibfnamefont {J.}~\bibnamefont {Virto}},\ }\href
  {\doibase 10.1007/JHEP05(2013)137} {\bibfield  {journal} {\bibinfo  {journal}
  {JHEP}\ }\textbf {\bibinfo {volume} {05}},\ \bibinfo {pages} {137} (\bibinfo
  {year} {2013})},\ \Eprint {http://arxiv.org/abs/1303.5794} {arXiv:1303.5794
  [hep-ph]} \BibitemShut {NoStop}%
\bibitem [{\citenamefont {Buras}\ and\ \citenamefont
  {Munz}(1995)}]{Buras:1994dj}%
  \BibitemOpen
  \bibfield  {author} {\bibinfo {author} {\bibfnamefont {A.~J.}\ \bibnamefont
  {Buras}}\ and\ \bibinfo {author} {\bibfnamefont {M.}~\bibnamefont {Munz}},\
  }\href {\doibase 10.1103/PhysRevD.52.186} {\bibfield  {journal} {\bibinfo
  {journal} {Phys. Rev. D}\ }\textbf {\bibinfo {volume} {52}},\ \bibinfo
  {pages} {186} (\bibinfo {year} {1995})},\ \Eprint
  {http://arxiv.org/abs/hep-ph/9501281} {arXiv:hep-ph/9501281} \BibitemShut
  {NoStop}%
\bibitem [{\citenamefont {Kruger}\ and\ \citenamefont
  {Sehgal}(1997)}]{Kruger:1996dt}%
  \BibitemOpen
  \bibfield  {author} {\bibinfo {author} {\bibfnamefont {F.}~\bibnamefont
  {Kruger}}\ and\ \bibinfo {author} {\bibfnamefont {L.}~\bibnamefont
  {Sehgal}},\ }\href {\doibase 10.1103/PhysRevD.55.2799} {\bibfield  {journal}
  {\bibinfo  {journal} {Phys. Rev. D}\ }\textbf {\bibinfo {volume} {55}},\
  \bibinfo {pages} {2799} (\bibinfo {year} {1997})},\ \Eprint
  {http://arxiv.org/abs/hep-ph/9608361} {arXiv:hep-ph/9608361} \BibitemShut
  {NoStop}%
\bibitem [{\citenamefont {Buchalla}\ \emph {et~al.}(1996)\citenamefont
  {Buchalla}, \citenamefont {Buras},\ and\ \citenamefont
  {Lautenbacher}}]{Buchalla:1995vs}%
  \BibitemOpen
  \bibfield  {author} {\bibinfo {author} {\bibfnamefont {G.}~\bibnamefont
  {Buchalla}}, \bibinfo {author} {\bibfnamefont {A.~J.}\ \bibnamefont {Buras}},
  \ and\ \bibinfo {author} {\bibfnamefont {M.~E.}\ \bibnamefont
  {Lautenbacher}},\ }\href {\doibase 10.1103/RevModPhys.68.1125} {\bibfield
  {journal} {\bibinfo  {journal} {Rev. Mod. Phys.}\ }\textbf {\bibinfo {volume}
  {68}},\ \bibinfo {pages} {1125} (\bibinfo {year} {1996})},\ \Eprint
  {http://arxiv.org/abs/hep-ph/9512380} {arXiv:hep-ph/9512380} \BibitemShut
  {NoStop}%
\bibitem [{\citenamefont {Deshpande}\ \emph {et~al.}(1989)\citenamefont
  {Deshpande}, \citenamefont {Trampetic},\ and\ \citenamefont
  {Panose}}]{Deshpande:1988bd}%
  \BibitemOpen
  \bibfield  {author} {\bibinfo {author} {\bibfnamefont {N.}~\bibnamefont
  {Deshpande}}, \bibinfo {author} {\bibfnamefont {J.}~\bibnamefont
  {Trampetic}}, \ and\ \bibinfo {author} {\bibfnamefont {K.}~\bibnamefont
  {Panose}},\ }\href {\doibase 10.1103/PhysRevD.39.1461} {\bibfield  {journal}
  {\bibinfo  {journal} {Phys. Rev. D}\ }\textbf {\bibinfo {volume} {39}},\
  \bibinfo {pages} {1461} (\bibinfo {year} {1989})}\BibitemShut {NoStop}%
\bibitem [{\citenamefont {Jezabek}\ and\ \citenamefont
  {Kuhn}(1989)}]{Jezabek:1988ja}%
  \BibitemOpen
  \bibfield  {author} {\bibinfo {author} {\bibfnamefont {M.}~\bibnamefont
  {Jezabek}}\ and\ \bibinfo {author} {\bibfnamefont {J.~H.}\ \bibnamefont
  {Kuhn}},\ }\href {\doibase 10.1016/0550-3213(89)90209-5} {\bibfield
  {journal} {\bibinfo  {journal} {Nucl. Phys. B}\ }\textbf {\bibinfo {volume}
  {320}},\ \bibinfo {pages} {20} (\bibinfo {year} {1989})}\BibitemShut
  {NoStop}%
\bibitem [{\citenamefont {Lim}\ \emph {et~al.}(1989)\citenamefont {Lim},
  \citenamefont {Morozumi},\ and\ \citenamefont {Sanda}}]{Lim:1988yu}%
  \BibitemOpen
  \bibfield  {author} {\bibinfo {author} {\bibfnamefont {C.}~\bibnamefont
  {Lim}}, \bibinfo {author} {\bibfnamefont {T.}~\bibnamefont {Morozumi}}, \
  and\ \bibinfo {author} {\bibfnamefont {A.}~\bibnamefont {Sanda}},\ }\href
  {\doibase 10.1016/0370-2693(89)91593-1} {\bibfield  {journal} {\bibinfo
  {journal} {Phys. Lett. B}\ }\textbf {\bibinfo {volume} {218}},\ \bibinfo
  {pages} {343} (\bibinfo {year} {1989})}\BibitemShut {NoStop}%
\bibitem [{\citenamefont {Misiak}(1993)}]{Misiak:1992bc}%
  \BibitemOpen
  \bibfield  {author} {\bibinfo {author} {\bibfnamefont {M.}~\bibnamefont
  {Misiak}},\ }\href {\doibase 10.1016/0550-3213(93)90235-H} {\bibfield
  {journal} {\bibinfo  {journal} {Nucl. Phys. B}\ }\textbf {\bibinfo {volume}
  {393}},\ \bibinfo {pages} {23} (\bibinfo {year} {1993})},\ \bibinfo {note}
  {[Erratum: Nucl.Phys.B 439, 461--465 (1995)]}\BibitemShut {NoStop}%
\bibitem [{\citenamefont {O'Donnell}\ and\ \citenamefont
  {Tung}(1991)}]{ODonnell:1991cdx}%
  \BibitemOpen
  \bibfield  {author} {\bibinfo {author} {\bibfnamefont {P.~J.}\ \bibnamefont
  {O'Donnell}}\ and\ \bibinfo {author} {\bibfnamefont {H.~K.}\ \bibnamefont
  {Tung}},\ }\href {\doibase 10.1103/PhysRevD.43.R2067} {\bibfield  {journal}
  {\bibinfo  {journal} {Phys. Rev. D}\ }\textbf {\bibinfo {volume} {43}},\
  \bibinfo {pages} {2067} (\bibinfo {year} {1991})}\BibitemShut {NoStop}%
\bibitem [{\citenamefont {Ali}\ \emph {et~al.}(1991)\citenamefont {Ali},
  \citenamefont {Mannel},\ and\ \citenamefont {Morozumi}}]{Ali:1991is}%
  \BibitemOpen
  \bibfield  {author} {\bibinfo {author} {\bibfnamefont {A.}~\bibnamefont
  {Ali}}, \bibinfo {author} {\bibfnamefont {T.}~\bibnamefont {Mannel}}, \ and\
  \bibinfo {author} {\bibfnamefont {T.}~\bibnamefont {Morozumi}},\ }\href
  {\doibase 10.1016/0370-2693(91)90306-B} {\bibfield  {journal} {\bibinfo
  {journal} {Phys. Lett. B}\ }\textbf {\bibinfo {volume} {273}},\ \bibinfo
  {pages} {505} (\bibinfo {year} {1991})}\BibitemShut {NoStop}%
\bibitem [{\citenamefont {Bobeth}\ \emph {et~al.}(2000)\citenamefont {Bobeth},
  \citenamefont {Misiak},\ and\ \citenamefont {Urban}}]{Bobeth:1999mk}%
  \BibitemOpen
  \bibfield  {author} {\bibinfo {author} {\bibfnamefont {C.}~\bibnamefont
  {Bobeth}}, \bibinfo {author} {\bibfnamefont {M.}~\bibnamefont {Misiak}}, \
  and\ \bibinfo {author} {\bibfnamefont {J.}~\bibnamefont {Urban}},\ }\href
  {\doibase 10.1016/S0550-3213(00)00007-9} {\bibfield  {journal} {\bibinfo
  {journal} {Nucl. Phys. B}\ }\textbf {\bibinfo {volume} {574}},\ \bibinfo
  {pages} {291} (\bibinfo {year} {2000})},\ \Eprint
  {http://arxiv.org/abs/hep-ph/9910220} {arXiv:hep-ph/9910220} \BibitemShut
  {NoStop}%
\bibitem [{\citenamefont {Chen}\ and\ \citenamefont
  {Geng}(2001)}]{Chen:2001zc}%
  \BibitemOpen
  \bibfield  {author} {\bibinfo {author} {\bibfnamefont {C.-H.}\ \bibnamefont
  {Chen}}\ and\ \bibinfo {author} {\bibfnamefont {C.~Q.}\ \bibnamefont
  {Geng}},\ }\href {\doibase 10.1103/PhysRevD.64.074001} {\bibfield  {journal}
  {\bibinfo  {journal} {Phys. Rev. D}\ }\textbf {\bibinfo {volume} {64}},\
  \bibinfo {pages} {074001} (\bibinfo {year} {2001})},\ \Eprint
  {http://arxiv.org/abs/hep-ph/0106193} {arXiv:hep-ph/0106193} \BibitemShut
  {NoStop}%
\bibitem [{\citenamefont {Wang}\ and\ \citenamefont
  {Xiao}(2012)}]{Wang:2012ab}%
  \BibitemOpen
  \bibfield  {author} {\bibinfo {author} {\bibfnamefont {W.-F.}\ \bibnamefont
  {Wang}}\ and\ \bibinfo {author} {\bibfnamefont {Z.-J.}\ \bibnamefont
  {Xiao}},\ }\href {\doibase 10.1103/PhysRevD.86.114025} {\bibfield  {journal}
  {\bibinfo  {journal} {Phys. Rev. D}\ }\textbf {\bibinfo {volume} {86}},\
  \bibinfo {pages} {114025} (\bibinfo {year} {2012})},\ \Eprint
  {http://arxiv.org/abs/1207.0265} {arXiv:1207.0265 [hep-ph]} \BibitemShut
  {NoStop}%
\bibitem [{\citenamefont {Efimov}\ and\ \citenamefont
  {Ivanov}(1989)}]{Efimov:1988yd}%
  \BibitemOpen
  \bibfield  {author} {\bibinfo {author} {\bibfnamefont {G.~V.}\ \bibnamefont
  {Efimov}}\ and\ \bibinfo {author} {\bibfnamefont {M.~A.}\ \bibnamefont
  {Ivanov}},\ }\href {\doibase 10.1142/S0217751X89000832} {\bibfield  {journal}
  {\bibinfo  {journal} {Int. J. Mod. Phys. A}\ }\textbf {\bibinfo {volume}
  {4}},\ \bibinfo {pages} {2031} (\bibinfo {year} {1989})}\BibitemShut
  {NoStop}%
\bibitem [{\citenamefont {Efimov}\ and\ \citenamefont
  {Ivanov}(1993)}]{Efimov:1993}%
  \BibitemOpen
  \bibfield  {author} {\bibinfo {author} {\bibfnamefont {G.~V.}\ \bibnamefont
  {Efimov}}\ and\ \bibinfo {author} {\bibfnamefont {M.~A.}\ \bibnamefont
  {Ivanov}},\ }\href@noop {} {\emph {\bibinfo {title} {{The Quark confinement
  model of hadrons}}}}\ (\bibinfo  {publisher} {IOP},\ \bibinfo {address}
  {Bristol},\ \bibinfo {year} {1993})\BibitemShut {NoStop}%
\bibitem [{\citenamefont {Ivanov}\ and\ \citenamefont
  {Santorelli}(1999)}]{Ivanov:1999ic}%
  \BibitemOpen
  \bibfield  {author} {\bibinfo {author} {\bibfnamefont {M.~A.}\ \bibnamefont
  {Ivanov}}\ and\ \bibinfo {author} {\bibfnamefont {P.}~\bibnamefont
  {Santorelli}},\ }\href {\doibase 10.1016/S0370-2693(99)00474-8} {\bibfield
  {journal} {\bibinfo  {journal} {Phys. Lett. B}\ }\textbf {\bibinfo {volume}
  {456}},\ \bibinfo {pages} {248} (\bibinfo {year} {1999})},\ \Eprint
  {http://arxiv.org/abs/hep-ph/9903446} {arXiv:hep-ph/9903446 [hep-ph]}
  \BibitemShut {NoStop}%
\bibitem [{\citenamefont {Branz}\ \emph {et~al.}(2010)\citenamefont {Branz},
  \citenamefont {Faessler}, \citenamefont {Gutsche}, \citenamefont {Ivanov},
  \citenamefont {Korner},\ and\ \citenamefont {Lyubovitskij}}]{Branz:2009cd}%
  \BibitemOpen
  \bibfield  {author} {\bibinfo {author} {\bibfnamefont {T.}~\bibnamefont
  {Branz}}, \bibinfo {author} {\bibfnamefont {A.}~\bibnamefont {Faessler}},
  \bibinfo {author} {\bibfnamefont {T.}~\bibnamefont {Gutsche}}, \bibinfo
  {author} {\bibfnamefont {M.~A.}\ \bibnamefont {Ivanov}}, \bibinfo {author}
  {\bibfnamefont {J.~G.}\ \bibnamefont {Korner}}, \ and\ \bibinfo {author}
  {\bibfnamefont {V.~E.}\ \bibnamefont {Lyubovitskij}},\ }\href {\doibase
  10.1103/PhysRevD.81.034010} {\bibfield  {journal} {\bibinfo  {journal} {Phys.
  Rev. D}\ }\textbf {\bibinfo {volume} {81}},\ \bibinfo {pages} {034010}
  (\bibinfo {year} {2010})},\ \Eprint {http://arxiv.org/abs/0912.3710}
  {arXiv:0912.3710 [hep-ph]} \BibitemShut {NoStop}%
\bibitem [{\citenamefont {Ivanov}\ \emph {et~al.}(2012)\citenamefont {Ivanov},
  \citenamefont {Korner}, \citenamefont {Kovalenko}, \citenamefont
  {Santorelli},\ and\ \citenamefont {Saidullaeva}}]{Ivanov:2011aa}%
  \BibitemOpen
  \bibfield  {author} {\bibinfo {author} {\bibfnamefont {M.~A.}\ \bibnamefont
  {Ivanov}}, \bibinfo {author} {\bibfnamefont {J.~G.}\ \bibnamefont {Korner}},
  \bibinfo {author} {\bibfnamefont {S.~G.}\ \bibnamefont {Kovalenko}}, \bibinfo
  {author} {\bibfnamefont {P.}~\bibnamefont {Santorelli}}, \ and\ \bibinfo
  {author} {\bibfnamefont {G.~G.}\ \bibnamefont {Saidullaeva}},\ }\href
  {\doibase 10.1103/PhysRevD.85.034004} {\bibfield  {journal} {\bibinfo
  {journal} {Phys. Rev. D}\ }\textbf {\bibinfo {volume} {85}},\ \bibinfo
  {pages} {034004} (\bibinfo {year} {2012})},\ \Eprint
  {http://arxiv.org/abs/1112.3536} {arXiv:1112.3536 [hep-ph]} \BibitemShut
  {NoStop}%
\bibitem [{\citenamefont {Gutsche}\ \emph {et~al.}(2012)\citenamefont
  {Gutsche}, \citenamefont {Ivanov}, \citenamefont {Korner}, \citenamefont
  {Lyubovitskij},\ and\ \citenamefont {Santorelli}}]{Gutsche:2012ze}%
  \BibitemOpen
  \bibfield  {author} {\bibinfo {author} {\bibfnamefont {T.}~\bibnamefont
  {Gutsche}}, \bibinfo {author} {\bibfnamefont {M.~A.}\ \bibnamefont {Ivanov}},
  \bibinfo {author} {\bibfnamefont {J.~G.}\ \bibnamefont {Korner}}, \bibinfo
  {author} {\bibfnamefont {V.~E.}\ \bibnamefont {Lyubovitskij}}, \ and\
  \bibinfo {author} {\bibfnamefont {P.}~\bibnamefont {Santorelli}},\ }\href
  {\doibase 10.1103/PhysRevD.86.074013} {\bibfield  {journal} {\bibinfo
  {journal} {Phys. Rev. D}\ }\textbf {\bibinfo {volume} {86}},\ \bibinfo
  {pages} {074013} (\bibinfo {year} {2012})},\ \Eprint
  {http://arxiv.org/abs/1207.7052} {arXiv:1207.7052 [hep-ph]} \BibitemShut
  {NoStop}%
\bibitem [{\citenamefont {Salam}(1962)}]{Salam:1962}%
  \BibitemOpen
  \bibfield  {author} {\bibinfo {author} {\bibfnamefont {A.}~\bibnamefont
  {Salam}},\ }\href {\doibase 10.1007/BF02733330} {\bibfield  {journal}
  {\bibinfo  {journal} {Nuovo Cim.}\ }\textbf {\bibinfo {volume} {25}},\
  \bibinfo {pages} {224} (\bibinfo {year} {1962})}\BibitemShut {NoStop}%
\bibitem [{\citenamefont {Weinberg}(1963)}]{Weinberg:1962}%
  \BibitemOpen
  \bibfield  {author} {\bibinfo {author} {\bibfnamefont {S.}~\bibnamefont
  {Weinberg}},\ }\href {\doibase 10.1103/PhysRev.130.776} {\bibfield  {journal}
  {\bibinfo  {journal} {Phys. Rev.}\ }\textbf {\bibinfo {volume} {130}},\
  \bibinfo {pages} {776} (\bibinfo {year} {1963})}\BibitemShut {NoStop}%
\bibitem [{\citenamefont {Ivanov}\ \emph {et~al.}(2015)\citenamefont {Ivanov},
  \citenamefont {K\"orner},\ and\ \citenamefont {Tran}}]{Ivanov:2015tru}%
  \BibitemOpen
  \bibfield  {author} {\bibinfo {author} {\bibfnamefont {M.~A.}\ \bibnamefont
  {Ivanov}}, \bibinfo {author} {\bibfnamefont {J.~G.}\ \bibnamefont
  {K\"orner}}, \ and\ \bibinfo {author} {\bibfnamefont {C.~T.}\ \bibnamefont
  {Tran}},\ }\href {\doibase 10.1103/PhysRevD.92.114022} {\bibfield  {journal}
  {\bibinfo  {journal} {Phys. Rev. D}\ }\textbf {\bibinfo {volume} {92}},\
  \bibinfo {pages} {114022} (\bibinfo {year} {2015})},\ \Eprint
  {http://arxiv.org/abs/1508.02678} {arXiv:1508.02678 [hep-ph]} \BibitemShut
  {NoStop}%
\bibitem [{\citenamefont {Ganbold}\ \emph {et~al.}(2015)\citenamefont
  {Ganbold}, \citenamefont {Gutsche}, \citenamefont {Ivanov},\ and\
  \citenamefont {Lyubovitskij}}]{Ganbold:2014pua}%
  \BibitemOpen
  \bibfield  {author} {\bibinfo {author} {\bibfnamefont {G.}~\bibnamefont
  {Ganbold}}, \bibinfo {author} {\bibfnamefont {T.}~\bibnamefont {Gutsche}},
  \bibinfo {author} {\bibfnamefont {M.~A.}\ \bibnamefont {Ivanov}}, \ and\
  \bibinfo {author} {\bibfnamefont {V.~E.}\ \bibnamefont {Lyubovitskij}},\
  }\href {\doibase 10.1088/0954-3899/42/7/075002} {\bibfield  {journal}
  {\bibinfo  {journal} {J. Phys. G}\ }\textbf {\bibinfo {volume} {42}},\
  \bibinfo {pages} {075002} (\bibinfo {year} {2015})},\ \Eprint
  {http://arxiv.org/abs/1410.3741} {arXiv:1410.3741 [hep-ph]} \BibitemShut
  {NoStop}%
\bibitem [{\citenamefont {Dubni\v{c}ka}\ \emph
  {et~al.}(2016{\natexlab{a}})\citenamefont {Dubni\v{c}ka}, \citenamefont
  {Dubni\v{c}kov\'{a}}, \citenamefont {Issadykov}, \citenamefont {Ivanov},
  \citenamefont {Liptaj},\ and\ \citenamefont {Sakhiyev}}]{Dubnicka:2016nyy}%
  \BibitemOpen
  \bibfield  {author} {\bibinfo {author} {\bibfnamefont {S.}~\bibnamefont
  {Dubni\v{c}ka}}, \bibinfo {author} {\bibfnamefont {A.~Z.}\ \bibnamefont
  {Dubni\v{c}kov\'{a}}}, \bibinfo {author} {\bibfnamefont {A.}~\bibnamefont
  {Issadykov}}, \bibinfo {author} {\bibfnamefont {M.~A.}\ \bibnamefont
  {Ivanov}}, \bibinfo {author} {\bibfnamefont {A.}~\bibnamefont {Liptaj}}, \
  and\ \bibinfo {author} {\bibfnamefont {S.~K.}\ \bibnamefont {Sakhiyev}},\
  }\href {\doibase 10.1103/PhysRevD.93.094022} {\bibfield  {journal} {\bibinfo
  {journal} {Phys. Rev. D}\ }\textbf {\bibinfo {volume} {93}},\ \bibinfo
  {pages} {094022} (\bibinfo {year} {2016}{\natexlab{a}})},\ \Eprint
  {http://arxiv.org/abs/1602.07864} {arXiv:1602.07864 [hep-ph]} \BibitemShut
  {NoStop}%
\bibitem [{\citenamefont {James}\ and\ \citenamefont
  {Roos}(1975)}]{James:1975dr}%
  \BibitemOpen
  \bibfield  {author} {\bibinfo {author} {\bibfnamefont {F.}~\bibnamefont
  {James}}\ and\ \bibinfo {author} {\bibfnamefont {M.}~\bibnamefont {Roos}},\
  }\href {\doibase 10.1016/0010-4655(75)90039-9} {\bibfield  {journal}
  {\bibinfo  {journal} {Comput. Phys. Commun.}\ }\textbf {\bibinfo {volume}
  {10}},\ \bibinfo {pages} {343} (\bibinfo {year} {1975})}\BibitemShut
  {NoStop}%
\bibitem [{\citenamefont {Buchalla}\ and\ \citenamefont
  {Buras}(1999)}]{Buchalla:1998ba}%
  \BibitemOpen
  \bibfield  {author} {\bibinfo {author} {\bibfnamefont {G.}~\bibnamefont
  {Buchalla}}\ and\ \bibinfo {author} {\bibfnamefont {A.~J.}\ \bibnamefont
  {Buras}},\ }\href {\doibase 10.1016/S0550-3213(99)00149-2} {\bibfield
  {journal} {\bibinfo  {journal} {Nucl. Phys. B}\ }\textbf {\bibinfo {volume}
  {548}},\ \bibinfo {pages} {309} (\bibinfo {year} {1999})},\ \Eprint
  {http://arxiv.org/abs/hep-ph/9901288} {arXiv:hep-ph/9901288} \BibitemShut
  {NoStop}%
\bibitem [{\citenamefont {Misiak}\ and\ \citenamefont
  {Urban}(1999)}]{Misiak:1999yg}%
  \BibitemOpen
  \bibfield  {author} {\bibinfo {author} {\bibfnamefont {M.}~\bibnamefont
  {Misiak}}\ and\ \bibinfo {author} {\bibfnamefont {J.}~\bibnamefont {Urban}},\
  }\href {\doibase 10.1016/S0370-2693(99)00150-1} {\bibfield  {journal}
  {\bibinfo  {journal} {Phys. Lett. B}\ }\textbf {\bibinfo {volume} {451}},\
  \bibinfo {pages} {161} (\bibinfo {year} {1999})},\ \Eprint
  {http://arxiv.org/abs/hep-ph/9901278} {arXiv:hep-ph/9901278} \BibitemShut
  {NoStop}%
\bibitem [{\citenamefont {Brod}\ \emph {et~al.}(2011)\citenamefont {Brod},
  \citenamefont {Gorbahn},\ and\ \citenamefont {Stamou}}]{Brod:2010hi}%
  \BibitemOpen
  \bibfield  {author} {\bibinfo {author} {\bibfnamefont {J.}~\bibnamefont
  {Brod}}, \bibinfo {author} {\bibfnamefont {M.}~\bibnamefont {Gorbahn}}, \
  and\ \bibinfo {author} {\bibfnamefont {E.}~\bibnamefont {Stamou}},\ }\href
  {\doibase 10.1103/PhysRevD.83.034030} {\bibfield  {journal} {\bibinfo
  {journal} {Phys. Rev. D}\ }\textbf {\bibinfo {volume} {83}},\ \bibinfo
  {pages} {034030} (\bibinfo {year} {2011})},\ \Eprint
  {http://arxiv.org/abs/1009.0947} {arXiv:1009.0947 [hep-ph]} \BibitemShut
  {NoStop}%
\bibitem [{\citenamefont {Issadykov}\ and\ \citenamefont
  {Ivanov}(2023)}]{Issadykov:2022iwp}%
  \BibitemOpen
  \bibfield  {author} {\bibinfo {author} {\bibfnamefont {A.}~\bibnamefont
  {Issadykov}}\ and\ \bibinfo {author} {\bibfnamefont {M.~A.}\ \bibnamefont
  {Ivanov}},\ }\href {\doibase 10.1142/S0217732323500062} {\bibfield  {journal}
  {\bibinfo  {journal} {Mod. Phys. Lett. A}\ }\textbf {\bibinfo {volume}
  {38}},\ \bibinfo {pages} {2350006} (\bibinfo {year} {2023})},\ \Eprint
  {http://arxiv.org/abs/2211.10683} {arXiv:2211.10683 [hep-ph]} \BibitemShut
  {NoStop}%
\bibitem [{\citenamefont {Li}\ and\ \citenamefont {Liu}(2023)}]{Li:2023mrj}%
  \BibitemOpen
  \bibfield  {author} {\bibinfo {author} {\bibfnamefont {Y.-S.}\ \bibnamefont
  {Li}}\ and\ \bibinfo {author} {\bibfnamefont {X.}~\bibnamefont {Liu}},\
  }\href {\doibase 10.1103/PhysRevD.108.093005} {\bibfield  {journal} {\bibinfo
   {journal} {Phys. Rev. D}\ }\textbf {\bibinfo {volume} {108}},\ \bibinfo
  {pages} {093005} (\bibinfo {year} {2023})},\ \Eprint
  {http://arxiv.org/abs/2309.08191} {arXiv:2309.08191 [hep-ph]} \BibitemShut
  {NoStop}%
\bibitem [{\citenamefont {Aaij}\ \emph
  {et~al.}(2016{\natexlab{b}})\citenamefont {Aaij} \emph
  {et~al.}}]{Aaij:2015oid}%
  \BibitemOpen
  \bibfield  {author} {\bibinfo {author} {\bibfnamefont {R.}~\bibnamefont
  {Aaij}} \emph {et~al.} (\bibinfo {collaboration} {LHCb Collaboration}),\
  }\href {\doibase 10.1007/JHEP02(2016)104} {\bibfield  {journal} {\bibinfo
  {journal} {JHEP}\ }\textbf {\bibinfo {volume} {02}},\ \bibinfo {pages} {104}
  (\bibinfo {year} {2016}{\natexlab{b}})},\ \Eprint
  {http://arxiv.org/abs/1512.04442} {arXiv:1512.04442 [hep-ex]} \BibitemShut
  {NoStop}%
\bibitem [{\citenamefont {Wehle}\ \emph {et~al.}(2017)\citenamefont {Wehle}
  \emph {et~al.}}]{Wehle:2016yoi}%
  \BibitemOpen
  \bibfield  {author} {\bibinfo {author} {\bibfnamefont {S.}~\bibnamefont
  {Wehle}} \emph {et~al.} (\bibinfo {collaboration} {Belle Collaboration}),\
  }\href {\doibase 10.1103/PhysRevLett.118.111801} {\bibfield  {journal}
  {\bibinfo  {journal} {Phys. Rev. Lett.}\ }\textbf {\bibinfo {volume} {118}},\
  \bibinfo {pages} {111801} (\bibinfo {year} {2017})},\ \Eprint
  {http://arxiv.org/abs/1612.05014} {arXiv:1612.05014 [hep-ex]} \BibitemShut
  {NoStop}%
\bibitem [{\citenamefont {Aaij}\ \emph
  {et~al.}(2015{\natexlab{c}})\citenamefont {Aaij} \emph
  {et~al.}}]{LHCb:2015wdu}%
  \BibitemOpen
  \bibfield  {author} {\bibinfo {author} {\bibfnamefont {R.}~\bibnamefont
  {Aaij}} \emph {et~al.} (\bibinfo {collaboration} {LHCb}),\ }\href {\doibase
  10.1007/JHEP09(2015)179} {\bibfield  {journal} {\bibinfo  {journal} {JHEP}\
  }\textbf {\bibinfo {volume} {09}},\ \bibinfo {pages} {179} (\bibinfo {year}
  {2015}{\natexlab{c}})},\ \Eprint {http://arxiv.org/abs/1506.08777}
  {arXiv:1506.08777 [hep-ex]} \BibitemShut {NoStop}%
\bibitem [{\citenamefont {Aaij}\ \emph
  {et~al.}(2018{\natexlab{d}})\citenamefont {Aaij} \emph
  {et~al.}}]{LHCb:2018jna}%
  \BibitemOpen
  \bibfield  {author} {\bibinfo {author} {\bibfnamefont {R.}~\bibnamefont
  {Aaij}} \emph {et~al.} (\bibinfo {collaboration} {LHCb}),\ }\href {\doibase
  10.1007/JHEP09(2018)146} {\bibfield  {journal} {\bibinfo  {journal} {JHEP}\
  }\textbf {\bibinfo {volume} {09}},\ \bibinfo {pages} {146} (\bibinfo {year}
  {2018}{\natexlab{d}})},\ \Eprint {http://arxiv.org/abs/1808.00264}
  {arXiv:1808.00264 [hep-ex]} \BibitemShut {NoStop}%
\bibitem [{\citenamefont {Arbey}\ \emph {et~al.}(2019)\citenamefont {Arbey},
  \citenamefont {Hurth}, \citenamefont {Mahmoudi}, \citenamefont
  {Mart\'\i{}nez~Santos},\ and\ \citenamefont {Neshatpour}}]{Arbey:2019duh}%
  \BibitemOpen
  \bibfield  {author} {\bibinfo {author} {\bibfnamefont {A.}~\bibnamefont
  {Arbey}}, \bibinfo {author} {\bibfnamefont {T.}~\bibnamefont {Hurth}},
  \bibinfo {author} {\bibfnamefont {F.}~\bibnamefont {Mahmoudi}}, \bibinfo
  {author} {\bibfnamefont {D.}~\bibnamefont {Mart\'\i{}nez~Santos}}, \ and\
  \bibinfo {author} {\bibfnamefont {S.}~\bibnamefont {Neshatpour}},\ }\href
  {\doibase 10.1103/PhysRevD.100.015045} {\bibfield  {journal} {\bibinfo
  {journal} {Phys. Rev. D}\ }\textbf {\bibinfo {volume} {100}},\ \bibinfo
  {pages} {015045} (\bibinfo {year} {2019})},\ \Eprint
  {http://arxiv.org/abs/1904.08399} {arXiv:1904.08399 [hep-ph]} \BibitemShut
  {NoStop}%
\bibitem [{\citenamefont {Kowalska}\ \emph {et~al.}(2019)\citenamefont
  {Kowalska}, \citenamefont {Kumar},\ and\ \citenamefont
  {Sessolo}}]{Kowalska:2019ley}%
  \BibitemOpen
  \bibfield  {author} {\bibinfo {author} {\bibfnamefont {K.}~\bibnamefont
  {Kowalska}}, \bibinfo {author} {\bibfnamefont {D.}~\bibnamefont {Kumar}}, \
  and\ \bibinfo {author} {\bibfnamefont {E.~M.}\ \bibnamefont {Sessolo}},\
  }\href {\doibase 10.1140/epjc/s10052-019-7330-2} {\bibfield  {journal}
  {\bibinfo  {journal} {Eur. Phys. J. C}\ }\textbf {\bibinfo {volume} {79}},\
  \bibinfo {pages} {840} (\bibinfo {year} {2019})},\ \Eprint
  {http://arxiv.org/abs/1903.10932} {arXiv:1903.10932 [hep-ph]} \BibitemShut
  {NoStop}%
\bibitem [{\citenamefont {Alok}\ \emph {et~al.}(2019)\citenamefont {Alok},
  \citenamefont {Dighe}, \citenamefont {Gangal},\ and\ \citenamefont
  {Kumar}}]{Alok:2019ufo}%
  \BibitemOpen
  \bibfield  {author} {\bibinfo {author} {\bibfnamefont {A.~K.}\ \bibnamefont
  {Alok}}, \bibinfo {author} {\bibfnamefont {A.}~\bibnamefont {Dighe}},
  \bibinfo {author} {\bibfnamefont {S.}~\bibnamefont {Gangal}}, \ and\ \bibinfo
  {author} {\bibfnamefont {D.}~\bibnamefont {Kumar}},\ }\href {\doibase
  10.1007/JHEP06(2019)089} {\bibfield  {journal} {\bibinfo  {journal} {JHEP}\
  }\textbf {\bibinfo {volume} {06}},\ \bibinfo {pages} {089} (\bibinfo {year}
  {2019})},\ \Eprint {http://arxiv.org/abs/1903.09617} {arXiv:1903.09617
  [hep-ph]} \BibitemShut {NoStop}%
\bibitem [{\citenamefont {Hiller}\ and\ \citenamefont
  {Schmaltz}(2014)}]{Hiller:2014yaa}%
  \BibitemOpen
  \bibfield  {author} {\bibinfo {author} {\bibfnamefont {G.}~\bibnamefont
  {Hiller}}\ and\ \bibinfo {author} {\bibfnamefont {M.}~\bibnamefont
  {Schmaltz}},\ }\href {\doibase 10.1103/PhysRevD.90.054014} {\bibfield
  {journal} {\bibinfo  {journal} {Phys. Rev. D}\ }\textbf {\bibinfo {volume}
  {90}},\ \bibinfo {pages} {054014} (\bibinfo {year} {2014})},\ \Eprint
  {http://arxiv.org/abs/1408.1627} {arXiv:1408.1627 [hep-ph]} \BibitemShut
  {NoStop}%
\bibitem [{\citenamefont {Hiller}\ \emph {et~al.}(2018)\citenamefont {Hiller},
  \citenamefont {Loose}, \citenamefont {Nis},\ and\ \citenamefont
  {Zic}}]{Hiller:2018wbv}%
  \BibitemOpen
  \bibfield  {author} {\bibinfo {author} {\bibfnamefont {G.}~\bibnamefont
  {Hiller}}, \bibinfo {author} {\bibfnamefont {D.}~\bibnamefont {Loose}},
  \bibinfo {author} {\bibnamefont {Nis}}, \ and\ \bibinfo {author}
  {\bibfnamefont {I.}~\bibnamefont {Zic}},\ }\href {\doibase
  10.1103/PhysRevD.97.075004} {\bibfield  {journal} {\bibinfo  {journal} {Phys.
  Rev. D}\ }\textbf {\bibinfo {volume} {97}},\ \bibinfo {pages} {075004}
  (\bibinfo {year} {2018})},\ \Eprint {http://arxiv.org/abs/1801.09399}
  {arXiv:1801.09399 [hep-ph]} \BibitemShut {NoStop}%
\bibitem [{\citenamefont {Crivellin}\ \emph {et~al.}(2017)\citenamefont
  {Crivellin}, \citenamefont {M\"uller},\ and\ \citenamefont
  {Ota}}]{Crivellin:2017zlb}%
  \BibitemOpen
  \bibfield  {author} {\bibinfo {author} {\bibfnamefont {A.}~\bibnamefont
  {Crivellin}}, \bibinfo {author} {\bibfnamefont {D.}~\bibnamefont {M\"uller}},
  \ and\ \bibinfo {author} {\bibfnamefont {T.}~\bibnamefont {Ota}},\ }\href
  {\doibase 10.1007/JHEP09(2017)040} {\bibfield  {journal} {\bibinfo  {journal}
  {JHEP}\ }\textbf {\bibinfo {volume} {09}},\ \bibinfo {pages} {040} (\bibinfo
  {year} {2017})},\ \Eprint {http://arxiv.org/abs/1703.09226} {arXiv:1703.09226
  [hep-ph]} \BibitemShut {NoStop}%
\bibitem [{\citenamefont {Ko}\ \emph {et~al.}(2017)\citenamefont {Ko},
  \citenamefont {Omura}, \citenamefont {Shigekami},\ and\ \citenamefont
  {Yu}}]{Ko:2017lzd}%
  \BibitemOpen
  \bibfield  {author} {\bibinfo {author} {\bibfnamefont {P.}~\bibnamefont
  {Ko}}, \bibinfo {author} {\bibfnamefont {Y.}~\bibnamefont {Omura}}, \bibinfo
  {author} {\bibfnamefont {Y.}~\bibnamefont {Shigekami}}, \ and\ \bibinfo
  {author} {\bibfnamefont {C.}~\bibnamefont {Yu}},\ }\href {\doibase
  10.1103/PhysRevD.95.115040} {\bibfield  {journal} {\bibinfo  {journal} {Phys.
  Rev. D}\ }\textbf {\bibinfo {volume} {95}},\ \bibinfo {pages} {115040}
  (\bibinfo {year} {2017})},\ \Eprint {http://arxiv.org/abs/1702.08666}
  {arXiv:1702.08666 [hep-ph]} \BibitemShut {NoStop}%
\bibitem [{\citenamefont {Kindra}\ and\ \citenamefont
  {Mahajan}(2018)}]{Kindra:2018ayz}%
  \BibitemOpen
  \bibfield  {author} {\bibinfo {author} {\bibfnamefont {B.}~\bibnamefont
  {Kindra}}\ and\ \bibinfo {author} {\bibfnamefont {N.}~\bibnamefont
  {Mahajan}},\ }\href {\doibase 10.1103/PhysRevD.98.094012} {\bibfield
  {journal} {\bibinfo  {journal} {Phys. Rev. D}\ }\textbf {\bibinfo {volume}
  {98}},\ \bibinfo {pages} {094012} (\bibinfo {year} {2018})},\ \Eprint
  {http://arxiv.org/abs/1803.05876} {arXiv:1803.05876 [hep-ph]} \BibitemShut
  {NoStop}%
\bibitem [{\citenamefont {Matias}(2012)}]{Matias:2012qz}%
  \BibitemOpen
  \bibfield  {author} {\bibinfo {author} {\bibfnamefont {J.}~\bibnamefont
  {Matias}},\ }\href {\doibase 10.1103/PhysRevD.86.094024} {\bibfield
  {journal} {\bibinfo  {journal} {Phys. Rev. D}\ }\textbf {\bibinfo {volume}
  {86}},\ \bibinfo {pages} {094024} (\bibinfo {year} {2012})},\ \Eprint
  {http://arxiv.org/abs/1209.1525} {arXiv:1209.1525 [hep-ph]} \BibitemShut
  {NoStop}%
\bibitem [{\citenamefont {Dubni\v{c}ka}\ \emph
  {et~al.}(2016{\natexlab{b}})\citenamefont {Dubni\v{c}ka}, \citenamefont
  {Dubni\v{c}kov\'{a}}, \citenamefont {Habyl}, \citenamefont {Ivanov},
  \citenamefont {Liptaj},\ and\ \citenamefont {Nurbakova}}]{Dubnicka:2015iwg}%
  \BibitemOpen
  \bibfield  {author} {\bibinfo {author} {\bibfnamefont {S.}~\bibnamefont
  {Dubni\v{c}ka}}, \bibinfo {author} {\bibfnamefont {A.~Z.}\ \bibnamefont
  {Dubni\v{c}kov\'{a}}}, \bibinfo {author} {\bibfnamefont {N.}~\bibnamefont
  {Habyl}}, \bibinfo {author} {\bibfnamefont {M.~A.}\ \bibnamefont {Ivanov}},
  \bibinfo {author} {\bibfnamefont {A.}~\bibnamefont {Liptaj}}, \ and\ \bibinfo
  {author} {\bibfnamefont {G.~S.}\ \bibnamefont {Nurbakova}},\ }\href {\doibase
  10.1007/s00601-015-1034-4} {\bibfield  {journal} {\bibinfo  {journal} {Few
  Body Syst.}\ }\textbf {\bibinfo {volume} {57}},\ \bibinfo {pages} {121}
  (\bibinfo {year} {2016}{\natexlab{b}})},\ \Eprint
  {http://arxiv.org/abs/1511.04887} {arXiv:1511.04887 [hep-ph]} \BibitemShut
  {NoStop}%
\bibitem [{\citenamefont {Wirbel}\ \emph {et~al.}(1985)\citenamefont {Wirbel},
  \citenamefont {Stech},\ and\ \citenamefont {Bauer}}]{Wirbel:1985ji}%
  \BibitemOpen
  \bibfield  {author} {\bibinfo {author} {\bibfnamefont {M.}~\bibnamefont
  {Wirbel}}, \bibinfo {author} {\bibfnamefont {B.}~\bibnamefont {Stech}}, \
  and\ \bibinfo {author} {\bibfnamefont {M.}~\bibnamefont {Bauer}},\ }\href
  {\doibase 10.1007/BF01560299} {\bibfield  {journal} {\bibinfo  {journal} {Z.
  Phys. C}\ }\textbf {\bibinfo {volume} {29}},\ \bibinfo {pages} {637}
  (\bibinfo {year} {1985})}\BibitemShut {NoStop}%
\bibitem [{\citenamefont {Blake}\ \emph {et~al.}(2017)\citenamefont {Blake},
  \citenamefont {Lanfranchi},\ and\ \citenamefont {Straub}}]{Blake:2016olu}%
  \BibitemOpen
  \bibfield  {author} {\bibinfo {author} {\bibfnamefont {T.}~\bibnamefont
  {Blake}}, \bibinfo {author} {\bibfnamefont {G.}~\bibnamefont {Lanfranchi}}, \
  and\ \bibinfo {author} {\bibfnamefont {D.~M.}\ \bibnamefont {Straub}},\
  }\href {\doibase 10.1016/j.ppnp.2016.10.001} {\bibfield  {journal} {\bibinfo
  {journal} {Prog. Part. Nucl. Phys.}\ }\textbf {\bibinfo {volume} {92}},\
  \bibinfo {pages} {50} (\bibinfo {year} {2017})},\ \Eprint
  {http://arxiv.org/abs/1606.00916} {arXiv:1606.00916 [hep-ph]} \BibitemShut
  {NoStop}%
\end{thebibliography}%

\end{document}